\documentclass[a4paper,fleqn,usenatbib]{mnras}

\usepackage{newtxtext,newtxmath}
\usepackage[T1]{fontenc}
\usepackage{ae,aecompl}

\usepackage{graphicx}	% Including figure files
\usepackage{amsmath}	% Advanced maths commands
\usepackage{amssymb}	% Extra maths symbols
\usepackage{pdflscape}

\title[The X-ray continuum time-lags and intrinsic coherence in AGN]{The X-ray continuum time-lags and intrinsic coherence in AGN}

\author[A. Epitropakis \& I. E. Papadakis]{
A. Epitropakis,$^{1,\,2}$\thanks{E-mail: epitrop@physics.uoc.gr}
and I. E. Papadakis$^{2,\,3}$
\\
$^{1}$Department of Physics, University of Crete, 71003 Heraklion, Greece\\
$^{2}$IESL, Foundation for Research and Technology-Hellas, GR-71110 Heraklion, Crete, Greece\\
$^{3}$Department of Physics and Institute of Theoretical and Computational Physics, University of Crete, 71003 Heraklion, Greece
}

\date{Accepted XXX. Received YYY; in original form ZZZ}

\pubyear{2016}

\begin{document}
\label{firstpage}
\pagerange{\pageref{firstpage}--\pageref{lastpage}}
\maketitle

% Abstract of the paper
\begin{abstract}
We present the results from a systematic analysis of the X-ray continuum (`hard') time-lags and intrinsic coherence between the $2-4\,\mathrm{keV}$ and various energy bands in the $0.3-10\,\mathrm{keV}$ range, for ten X-ray bright and highly variable active galactic nuclei (AGN). We used all available archival \textit{XMM-Newton} data, and estimated the time-lags following Epitropakis \& Papadakis (2016). By performing extensive numerical simulations, we arrived at useful guidelines for computing intrinsic coherence estimates that are minimally biased, have known errors, and are (approximately) Gaussian distributed. Owing to the way we estimated the time-lags and intrinsic coherence, we were able to do a proper model fitting to the data. Regarding the continuum time-lags, we are able to demonstrate that they have a power-law dependence on frequency, with a slope of $-1$, and that their amplitude scales with the logarithm of the light-curve mean-energy ratio. We also find that their amplitude increases with the square root of the X-ray Eddington ratio. Regarding the intrinsic coherence, we found that it is approximately constant at low frequencies. It then decreases exponentially at frequencies higher than a characteristic `break frequency.' Both the low-frequency constant intrinsic-coherence value and the break frequency have a logarithmic dependence on the light-curve mean-energy ratio. Neither the low-frequency constant intrinsic-coherence value, nor the break frequency exhibit a universal scaling with either the central black hole mass, or the the X-ray Eddington ratio. Our results could constrain various theoretical models of AGN X-ray variability.
\end{abstract}

% Select between one and six entries from the list of approved keywords.
% Don't make up new ones.
\begin{keywords}
galaxies: active -- X-rays: galaxies -- accretion, accretion discs -- galaxies: Seyfert -- relativistic processes
\end{keywords}

%%%%%%%%%%%%%%%%%%%%%%%%%%%%%%%%%%%%%%%%%%%%%%%%%%

%%%%%%%%%%%%%%%%% BODY OF PAPER %%%%%%%%%%%%%%%%%%

\section{Introduction} \label{sec1}

According to the currently accepted paradigm, active galactic nuclei (AGN) contain a central, super-massive ($M_{\mathrm{BH}}\sim10^{6-9}\,\mathrm{M}_{\odot}$) black hole (BH), onto which matter accretes in a disc-like configuration. A fraction of the  low-energy photons emitted by the disc is assumed to be Compton up-scattered by a population of high-energy ($\sim100\,\mathrm{keV}$) electrons, which is often referred to as the X-ray corona. The Compton up-scattered disc photons form a power-law spectrum that is observed in the X-ray spectra of AGN. We will henceforth refer to this source as the X-ray source, and to its emission as the X-ray continuum emission.

In addition to spectral studies, X-ray variability studies can also provide valuable information that can be used to understand the nature of the X-ray source in AGN, which remains largely unknown. One particular, and rather powerful, variability analysis tool is the estimation of `time-lags' (delays) between temporal variations of the X-ray continuum emission in different energy bands (we will henceforth refer to these time-lags as the continuum time-lags). Such studies were first performed for X-ray binaries \citep[XRBs; e.g.][]{1989Natur.342..773M,1996MNRAS.280..227N,1999ApJ...510..874N,2000A&A...357L..17P}, which are also thought to be compact accreting systems, where the central BH has a mass of $\sim10\,\mathrm{M}_{\odot}$. Several characteristics of continuum time-lags in XRBs have since been established: a) variations in hard energy-bands are delayed with respect to variations in softer energy-bands, b) the time-lags have an approximately power-law dependence on temporal frequency (with their magnitude decreasing with increasing frequency), and c) the magnitude of the continuum time-lags at a given frequency has an approximately log-linear dependence on the energy separation of the light curves. Continuum time-lags with similar characteristics were later reported in several AGN as well \citep[e.g.][]{2001ApJ...554L.133P,2004MNRAS.348..783M,2006MNRAS.372..401A,2008MNRAS.388..211A,2009ApJ...700.1042S}.

Apart from the time-lags, an additional (potentially useful) tool in understanding the nature of the X-ray variability in compact accreting objects is the so-called coherence function, which is a measure of the degree of correlation between variations in two light curves as a function of temporal frequency \citep[][henceforth VN97]{1997ApJ...474L..43V}. When correcting for the effect of Poisson noise, the intrinsic coherence in XRBs is generally observed to be frequency- and energy-dependent, remaining close to unity for a wide range of frequencies and for light curves with a small energy separation (VN97). This behaviour is observed in AGN as well, although, contrary to time-lags, quantitative studies of the energy- and frequency-dependence of the intrinsic coherence are limited. This is partly because the methods of intrinsic coherence estimation have not been established as well as the time-lag estimation methods, and because its interpretation is less straight-forward.

The main aim of our work is to perform a systematic study of the energy- and frequency-dependence of the continuum time-lags and of the intrinsic coherence in AGN. To this end, we chose a sample of ten X-ray bright and variable AGN that have been observed many times by \textit{XMM-Newton}. We relied on the work of \citet[][EP16 hereafter]{2016A&A...591A.113E} to calculate time-lags that are minimally biased, have known errors, and are approximately Gaussian distributed. Following their work, in this paper we also present the results from an extensive study of the statistical properties of the traditional, Fourier-based intrinsic coherence estimator. We provide practical guidelines that can be used to compute intrinsic coherence estimates that are minimally biased, have known errors, and are approximately Gaussian distributed.

We used all the existing \textit{XMM-Newton} archival data for these objects to estimate the time-lags and intrinsic coherence between light curves in various energy bands. Our results provide a quantitative description of the dependence of the time-lags and intrinsic coherence on frequency and energy in AGN. We also provide results regarding their scaling with BH mass and (X-ray) Eddington ratio. Our results could be used to constrain theoretical models for the X-ray variability in AGN.

\section{Observations and data reduction} \label{sec2}

\begin{table*}
\caption{\textit{XMM-Newton} observations log. Sources are listed in order of decreasing net exposure.}
\label{table1}      
\centering          
\begin{tabular}{c c c | c c c}     % 7 columns 
\hline\hline       
                      % To combine 4 columns into a single one 
(1) & (2) & (3) & (1) & (2) & (3) \\
Source & Obs. ID & Exp. & Source & Obs. ID & Exp. \\
 & & (ksec) & & & (ksec) \\
\hline
\underline{1H 0707--495} & & & \underline{Mrk 766} & & \\
$z=0.040568$ & 0110890201 & 40.6 & $z=0.012929$ & 0109141301 & 128.5 \\
$M_{\mathrm{BH}}=5.2\pm3.2M_6$ & 0148010301 & 76.4 & $M_{\mathrm{BH}}=1.76^{+1.56}_{-1.40}M_6$ & 0304030101 & 94.8 \\
\citet{2016ApJ...819L..19P} & 0506200201 & 38.6 & \citet{2009ApJ...705..199B} & 0304030301 & 98.4 \\
$L_{2-10}=1.4L_{43}$ & 0506200301 & 38.6 & $L_{2-10}=1.1L_{43}$ & 0304030401 & 94.1 \\
$\lambda_\mathrm{X}=0.04\pm0.01$ & 0506200401 & 40.6 & $\lambda_\mathrm{X}=0.042\pm0.036$ & 0304030501 & 94.2 \\
 & 0506200501 & 40.8 & & 0304030601 & 85.2 \\
 & 0511580101 & 112.0 & & 0304030701 & 29.1 \\
 & 0511580201 & 99.6 & \underline{Ark 564} &  &  \\
 & 0511580301 & 85.7 & $z=0.024684$ & 0006810101 & 10.6 \\
 & 0511580401 & 81.3 & $M_{\mathrm{BH}}=2.32\pm0.41M_6\,^{a}$ & 0206400101 & 98.9 \\
 & 0554710801 & 59.6 & $L_{2-10}=2.2L_{43}$ & 0670130201 & 59.0 \\
 & 0653510301 & 111.9 & $\lambda_\mathrm{X}=0.06\pm0.01$ & 0670130301 & 55.4 \\
 & 0653510401 & 122.7 & & 0670130401 & 56.1 \\
 & 0653510501 & 115.2 & & 0670130501 & 66.8 \\
 & 0653510601 & 113.6 & & 0670130601 & 60.4 \\
\underline{MCG--6-30-15} &  & & & 0670130701 & 47.1 \\
$z=0.007749$ & 0029740101 & 80.5 & & 0670130801 & 57.7 \\
$M_{\mathrm{BH}}=1.6\pm0.4M_6$ & 0029740701 & 123.0 & & 0670130901 & 55.4 \\
\citet{2016ApJ...830..136B} & 0029740801 & 124.1 & \underline{IRAS 13224--3809} &  &  \\
$L_{2-10}=0.6L_{44}$ & 0111570101 & 43.1 & $z=0.065799$ & 0110890101 & 60.8 \\
$\lambda_\mathrm{X}=0.025\pm0.006$ & 0111570201 & 52.9 & $M_{\mathrm{BH}}=5.75\pm0.82M_6$ & 0673580101 & 57.0 \\
 & 0693781201 & 131.6 & \citet{2005ApJ...618L..83Z} & 0673580201 & 86.7 \\
 & 0693781301 & 130.0 & $L_{2-10}=0.7L_{43}$ & 0673580301 & 84.3 \\
 & 0693781401 & 48.4 & $\lambda_\mathrm{X}=0.0058\pm0.0008$ & 0673580401 & 114.4 \\
\underline{NGC 4051} &  & &  \\
$z=0.002336$ & 0109141401 & 105.9 & \underline{MCG--5-23-16} &  &  \\
$M_{\mathrm{BH}}=1.73^{+0.55}_{-0.52}M_6$ & 0157560101 & 49.9 & $z=0.008486$ & 0112830401 & 21.6 \\
\citet{2010ApJ...721..715D} & 0606320101 & 45.2 & $M_{\mathrm{BH}}=7.9\pm0.4M_6$ & 0302850201 & 110.7 \\
$L_{2-10}=0.04L_{43}$ & 0606320201 & 44.0 & Oliva et al. (1995) & 0727960101 & 127.5 \\
$\lambda_\mathrm{X}=0.0016\pm0.0005$ & 0606320301 & 24.6 & $L_{2-10}=1.7L_{43}$ & 0727960201 & 133.2 \\
 & 0606320401 & 24.1 & $\lambda_\mathrm{X}=0.014\pm0.007$ &  &  \\
 & 0606321301 & 30.1 &  \\
 & 0606321401 & 39.2 & \underline{NGC 7314} &  &  \\
 & 0606321501 & 35.6 & $z=0.004763$ & 0111790101 & 43.2 \\
 & 0606321601 & 41.4 & $M_{\mathrm{BH}}=0.87\pm0.45M_6$ & 0311190101 & 82.0 \\
 & 0606321701 & 38.3 & \citet{2013MNRAS.430L..49M} & 0725200101 & 125.3 \\
 & 0606321801 & 21.0 & $L_{2-10}=0.1L_{43}$ & 0725200301 & 130.5 \\
 & 0606322001 & 23.8 & $\lambda_\mathrm{X}=0.008\pm0.004$ &  &  \\
 & 0606322101 & 37.6 &  \\
 & 0606322201 & 36.3 & \underline{Mrk 335} &  &  \\
 & 0606322301 & 42.2 & $z=0.025785$ & 0101040101 & 31.6 \\
\underline{PKS 0558-403} &  & & $M_{\mathrm{BH}}=26\pm8M_6$ & 0306870101 & 126.5 \\
$z=0.137200$ & 0117710601 & 15.9 & \citet{2012ApJ...744L...4G} & 0510010701 & 16.7 \\
$M_{\mathrm{BH}}=250^{+50}_{-190}$ & 0117710701 & 19.4 & $L_{2-10}=1.4L_{43}$ & 0600540501 & 80.7 \\
\citet{2010ApJ...717.1243G} & 0555170201 & 113.7 & $\lambda_\mathrm{X}=0.004\pm0.001$ & 0600540601 & 114.1 \\
$L_{2-10}=70L_{43}$ & 0555170301 & 120.5 & & &  \\
$\lambda_\mathrm{X}=0.018\pm0.009$ & 0555170401 & 123.3 & & &  \\
 & 0555170501 & 124.1 & & &  \\
 & 0555170601 & 115.3 & & &  \\
\hline
\end{tabular}
\\
$^a$ Estimated using equation 5 in \citet{2006ApJ...641..689V}, for the $\mathrm{FWHM}(\mathrm{H}\beta)$ and $\lambda L_\lambda(5100\,\AA)$ values in \citet{2004ApJ...602..635R}
\\
\end{table*}

Table \ref{table1} lists the details of the \textit{XMM-Newton} observations we used. Column 1 lists the source name, redshift, $z$ (taken from the NASA/IPAC Extragalactic Database (NED)), central BH mass, $M_{\mathrm{BH}}$, estimate in units of $M_6=10^6\,\mathrm{M}_{\odot}$ (along with the respective reference below the listed value), and the mean $2-10\,\mathrm{keV}$ luminosity, $L_{2-10}$, in units of $L_{43}=10^{43}\,\mathrm{erg}\,\mathrm{s}^{-1}$. The luminosity was determined using the mean $2-10\,\mathrm{keV}$ fluxes listed in the \textit{RXTE} AGN Timing \& Spectral Database, and the respective luminosity distance values listed in the NED (assuming a $\Lambda$-CDM cosmology with $H_0=73\,\mathrm{km}\,\mathrm{sec}^{-1}\,\mathrm{Mpc}^{-1}$, $\Omega_{\mathrm{m}}=0.27$, and $\Omega_\Lambda=0.73$). The only exception is IRAS 13324--3809, which is not listed in the former database, for which we used the mean flux reported by \citet{2002A&A...390...65D}. In the same column we also list the ratio of the $2-10\,\mathrm{keV}$ luminosity over the Eddington luminosity (henceforth, the X-ray Eddington ratio, $\lambda_\mathrm{X}$). Columns 2 and 3 of the same figure show the identification number (ID) of each observation and net exposure in units of $\mathrm{ksec}$, respectively.

We processed data from the \textit{XMM-Newton} satellite using the Scientific Analysis System \citep[SAS, v. 14.0.0;][]{2004ASPC..314..759G}. We only used EPIC-pn \citep[][]{2001A&A...365L..18S} data. Source and background light curves were extracted from circular regions on the CCD. The source regions had a fixed radius of 800 pixels ($40^{\prime\prime}$) centred on the source coordinates listed on the NASA/IPAC Extragalactic Database. The positions and radii of the background regions were determined by placing them sufficiently far from the location of the source, but within the boundaries of the same CCD chip.

Source and background light curves with a bin size of $10\,\mathrm{sec}$ were extracted, using SAS command evselect, in the following energy bands: $0.3-0.5$, $0.5-0.7$, $0.7-1$, $0.3-1$, $1-2$, $2-4$, $4-5$, $5-7$, and $7-10\,\mathrm{keV}$. We included the criteria PATTERN==0--4 and FLAG==0 in the extraction process, which select only single- and double-pixel events and reject `bad' pixels from the edges of the detector CCD chips. Periods of high flaring background activity owing to solar activity were determined by observing the $10-12\,\mathrm{keV}$ light curves (which contain very few source photons) extracted from the whole surface of the detector, and subsequently excluded during the source and background light curve extraction process.

We checked all source light curves for pile-up using the SAS task epatplot, and found that only observations 0670130201, 0670130501, and 0670130901 of Ark 564 are affected. For those observations we used annular instead of circular source regions with inner radii of 280, 200, and 250 pixels (the outer radii were held at 800 pixels), respectively, which we found to adequately reduce the effects of pile-up.

The background light curves were then subtracted from the corresponding source light curves using the SAS command epiclccorr. Most of the resulting light curves were continuously sampled, except for a few cases that contained a small ($\lesssim5$ per cent of the total number of points in the light curve) number of missing points. These were either randomly distributed throughout the duration of an observation, or appeared in groups of $\lesssim100$ points. We replaced the missing points by linear interpolation, with the addition of the appropriate Poisson noise.

%%%%%%%%%%%%%%. SECTION 3 %%%%%%%%%%%%%%%%%
%%%%%%%%%%%%%%%%%%%%%%%%%%%%%%%%
\section{Time-lag estimation} \label{sec3}

%%%%%%%%%%%%%%%%%%%% TABLE 2 %%%%%%%%%%%%%%%%%%%%%
\begin{table*}
\caption{The number of light curve segments, $m$, mean count rate in each energy band, and the frequency $\nu_{\mathrm{crit}}$ ($\nu_{\mathrm{max}}$) below which time-lags (intrinsic coherence) can be reliably estimated.}
\label{table2}
%\centering
\begin{tabular}{c c c c  | c c c c }
\hline\hline
(1) & (2) & (3) & (4) & (1) & (2) & (3) & (4) \\
Source & $E$ ($\overline{E}$) & Mean c.r. & $\nu_{\mathrm{crit}}/\nu_{\mathrm{max}}$ & Source & $E$ ($\overline{E}$) & Mean c.r. & $\nu_{\mathrm{crit}}$/$\nu_{\mathrm{max}}$ \\
& ($\mathrm{keV}$) & ($\mathrm{cts/sec}$) &  ($10^{-4}\,\mathrm{Hz}/10^{-4}\,\mathrm{Hz}$) &  & ($\mathrm{keV}$) & ($\mathrm{cts/sec}$) & ($10^{-4}\,\mathrm{Hz}/10^{-4}\,\mathrm{Hz}$)\\
\hline
& $0.3-0.5$ (0.40) & 1.454 & 13.9/6.8 & & $0.3-0.5$ (0.40) & 14.080 & 13.0/6.1 \\
& $0.5-0.7$ (0.60) & 1.134 & 15.1/6.9 & & $0.5-0.7$ (0.60) & 10.388 & 14.0/6.2 \\
& $0.7-1.0$ (0.85) & 1.003 & 17.5/7.2 & & $0.7-1.0$ (0.85) & 9.362 & 16.4/6.4 \\
\underline{1H 0707--495} & $1.0-2.0$ (1.50) & 0.528 & 22.0/7.5 & \underline{Ark 564} & $1.0-2.0$ (1.50) & 9.149 & 19.6/6.6 \\
$m=51$ & $2.0-4.0$ (3.00) & 0.106 &  & $m=22$ & $2.0-4.0$ (3.00) & 2.115 &  \\
 & $4.0-5.0$ (4.50) & 0.018 & 9.0/1.8 & & $4.0-5.0$ (4.50) & 0.329 & 10.1/1.8 \\
& $5.0-7.0$ (6.00) & 0.020 & 7.2/1.8 & & $5.0-7.0$ (6.00) & 0.326 & 11.4/1.8 \\
& $7.0-10$ (8.50) & 0.004 & 2.6/--- & & $7.0-10$ (8.50) & 0.119 & 4.3/0.9 \\
\hline
& $0.3-0.5$ (0.40) & 5.909 & 11.8/7.0 & & $0.3-0.5$ (0.40) & 0.824 & 4.2/2.7 \\
& $0.5-0.7$ (0.60) & 4.826 & 12.8/8.2 & & $0.5-0.7$ (0.60) & 0.560 & 4.4/2.7 \\
& $0.7-1.0$ (0.85) & 3.538 & 14.1/8.4 & & $0.7-1.0$ (0.85) & 0.429 & 6.7/3.1 \\
\underline{MCG--6-30-15} & $1.0-2.0$ (1.50) & 7.019 & 17.2/8.9 & \underline{IRAS} & $1.0-2.0$ (1.50) & 0.233 & 8.8/3.3 \\
$m=34$ & $2.0-4.0$ (3.00) & 3.326 &  & \underline{13224-3809} & $2.0-4.0$ (3.00) & 0.051 &   \\
 & $4.0-5.0$ (4.50) & 0.752 & 12.1/5.4 & $m=18$ & $4.0-5.0$ (4.50) & 0.010 & 2.0/0.8 \\
& $5.0-7.0$ (6.00) & 0.889 & 11.2/4.1 & & $5.0-7.0$ (6.00) & 0.012 & 2.2/0.8 \\
& $7.0-10$ (8.50) & 0.370 & 8.4/1.8 & & $7.0-10$ (8.50) & 0.003 & --- \\
\hline
& $0.3-0.5$ (0.40) & 5.447 & 28.0/14.2 & &   \\
& $0.5-0.7$ (0.60) & 3.671 & 28.4/14.2 & & $0.3-1.0$ (0.65) & 0.603 & 1.8/0.8 \\
& $0.7-1.0$ (0.85) & 2.422 & 32.9/14.7 & &   \\
\underline{NGC 4051} & $1.0-2.0$ (1.50) & 2.789 & 37.4/15.2 & \underline{MCG--5-23-16} & $1.0-2.0$ (1.50) & 5.642 & 4.4/3.2 \\
$m=25$ & $2.0-4.0$ (3.00) & 1.177 &  & $m=18$ & $2.0-4.0$ (3.00) & 6.860 &  \\
& $4.0-5.0$ (4.50) & 0.298 & 16.5/5.1 &  & $4.0-5.0$ (4.50) & 1.928 & 3.5/2.6 \\
& $5.0-7.0$ (6.00) & 0.383 & 12.8/5.5 & & $5.0-7.0$ (6.00) & 2.484 & 3.7/2.1 \\
& $7.0-10$ (8.50) & 0.159 & 7.1/3.1 & & $7.0-10$ (8.50) & 1.188 & 3.2/1.5 \\
\hline
& $0.3-0.5$ (0.40) & 5.059 & 5.1/2.9 & & $0.3-0.5$ (0.40) & 0.079 & 1.9/--- \\
& $0.5-0.7$ (0.60) & 3.503 & 5.3/2.9 & & $0.5-0.7$ (0.60) & 0.096 & 4.1/0.9 \\
& $0.7-1.0$ (0.85) & 3.350 & 6.3/3.0 & & $0.7-1.0$ (0.85) & 0.284 & 7.6/2.5 \\
\underline{PKS 0558--504} & $1.0-2.0$ (1.50) & 3.980 & 6.8/3.1 & \underline{NGC 7314} & $1.0-2.0$ (1.50) & 2.075 & 15.5/9.1 \\
$m=28$ & $2.0-4.0$ (3.00) & 1.207 &  & $m=18$ & $2.0-4.0$ (3.00) & 1.621 &  \\
 & $4.0-5.0$ (4.50) & 0.224 & 2.8/1.5 & & $4.0-5.0$ (4.50) & 0.406 & 10.1/3.3 \\
& $5.0-7.0$ (6.00) & 0.241 & 2.2/0.8 & & $5.0-7.0$ (6.00) & 0.505 & 10.8/3.4 \\
& $7.0-10$ (8.50) & 0.109 & 2.1/--- & & $7.0-10$ (8.50) & 0.236 & 0.6/1.7 \\
\hline
& $0.3-0.5$ (0.40) & 4.097 & 9.2/6.9 & & $0.3-0.5$ (0.40) & 3.777 & 4.2/2.7 \\
& $0.5-0.7$ (0.60) & 2.755 & 11.5/7.4 & & $0.5-0.7$ (0.60) & 2.584 & 5.3/2.9 \\
& $0.7-1.0$ (0.85) & 2.165 & 11.8 7.5 & & $0.7-1.0$ (0.85) & 2.257 & 5.7/3.0 \\
\underline{Mrk 766} & $1.0-2.0$ (1.50) & 3.322 & 13.8/7.9 & \underline{Mrk 335} & $1.0-2.0$ (1.50) & 2.689 & 6.0/3.0 \\
 $m=27$ & $2.0-4.0$ (3.00) & 1.284 &  & $m=16$ & $2.0-4.0$ (3.00) & 0.881 & \\
& $4.0-5.0$ (4.50) & 0.270 & 7.2/3.1 & & $4.0-5.0$ (4.50) & 0.178 & 3.6/2.1 \\
& $5.0-7.0$ (6.00) & 0.320 & 7.8/1.8 & & $5.0-7.0$ (6.00) & 0.214 & 3.4/1.5 \\
& $7.0-10$ (8.50) & 0.132 & 4.8/0.9 & & $7.0-10$ (8.50) & 0.091& 2.5/ --- \\
 \hline
\end{tabular}
\end{table*}

We calculated the time-lag estimates between light curves in seven energy bands, and the light curves in the $2-4\,\mathrm{keV}$ energy band (henceforth, the reference band; the reason for this particular reference band choice is explained in Section \ref{sec4}). The energy bands, along with their mean energy, $\overline{E}$, are listed in column 2 of Table \ref{table2}. In the case of MCG--5-23-16, which has a very low count rate at energies $\lesssim1\,\mathrm{keV}$ (owing to the fact that it is an absorbed AGN), we used light curves in the entire $0.3-1\,\mathrm{keV}$ energy band. We chose the energy bands to be as narrow as possible to maximise energy resolution, while at the same time maintaining a relatively high mean count rate to minimise Poisson noise effects. We also considered the $0.3-1$ vs. $2-4\,\mathrm{keV}$ band time-lags to determine the frequency range over which we fitted the observed time-lags at low frequencies (see Section \ref{sec31}). We used light curves with a bin size of $\Delta t=10\,\mathrm{s}$. The time-lags were estimated following the prescription of EP16 to ensure that they (approximately) follow a Gaussian distribution with know errors. We provide below a short description of our methodology.

We first partitioned all available light curves in each energy band into $m$ segments of duration $T=20\,\mathrm{ksec}$ (the number of segments is listed in column 1 of Table \ref{table2}). For a given pair of segments we calculated the so-called cross-periodogram at the frequencies $\nu_p=p/N\Delta t$, where $p=1,2,\ldots,N/2$ ($N=2000$ is the total number of points in each segment). The cross-periodogram is an estimator of the cross-spectrum (CS), which, in turn, is a measure of the correlation between two random signals \citep[][henceforth P81]{Priestley:81}. Our final estimate for the CS, $\hat{C}_{xy}(\nu_p)$, was obtained by averaging the $m$ individual cross-periodograms at each frequency. We did not average $\hat{C}_{xy}(\nu_p)$ over neighbouring frequencies, as this can introduce a bias at low frequencies (EP16). We only considered frequencies $\le\nu_{\mathrm{Nyq}}/2$ ($=2.5\times10^{-2}\,\mathrm{Hz}$ in our case) to minimise the effects of light-curve binning on the time-lag estimates.

The time-lag at each frequency is defined as the argument of the CS, divided by the angular frequency (P81). Following standard practice, we thus used
\noindent
\begin{equation} \label{eq1}
\hat{\tau}_{xy}(\nu_p)\equiv\frac{1}{2\pi\nu_p}\mathrm{arg}[\hat{C}_{xy}(\nu_p)]
\end{equation}
\noindent
and
\noindent
\begin{equation} \label{eq2}
\hat{\sigma}_{\hat{\tau}}(\nu_p)\equiv\frac{1}{2\pi\nu_p}\frac{1}{\sqrt{2m}}\sqrt{\frac{1-\hat{\gamma}^2_{xy}(\nu_p)}{\hat{\gamma}^2_{xy}(\nu_p)}}
\end{equation}
\noindent
as our estimates of the time-lags and their corresponding error, respectively. The quantity $\hat{\gamma}^2_{xy}(\nu_p)$ is the so-called coherence estimate, which is defined as (P81; VN97)
\noindent
\begin{equation} \label{eq3}
\hat{\gamma}^2_{xy}(\nu_p)\equiv\frac{|\hat{C}_{xy}(\nu_p)|^2}{\hat{P}_x(\nu_p)\hat{P}_y(\nu_p)}.
\end{equation}
\noindent
$\hat{P}_x(\nu_p)$ and $\hat{P}_y(\nu_p)$ are the traditional periodograms of the two light curves, which are also calculated by averaging over $m$ segments. The coherence function between two random processes is a measure of the degree of linear correlation between their corresponding sinusoidal components at each frequency. As we explain in detail in Appendix \ref{appa}, the coherence estimate defined by equation \ref{eq3} is a biased estimator of the intrinsic coherence of the measured processes. Nevertheless, its estimation plays a crucial role in the determination of reliable time-lag estimates, as demonstrated by EP16.

%%%%%%%%%%%%%%%%%%% FIG1 %%%%%%%%%%%%%%%%%%%%%%%%%
\begin{figure}
\centering
\includegraphics[width=\hsize]{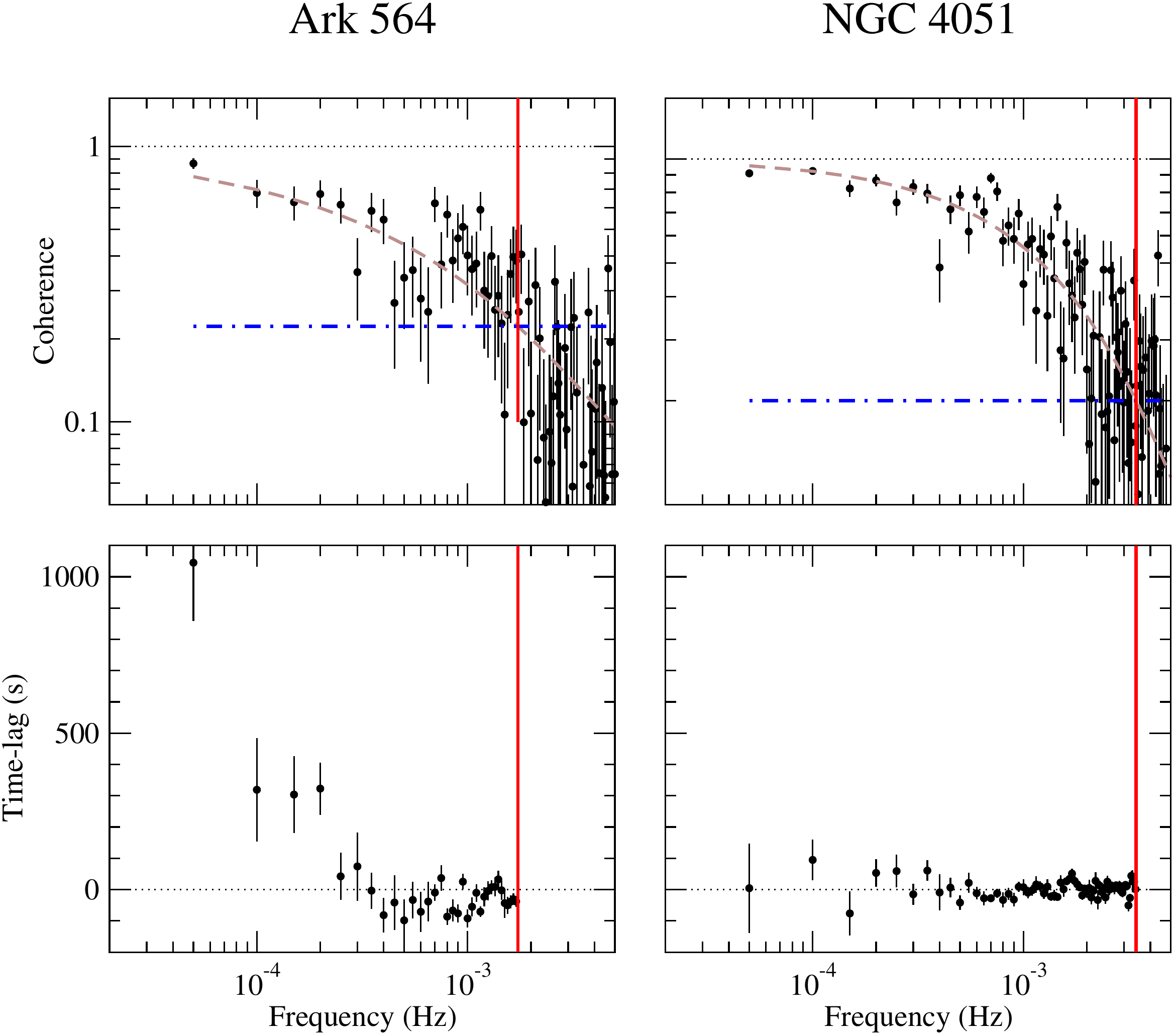}
\caption{The $0.3-1$ vs. $2-4\,\mathrm{keV}$ coherence and time-lag spectra (top and bottom panels, respectively) of Ark 564 (left column), and NGC 4051 (right column). The dashed brown lines in the top panels shows the best-fitting model to the sample coherence, the horizontal blue dotted-dashed lines in the top panels indicate the constant coherence value $1.2/(1+0.2m)$, and the continuous red vertical lines indicates $\nu_{\rm crit}$ (see the text for more details).}
\label{fig1}
\end{figure}

%%%%%%%%%%%%%%%%%%% FIG2 %%%%%%%%%%%%%%%%%%%%%%%%%
\begin{figure}
\centering
\includegraphics[width=\hsize]{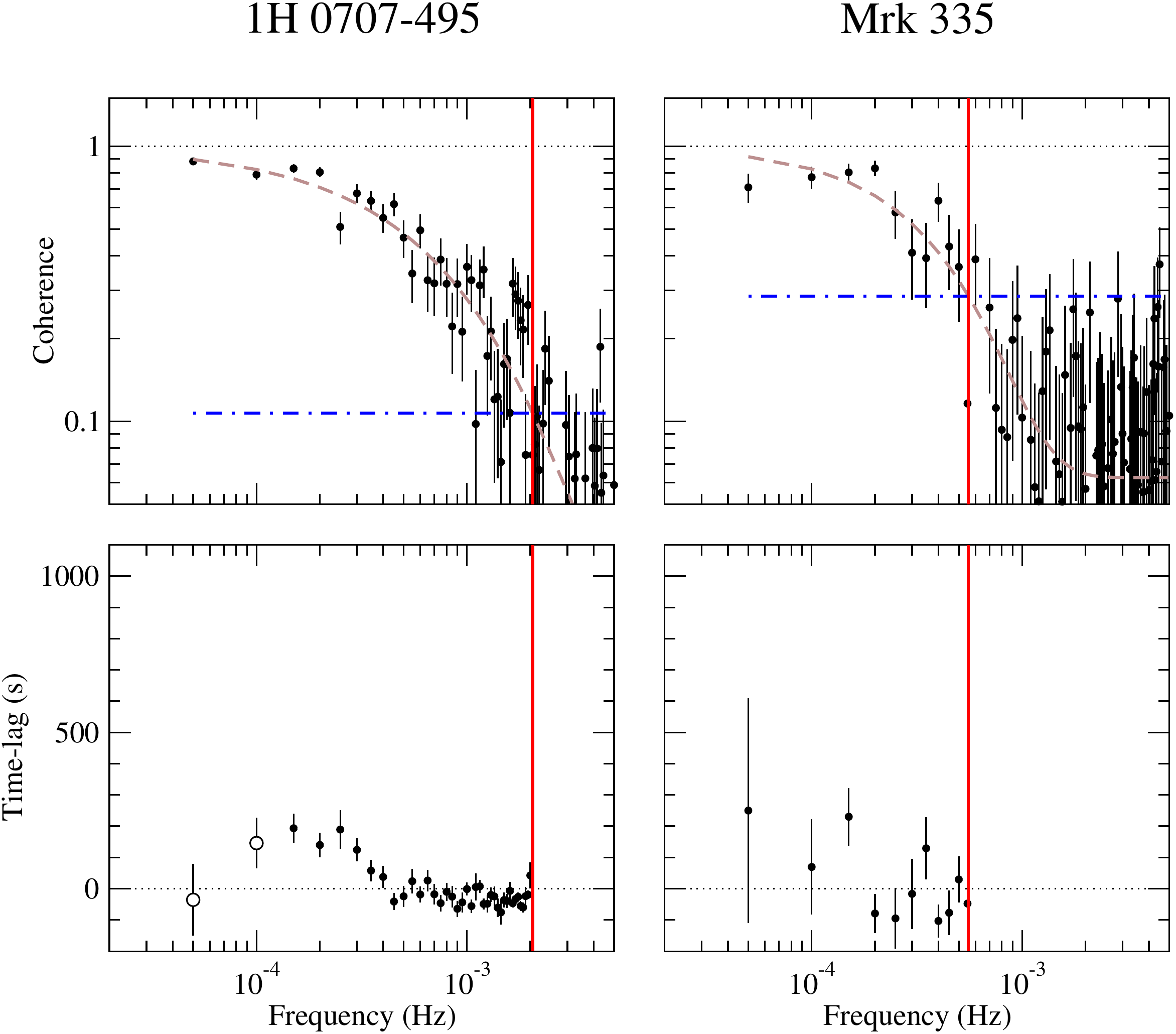}
\caption{As in Fig. \ref{fig1}, for 1H 0707--495 (left column), and Mrk 335 (right column).}
\label{fig2}
\end{figure}

Figure \ref{fig1} shows the $0.3-1$ vs. $2-4\,\mathrm{keV}$ coherence and time-lag estimates (top and bottom panels, respectively) of Ark 564 and NGC 4051. Both sources are X-ray bright and highly variable. Figure \ref{fig2} shows the same results for 1H 0707--495 and Mrk 335 (two sources that are fainter, and, in the case of Mrk 355, less variable). They were calculated using equations \ref{eq3} and \ref{eq1}, respectively. The coherence estimates decrease with increasing frequency in all cases. EP16 showed that, in the presence of measurement errors, the coherence estimates converge to the constant $1/m$ at frequencies where the amplitude of the experimental noise dominates over the amplitude of the intrinsic variations. In fact, EP16 showed that, if the measured processes are intrinsically coherent (i.e. the intrinsic coherence function is equal to unity at all frequencies), $\hat{\gamma}^2_{xy}(\nu_p)$ is well-fitted by a function of the form
\noindent
\begin{equation} \label{eq4}
\hat\gamma^2_{xy}(\nu)=\left(1-\frac{1}{m}\right)\mathrm{exp}[-(\nu/\nu_0)^q]+\frac{1}{m},
\end{equation}
\noindent
where $\nu_0$ and $q$ are constants. The brown dashed lines in the top panels of Figs. \ref{fig1} and \ref{fig2} show the best-fitting models to the coherence estimates.

The error of the time-lag estimates increases as the coherence decreases. Therefore, we expect that Poisson noise will severely affect the reliability of the time-lag estimates above a certain critical frequency, $\nu_{\mathrm{crit}}$. According to EP16, $\nu_{\mathrm{crit}}$ is the frequency at which the mean sample coherence function becomes equal to $1.2/(1+0.2m)$. At higher frequencies the analytic error prescription (equation \ref{eq2}) underestimates the true scatter of the time-lag estimates around their mean, their distribution becomes uniform, and their mean value converges to zero, irrespective of the intrinsic time-lag spectrum. At frequencies lower than $\nu_{\mathrm{crit}}$, and as long as $m\gtrsim20$, the time-lag estimates are unbiased, equation \ref{eq2} provides a reliable estimate of their true scatter around the mean, and their distribution is approximately Gaussian.

The horizontal blue dotted-dashed lines in the upper panels of Figs. \ref{fig1} and \ref{fig2} indicate the constant value of $1.2/(1+0.2m)$, and the vertical red lines indicate $\nu_{\rm crit}$, i.e. the frequency at which the best-fitting coherence model is equal to this value. EP16 showed that, for a given intrinsic PSD, $\nu_{\mathrm{crit}}$ decreases with decreasing S/N of the light curves (in particular, the one with the smaller mean count rate). As the S/N decreases, the frequency range over which we can obtain realiable time-lag estimates decreases. Therefore, it is not surprising that the critical frequency is highest (lowest) in the case of NGC 4051 (Mrk 355), respectively. However, S/N is not the only parameter that determines $\nu_{\rm crit}$. For example, despite the fact that the mean count rate of the $0.3-1$ and $2-4\,\mathrm{keV}$ light curves is significantly higher in the case of Mrk 335, $\nu_{\mathrm{crit},\,\text{1H 0707--495}}>\nu_{\mathrm{crit},\,\text{Mrk 335}}$. This is because 1H 0707--495 is much more variable. Consequently, the amplitude of the intrinsic variations is higher than the amplitude of the Poisson noise variations in the case of 1H 0707--495, even at frequencies that are four times higher than $\nu_{\mathrm{crit},\,\text{Mrk 335}}$.

We fitted the coherence estimates of each source (at all energy bands) to the exponential function given by equation \ref{eq4}. We then equated the best-fitting model to the constant $1.2/(1+0.2m)$ to estimate $\nu_{\mathrm{crit}}$. These values are listed in column 4 of Table \ref{table2}. The observed time-lag spectra, for all the sources in our sample, are shown in Figs. \ref{figb1}, \ref{figb3}, \ref{figb5}, \ref{figb7}, \ref{figb9}, \ref{figb11}, \ref{figb13}, \ref{figb15}, \ref{figb17}, and \ref{figb19} in Appendix \ref{appb}. The time-lag estimates in these figures are plotted at frequencies $\le\nu_{\mathrm{crit}}$ in each case. The time-lags were estimated such that a positive time-lag value indicates that variations in the reference band are delayed with respect to variations in the other energy band (and vice-versa).

The low frequency time-lags between the reference band and those at lower (higher) energies are positive (negative). This shows that X-ray continuum variations in hard energy-bands are always delayed with respect to variations in softer energy-bands. In all cases, the low frequency time-lag amplitude increases with increasing energy separation (the limits in the vertical axis are the same for all sample time-lag spectra in each figure). The frequency range of the $7-10$ vs. $2-4\,\mathrm{keV}$ time-lags is the smallest among all sample time-lag spectra. This is because the count rate of the $7-10\,\mathrm{keV}$ light curves is very small. We could not estimate the soft band time-lags seperately in the case of MCG--5-23-16, because the count rate of the corresponding light curves is almost zero (because of absorption). For this source we hence utilised the entire $0.3-1\,\mathrm{keV}$ energy band, and estimated the corresponding $0.3-1$ vs. $2-4\,\mathrm{keV}$ time-lags. The $0.3-0.5$ vs. $2-4\,\mathrm{keV}$ time-lags of NGC 7314 are poorly determined for the same reason. On the other hand, the hard band time-lags are poorly determined in IRAS 13224--3809, because the source is not particularly bright and has a very soft energy spectrum, hence the count rate at energies $\gtrsim4\,\mathrm{keV}$ is very small.

%#######################TABLE 3 %%%%%%%%%%%%%%%%%%%%
\begin{table*}
\caption{Best-fitting results for the power-law time-lag model. The model is defined by equation \ref{eq5}, and the best-fitting models are shown as red dashed lines in the relevant Appendix \ref{appb} figures.}
\label{table3}
\centering
\begin{tabular}{c c c c | c c c c}
\hline\hline
(1) & (2) & (3) & (4) & (1) & (2) & (3) & (4) \\
Source & $\overline{E}/(3\,\mathrm{keV})$ & $[\nu_{\rm low},\nu_{\rm high}]$ & $A(\overline{E},3\,\mathrm{keV})$ & Source & $\overline{E}/(3\,\mathrm{keV})$ &  $[\nu_{\rm low},\nu_{\rm high}]$ & $A(\overline{E},3\,\mathrm{keV})$ \\
& & ($10^{-4}\,\mathrm{Hz}$) & ($\mathrm{sec}$) & & & ($10^{-4}\,\mathrm{Hz}$) & ($\mathrm{sec}$) \\
\hline
& 0.13 & $[1.5,5.5]$ & $481\pm139$ & & 0.13 & $[0.5,4.5]$ & $451\pm86$ \\
& 0.20 & $[1.5,5.5]$ & $488\pm133$ & & 0.20 & $[0.5,4.5]$ & $368\pm73$ \\
\underline{1H 0707--495} & 0.28 & $[1.5,5.5]$ & $331\pm100$ & \underline{Ark 564} & 0.28 & $[1.5,4.5]$ & $300\pm61$ \\
$s=1.7\pm0.3$ & 0.50 & $[1.5,5.5]$ & $88\pm48$ & $s=1.4\pm0.1$ & 0.50 & $[1.5,4.5]$ & $193\pm37$ \\
$\chi^2_{\mathrm{min}}/\mathrm{dof}=74.5/55$ & 1.50 & $[1.5,5.5]$ & $-46\pm117$ & $\chi^2_{\mathrm{min}}/\mathrm{dof}=89.44/54$ & 1.50 & $[0.5,4.5]$ & $-246\pm59$ \\
& 2.00 & $[1.5,5.5]$ & $-34\pm132$ & & 2.00 & $[0.5,4.5]$ & $-330\pm68$ \\
& 2.83 & $[1.5,2.5]$ & $-635\pm399$ & & 2.83 & $[0.5,4.0]$ & $-480\pm114$ \\
\hline
& 0.13 & $[1.5,6.5]$ & $287\pm95$ & & 0.13 & $[0.5,2.5]$ & $208\pm75$ \\
& 0.20 & $[1.5,6.5]$ & $280\pm90$ & & 0.20 & $[0.5,2.5]$ & $170\pm68$ \\
\underline{MCG--6-30-15} & 0.28 & $[1.5,6.5]$ & $168\pm63$ & \underline{IRAS 13224--3809} & 0.28 & $[0.5,2.5]$ & $126\pm52$ \\
$s=1.3\pm0.3$ & 0.50 & $[1.5,6.5]$ & $77\pm33$ & $s=1.4\pm0.4$ & 0.50 & $[0.5,2.5]$ & $43\pm31$ \\
$\chi^2_{\mathrm{min}}/\mathrm{dof}=66.3/69$ & 1.50 & $[1.5,5.5]$ & $-51\pm48$ & $\chi^2_{\mathrm{min}}/\mathrm{dof}=14.6/21$ & 1.50 & $[0.5,2.0]$ & $-16\pm96$ \\
& 2.00 & $[1.5,6.5]$ & $-74\pm52$ & & 2.00 & $[1.5,2.0]$ & $-194\pm113$ \\
& 2.83 & $[1.5,6.5]$ & $-211\pm87$ & & 2.83 & --- & --- \\
\hline
& 0.13 & $[0.5,4.0]$ & $63\pm36$ \\
& 0.20 & $[0.5,4.0]$ & $49\pm33$ & & 0.22 & $[0.5,1.5]$ & $230\pm149$ \\
\underline{NGC 4051} & 0.28 & $[0.5,4.0]$ & $28\pm34$ & \underline{MCG--5-23-16} \\
$s=0.9\pm0.4$ & 0.50 & $[0.5,4.0]$ & $11\pm24$ & $s=1.0\pm0.3$ & 0.50 & $[0.5,4.0]$ & $31\pm32$ \\
$\chi^2_{\mathrm{min}}/\mathrm{dof}=49.5/48$ & 1.50 & $[0.5,4.0]$ & $-16\pm29$ & $\chi^2_{\mathrm{min}}/\mathrm{dof}=25.6/25$ & 1.50 & $[0.5,3.5]$ & $-140\pm36$ \\
& 2.00 & $[0.5,4.0]$ & $-30\pm34$ & & 2.00 & $[0.5,3.5]$ & $-136\pm38$ \\
& 2.83 & $[0.5,4.0]$ & $-94\pm50$ & & 2.83 & $[0.5,3.0]$ & $-228\pm54$ \\
\hline
& 0.13 & $[0.5,4.0]$ & $5\pm64$ & & 0.13 & $[0.5,1.5]$ & $117\pm212$ \\
& 0.20 & $[0.5,4.0]$ & $58\pm52$ & & 0.20 & $[0.5,4.0]$ & $203\pm92$ \\
\underline{PKS 0558--504} & 0.28 & $[0.5,4.0]$ & $66\pm40$ & \underline{NGC 7314} & 0.28 & $[0.5,6.5]$ & $212\pm57$ \\
$s=1.1\pm0.4$ & 0.50 & $[0.5,4.0]$ & $26\pm29$ & $s=0.6\pm0.2$ & 0.50 & $[0.5,6.5]$ & $81\pm23$ \\
$\chi^2_{\mathrm{min}}/\mathrm{dof}=29.7/37$ & 1.50 & $[0.5,2.5]$ & $-73\pm53$ & $\chi^2_{\mathrm{min}}/\mathrm{dof}=106.2/61$ & 1.50 & $[0.5,6.5]$ & $-9\pm34$ \\
& 2.00 & $[0.5,2.0]$ & $-74\pm71$ & & 2.00 & $[0.5,6.5]$ & $-66\pm37$ \\
& 2.83 & $[0.5,2.0]$ & $-571\pm113$ & & 2.83 & $[0.5,6.5]$ & $-149\pm58$ \\
\hline
& 0.13 & $[0.5,3.5]$ & $30\pm38$ & & 0.13 & $[0.5,2.5]$ & $27\pm112$ \\
& 0.20 & $[0.5,3.5]$ & $46\pm34$ & & 0.20 & $[0.5,2.5]$ & $42\pm98$ \\
\underline{Mrk 766} & 0.28 & $[0.5,3.5]$ & $21\pm27$ & \underline{Mrk 335} & 0.28 & $[0.5,2.5]$ & $121\pm79$ \\
$s=-0.1\pm0.3$ & 0.50 & $[0.5,3.5]$ & $15\pm19$ & $s=0.9\pm0.5$ & 0.50 & $[0.5,2.5]$ & $174\pm57$ \\
$\chi^2_{\mathrm{min}}/\mathrm{dof}=61.4/41$ & 1.50 & $[0.5,3.5]$ & $-173\pm59$ & $\chi^2_{\mathrm{min}}/\mathrm{dof}=46.8/21$ & 1.50 & $[0.5,2.5]$ & $-23\pm91$ \\
& 2.00 & $[0.5,3.5]$ & $-127\pm51$ & & 2.00 & $[0.5,2.5]$ & $-95\pm91$ \\
& 2.83 & $[0.5,3.5]$ & $-96\pm66$ & & 2.83 & $[0.5,2.5]$ & $-237\pm156$ \\
\hline
\end{tabular}
\vspace{1cm}
\end{table*}

%%%%%%%%%%%%%%%%%%% FIG3 %%%%%%%%%%%%%%%%%%%%%%%%%
\begin{figure*}
 \includegraphics[width=400pt]{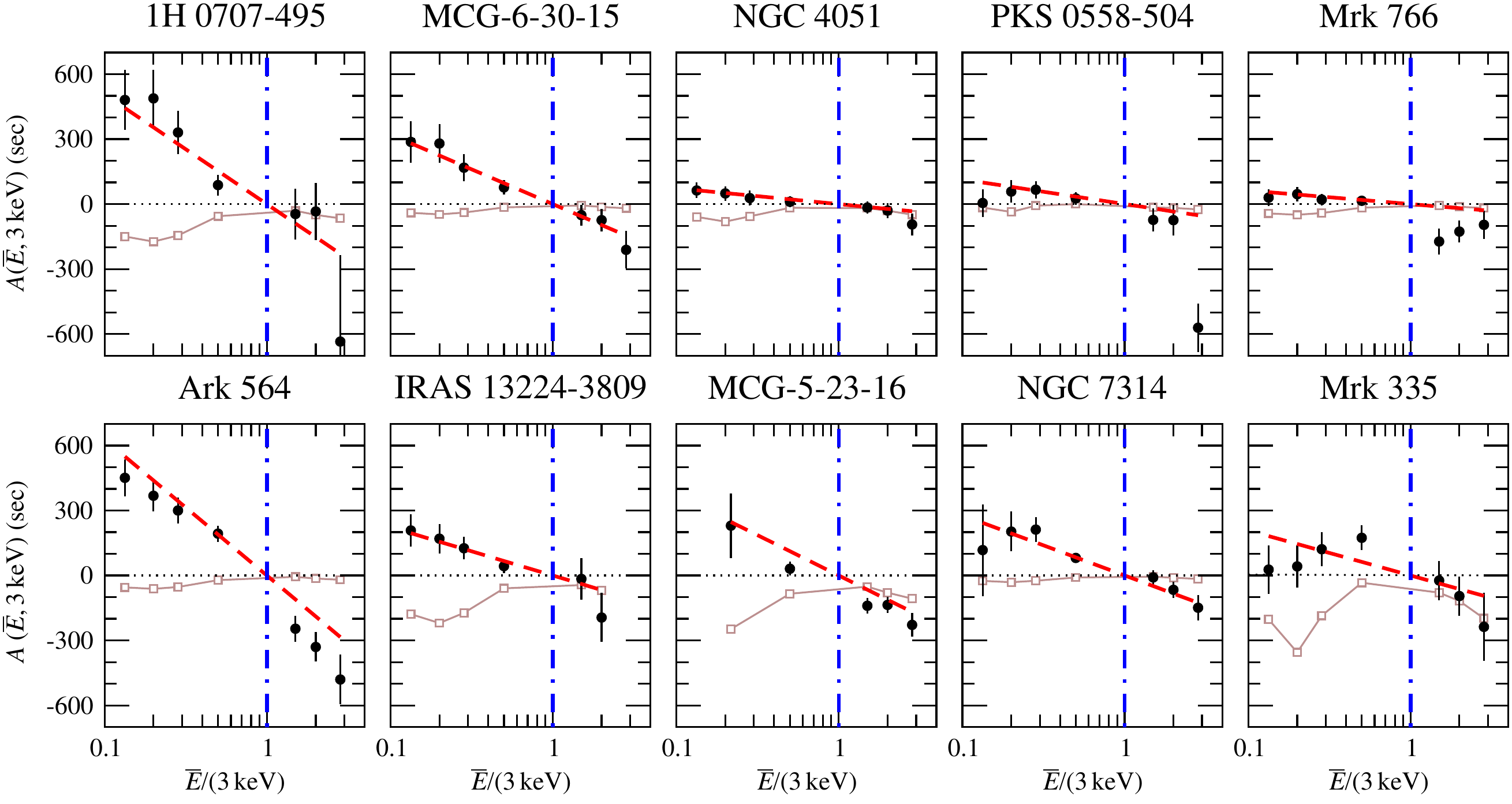}
 \caption{Plots of the best-fitting time-lag amplitude, $A(\overline{E},3\,\mathrm{keV})$ (black points), as a function of $\overline{E}/(3\,\mathrm{keV})$. The vertical blue dotted-dashed lines indicate $\overline{E}/(3\,\mathrm{keV})=1$, while the red dashed lines show the best-fitting model for the $A(\overline{E},3\,\mathrm{keV})$ vs. $\overline{E}/(3\,\mathrm{keV})$ data (see the text for more details). The open brown squares correspond to model X-ray reverberation time-lags at $10^{-4}\,\mathrm{Hz}$ as a function of $\overline{E}/(3\,\mathrm{keV})$ (see Section \ref{sec513} for details).}
 \label{fig3}
\end{figure*}

%%%%%%%%%%%%SECTION 3.1%%%%%%%%%%%%%%%%%
%%%%%%%%%%%%%%%%%%%%%%%%%%%%%
\subsection{Modelling the sample time-lag spectra} \label{sec31}

The statistical properties of the sample time-lag spectra plotted in the relevant Appendix \ref{appb} figures are appropriate for model fitting, using traditional $\chi^2$ minimisation techniques. We fitted the sample time-lag spectra with a power-law function of the form
\noindent
\begin{equation}
\label{eq5}
\hat{\tau}(\nu;\overline{E},3\,\mathrm{keV})=A(\overline{E},3\,\mathrm{keV})(\nu/10^{-4}\,\mathrm{Hz})^{-s}\,\mathrm{sec},
\end{equation}
\noindent
where $\overline{E}$ is the mean energy of each band (listed in column 2 of Table \ref{table2}), $3\,\mathrm{keV}$ is the mean energy of the reference band, $s$ is the power-law slope, and $A(\overline{E},3\,\mathrm{keV})$ is the (energy dependent) amplitude at $10^{-4}\,\mathrm{Hz}$.

We fitted the model in a limited frequency range between $\nu_{\rm low}$ and $\nu_{\rm high}$. These values are listed in column 3 of Table \ref{table3}, and are indicated by the vertical blue dotted-dashed lines in the $0.3-1$ vs. $2-4\,\mathrm{keV}$ time-lag panel of the same figures. The points plotted with filled circles in all panels of the same figures indicate the time-lag estimates used for the model-fitting procedure. The low frequency limit, $\nu_{\rm low}$, is the lowest sampled frequency ($5\times10^{-5}\,\mathrm{Hz}$) in all sources except 1H 0707--495 and MCG--6-30-15. For these sources, we observe a low frequency turn-over in the sample time-lag spectra (see Figs. \ref{figb1} and \ref{figb3}). This turn-over is more pronounced in the soft band time-lags. We decided to ignore the time-lags at these low frequencies, because the best-fitting results change significantly depending on whether we keep them or not. At high frequencies the sample time-lag spetra may change sign, most probably because of the presence of so-called X-ray reverberation time-lags. Since we are interested in studying the continuum time-lags, we decided to fit the sample time-lag spectra only at frequencies where the time-lags are predominately positive or negative (at energies lower or higher than the reference band, respectively). We defined $\nu_{\rm high}$ as the frequency above which the probability that all $0.3-1$ vs. $2-4\,\mathrm{keV}$ time-lag estimates in the range $[\nu_{\rm low},\nu_{\rm high}]$ are positive is smaller than 0.01. This probability was calculated by assuming that the time-lag estimates have a Gaussian distribution (with a mean and standard deviation given by equation \ref{eq1} and \ref{eq2}, respectively), and are independent at each frequency. In this case, the aforementioned probability is equal to the product of the integrated (Gaussian) probability distribution functions over the interval $[0,\infty)$ of all the time-lag estimates in the range $[\nu_{\rm low},\nu_{\rm high}]$.

For each source we fitted all available sample time-lag spectra simultaneously. We left $A(\overline{E},3\,\mathrm{keV})$ as a free parameter, and kept the slope, $s$, fixed at the same value for all time-lag spectra. We determined the best-fitting parameter values by locating the minimum of the $\chi^2$ function, $\chi^2_{\mathrm{min}}$, using the Levenberg-Marquardt method. The 68 per cent (95 per cent) confidence intervals of the best-fitting model parameters were determined by the standard $\Delta\chi^2=1$ ($\Delta\chi^2=4$) method for one independent parameter. Unless otherwise mentioned, best-fitting parameters will henceforth be quoted at the 68 per cent confidence level.

%%%%%%%%%%%%SECTION 3.2%%%%%%%%%%%%%%%%%
%%%%%%%%%%%%%%%%%%%%%%%%%%%%%
\subsection{The best-fitting results} \label{sec32}

The best-fitting results are listed in Table \ref{table3}, and the best-fitting models are shown as red dashed lines in the relevant Appendix \ref{appb} figures. The best-fitting models describe well the overall shape of the low-frequency sample time-lag spectra. The $\chi^2_{\mathrm{min}}$ values in some cases (Ark 564, NGC 7314, and Mrk 335) imply that the power-law model does not fit the data well (the null hypothesis probability, $p_{\rm null}$, is smaller than 1 per cent). However, it is not easy to judge the quality of the fits in our case. Although the time-lag estimates should be uncorrelated at each frequency, the fact that the light curves in the various energy bands are correlated implies that (within each source) the time-lags between the reference band and different energy bands should be also be correlated to some extent. In this case, the actual number of degrees of freedom should be smaller than the numbers listed in Table \ref{table3}. This would imply that the model fit may not be acceptable even in other sources as well, however, as we argue below, we do not believe this is the case.

We fitted the individual sample time-lag spectra of each source with the model defined by equation \ref{eq5}. The fit was acceptable in all cases ($p_{\rm null}>0.01$). The best-fitting slope values were consistent with the corresponding weighted-mean value for each source, hence the hypothesis of a constant (i.e. energy independent) slope is likely to be true. We could consider the best-fitting results from these fits, however the best-fitting amplitudes were poorly determined in that case. In fact, it was for this reason that we decided to fit all sample time-lag spectra simultaneously for each source: The best-fitting parameter values are consistent (within the errors) in both cases, but the errors are smaller when we fit all time-lag spectra simultaneously. We conclude that a power-law time-lag model, with the same slope at all energies, fits the sample time-lag spectra well.

Figure \ref{fig3} shows the power-law amplitude, $A(\overline{E},3\,\mathrm{keV})$, plotted as a function of the light curve mean-energy ratio, $\overline{E}/(3\,\mathrm{keV})$. The logarithm of this ratio is a measure of the energy separation between the light curves. The amplitude's sign `flips' from positive to negative when $\overline{E}<3\,\mathrm{keV}$ and $\overline{E}>3\,\mathrm{keV}$, respectively. This behaviour is the result of the fact that hard energy-band variations are delayed with respect to variations in softer energy-bands. The plots in Fig. \ref{fig3} show that, in all sources, the power-law time-lag model amplitude increases with increasing energy separation. To quantify this trend we fitted the data plotted in the panels of Fig. \ref{fig3} with the following model:
\noindent
\begin{equation}
\label{eq6}
A(\overline{E},3\,\mathrm{keV})=-A_0\log{(\overline{E}/3\,\mathrm{keV})}\,\mathrm{sec}.
\end{equation}
\noindent
Equation \ref{eq6} describes a function that becomes zero when $\overline{E}=3\,\mathrm{keV}$, increases in magnitude with increasing $|\log(\overline{E}/3\,\mathrm{keV})|$, and whose sign shifts from positive to negative when $\overline{E}<3\,\mathrm{keV}$ and $\overline{E}>3\,\mathrm{keV}$, respectively (as seen in the sample time-lag spectra). The amplitude $A_0$ corresponds to the power-law time-lag amplitude (at $10^{-4}\,\mathrm{Hz}$) between the reference band and an energy band with $\overline{E}=0.3\,\mathrm{keV}$ (or $30\,\mathrm{keV}$).

Our best-fitting results are listed in Table \ref{table4}, and the best-fitting models are shown as red dashed lines in Fig. \ref{fig3}. The model fits the data well, except for PKS 0558--504, where the $7-10$ vs. $2-4\,\mathrm{keV}$ power law time-lag amplitude appears to be significantly higher than for other energy bands. Perhaps the more significant discrepancy between the model and the data appears in Ark 564: a log-linear relation between the time-lag amplitude and energy may be just a first-order approximation in this case. Just like in PKS 0558--504, Mrk 766, and Mrk 335, the `amplitude vs. energy' plot of Ark 564 suggests that the energy dependence is less (more) steep than what the model defined by equation \ref{eq6} predicts when $\overline{E}<3\,\mathrm{keV}$ ($\overline{E}>3\,\mathrm{keV}$), respectively (although the errors of the time-lag amplitudes are larger for the former sources compared to Ark 564).

%%%%%%%%%%%%%%%%%%% TABLE4 %%%%%%%%%%%%%%%%%%%%%%%%%

\begin{table}
\centering
 \caption{Best-fitting results for the $A(\overline{E},3\,\mathrm{keV})$ vs. $\overline{E}/(3\,\mathrm{keV})$ data. The model is defined by equation \ref{eq6}, and the best-fitting models are shown as red dashed lines in Fig. \ref{fig3}.}
 \label{table4}
 \begin{tabular}{ccc}
  \hline
  Source & $A_0$ & $\chi^2_{\mathrm{min}}/\mathrm{dof}$ \\
& ($\mathrm{sec}$) & \\
  \hline
  \hline
  1H0707--495 & $507\pm77$ & $5.2/6$ \\
  MCG--6-30-15 & $321\pm26$ & $1.6/6$ \\
  NGC 4051 & $73\pm14$ & $1.9/6$ \\
  PKS 0558--504 & $109\pm77$ & $24.9/6$ \\
  Mrk 766 & $63\pm37$ & $13.8/6$ \\
  Ark 564 & $618\pm81$ & $15.3/6$ \\
  IRAS 13224--3809 & $223\pm30$ & $2.0/5$ \\
  MCG--5-23-16 & $372\pm108$ & $12.3/4$ \\
  NGC 7314 & $278\pm33$ & $3.2/6$ \\
  Mrk 335 & $190\pm78$ & $7.8/6$ \\
  \hline
 \end{tabular}
\end{table}

%%%%%%%%%%%%%%%%%% SECTION 4 %%%%%%%%%%%%%%%
%%%%%%%%%%%%%%%%%%%%%%%%%%%%%%%%%%%%%%%%
\section{Intrinsic coherence estimation} \label{sec4}

We discuss in detail the estimation of the intrinsic coherence between two light curves in Appendix \ref{appa}. We followed the prescription described in Section \ref{seca5}, and estimated the sample intrinsic coherence function between the same light curves that we used to estimate the time-lag spectra. The results are plotted in Figs. \ref{figb2}, \ref{figb4}, \ref{figb6}, \ref{figb8}, \ref{figb10}, \ref{figb12}, \ref{figb14}, \ref{figb16}, \ref{figb18}, and \ref{figb20} in Appendix \ref{appb}. We first calculated the intrinsic coherence estimates up to $\nu_{\rm crit}$. The vertical, blue dotted-dashed lines in the panels of the same figures indicate $\nu_{\rm max}$ (estimated as explained in Section \ref{seca2}; these values are listed in column 4 of Table \ref{table1}). The intrinsic coherence estimate (as defined by equation \ref{eqa1}) at frequencies below $\nu_{\rm max}$ should be an unbiased estimator of the intrinsic coherence. Their distribution should (roughly) follow a Gaussian distribution, and their error (as defined by equation \ref{eqa3}) should be representative of their intrinsic scatter around the mean, provided they are corrected as explained in Section \ref{seca3}. In addition to those cases where we could not reliably estimate time-lags, $\nu_{\rm max}$ turned out to be smaller than the lowest sampled frequency in a few other cases, owing to the very low count rate of the respective light curves (e.g. the $7-10$ vs. $2-4\,\mathrm{keV}$ sample intrinsic coherence function of 1H 0707--495 and Mrk 335).

In many sources, the sample intrinsic coherence function is not equal to one, even at the lowest sampled frequencies, and they decrease rapidly with increasing frequency. We stress that, in this case, the loss of coherence at high frequencies is not due to the presence of experimental noise in the light curves. The intrinsic coherence amplitude appears to be energy dependent. For example, the $1-2$ vs. $2-4\,\mathrm{keV}$ sample intrinsic coherence function of MCG--6-30-15 (see Fig. \ref{figb4}) is almost equal to one at all sampled frequencies but, clearly, the $0.3-0.5$ vs. $2-4\,\mathrm{keV}$ sample intrinsic coherence is not equal to one, even at the lowest sampled frequency, and it decreases rapidly with increasing frequency. In fact, the $0.3-0.5$ vs. $7-10\,\mathrm{keV}$ sample intrinsic coherence function (which we do not show here) is even smaller in amplitude.

Since time-lag estimation is less accurate when the coherence is low, we decided to choose the $2-4\,\mathrm{keV}$ band as our reference band (as opposed to the lowest energy band, which is the usual choice) to estimate both the time-lags and the intrinsic coherence. This band has a mean energy that is around the middle of the total available {\it XMM-Newton} EPIC-pn energy range, and therefore the energy separation between $2-4\,\mathrm{keV}$ and the lowest/highest energy bands we considered is somewhat balanced. In addition, the $2-4\,\mathrm{keV}$ band is more representative of the X-ray continuum emission, as it is expected to be less affected by components originating from X-ray reflection, or the presence of a warm absorber, compared to other bands.

%%%%%%%%%%%%%%%%%% SECTION 4.1 %%%%%%%%%%%%%%%
\subsection{Modelling the sample intrinsic coherence} \label{sec41}

%%%%%%%%%%%%%%%%%%%% TABLE 5 %%%%%%%%%%%%%%%%%
\begin{table*}
\caption{Best-fitting results for the intrinsic coherence model. The model is defined by equation \ref{eq7}, and the best-fitting models are shown as red dashed lines in the relevant Appendix \ref{appb} figures.}
\label{table5}
\centering
\begin{tabular}{c c c c c | c c c c c}
\hline\hline
(1) & (2) & (3) & (4) & (5) & (1) & (2) & (3) & (4) & (5) \\
Source & $\overline{E}/(3\,\mathrm{keV})$ & $C(\overline{E},3\,\mathrm{keV})$ & $\nu_{\mathrm{b}}(\overline{E},3\,\mathrm{keV})$ & $\chi^2_{\mathrm{min}}/\mathrm{dof}$ & Source & $\overline{E}/(3\,\mathrm{keV})$ & $C(\overline{E},3\,\mathrm{keV})$ & $\nu_{\mathrm{b}}(\overline{E},3\,\mathrm{keV})$ & $\chi^2_{\mathrm{min}}/\mathrm{dof}$ \\
& & & ($10^{-3}\,\mathrm{Hz}$) & & & & & ($10^{-3}\,\mathrm{Hz}$) & \\
\hline
& 0.13 & $0.91\pm0.03$ & $0.9^{+0.2}_{-0.1}$ & 14.4/11 & & 0.13 & $0.94\pm0.06$ & $0.38^{+0.08}_{-0.06}$ & 6.8/10 \\
& 0.20 & $0.93\pm0.02$ & $1.1\pm0.2$ & 19.0/11 & & 0.20 & $0.94^{+0.05}_{-0.04}$ & $0.6\pm0.1$ & 11.1/10 \\
& 0.28 & $0.94\pm0.02$ & $2.2^{+0.6}_{-0.4}$ & 15.3/12 & & 0.28 & $0.93\pm0.03$ & $1.2^{+0.4}_{-0.3}$ & 9.2/10 \\
\underline{1H} & 0.50 & $0.99\pm0.01$ & $11^{+15}_{-4}$ & 5.0/12 & \underline{Ark 564} & 0.50 & $0.98\pm0.01$ & $4.5^{+3.3}_{-1.3}$ & 15.1/11 \\
\underline{0707-495} & 1.50 & $0.95^{+0.05}_{-0.04}$ & $>1.6$ (0.8) & 0.8/1 & & 1.50 & $0.96^{+0.04}_{-0.03}$ & $>1.1$ (0.7) & 1.9/1 \\
& 2.00 & $0.86^{+0.07}_{-0.04}$ & $>1.2$ (0.5) & 0.3/1 & & 2.00 & $0.95\pm0.05$ & $>0.6$ (0.4) & 4.5/1 \\
& 2.83 & --- & --- & --- &  & 2.83 & --- & --- & --- \\
\hline
& 0.13 & $0.92\pm0.03$ & $1.2\pm0.2$ & 13.4/12 & & 0.13 & $1.00^{+\text{---}}_{-0.08}$ & $0.27^{+0.10}_{-0.04}$ & 0.6/3 \\
& 0.20 & $0.93\pm0.02$ & $1.8^{+0.4}_{-0.3}$ & 9.2/14 & & 0.20 & $1.00^{+\text{---}}_{-0.08}$ & $0.32^{+0.15}_{-0.05}$ & 0.6/3 \\
& 0.28 & $0.95\pm0.01$ & $3.2^{+1.1}_{-0.7}$ & 13.0/14 & & 0.28 & $1.00^{+\text{---}}_{-0.02}$ & $0.50^{+0.12}_{-0.08}$ & 2.4/4 \\
\underline{MCG} & 0.50 & $0.977^{+0.013}_{-0.003}$ & $>10.3$ (8.0) & 13.3/15 & \underline{IRAS} & 0.50 & $1.00^{+\text{---}}_{-0.01}$ & $2.2^{+1.8}_{-0.6}$ & 5.0/4 \\
\underline{--6-30-15} & 1.50 & $1.000^{+\text{---}}_{-0.009}$ & $4.3^{+3.5}_{-1.0}$ & 7.2/8 & \underline{13224-3809} & 1.50 & --- & --- & --- \\
& 2.00 & $0.96\pm0.02$ & $>3.3$ (2.1) & 2.2/6 & & 2.00 & --- & --- & --- \\
& 2.83 & $0.98^{+0.02}_{-0.04}$ & $>1.0$ (0.8) & 0.3/1 & & 2.83 & --- & --- & --- \\
\hline
& 0.13 & $0.91\pm0.02$ & $4.1^{+0.9}_{-0.7}$ & 53.2/26 \\
& 0.20 & $0.92^{+0.02}_{-0.01}$ & $5.2^{+1.5}_{-0.9}$ & 44.1/26 & & 0.22 & --- & --- & --- \\
& 0.28 & $0.89\pm0.02$ & $11^{+9}_{-3}$ & 43.2/27 \\
\underline{NGC} & 0.50 & $0.95\pm0.01$ & $22^{+31}_{-8}$ & 36.2/28 & \underline{MCG} & 0.50 & $0.99\pm0.01$ & $>4.2$ (2.7) & $1.5/4$ \\
\underline{4051} & 1.50 & $0.96^{+0.04}_{-0.01}$ & $>2.6$ (2.1) & 9.8/8 & \underline{--5-23-16} & 1.50 & $1.000^{+\text{---}}_{-0.009}$ & $>6.3$ (3.2) & 0.6/3 \\
& 2.00 & $0.93^{+0.02}_{-0.01}$ & $>5.9$ (3.0) & 4.6/8 & & 2.00 & $1.00^{+\text{---}}_{-0.01}$ & $>3.9$ (2.4) & 0.4/2 \\
& 2.83 & $0.89\pm0.03$ & $>4.4$ (1.7) & 0.7/4 & & 2.83 & $0.97^{+0.03}_{-0.02}$ & $>2.0$ (1.0) & $2.5/1$ \\
\hline
& 0.13 & $0.72^{+0.09}_{-0.07}$ & $>0.7$ (0.4) & 0.9/3 & & 0.13 & --- & --- & --- \\
& 0.20 & $0.84^{+0.07}_{-0.06}$ & $>0.8$ (0.5) & 1.2/3 & & 0.20 & --- & --- & --- \\
& 0.28 & $0.93\pm0.04$ & $2^{+55}_{-1}$ & 0.4/4 & & 0.28 & $1.00^{+\text{---}}_{-0.04}$ & $1.5^{+4.5}_{-0.5}$ & 0.7/3 \\
\underline{PKS} & 0.50 & $0.97\pm0.02$ & $>2.0$ (1.4) & 0.9/4 & \underline{NGC} & 0.50 & $0.993^{+0.005}_{-0.003}$ & $>34.9$ (13.1) & 13.6/16 \\
\underline{0558-504} & 1.50 & --- & --- & --- & \underline{7314} & 1.50 & $0.995^{+0.005}_{-0.011}$ & $>5.4$ (3.1) & 2.6/4 \\
& 2.00 & --- & --- & --- & & 2.00 & $0.98^{+0.02}_{-0.01}$ & $>2.9$ (1.8) & 0.5/4 \\
& 2.83 & --- & --- & --- & & 2.83 & $0.95^{+0.05}_{-0.03}$ & $>1.3$ (0.8) & 0.7/1 \\
\hline
& 0.13 & $0.71\pm0.06$ & $2.1^{+3.1}_{-0.8}$ & 10.2/11 & & 0.13 & $0.75^{+0.06}_{-0.05}$ & $>2.1$ (0.7) & 3.5/3 \\
& 0.20 & $0.73^{+0.03}_{-0.02}$ & $>7.1$ (2.9) & 7.8/12 & & 0.20 & $0.82^{+0.05}_{-0.04}$ & $>3.5$ (1.1) & 2.4/3\\
& 0.28 & $0.82^{+0.03}_{-0.02}$ & $>7.6$ (3.5) & 14.5/13 & & 0.28 & $0.90\pm0.03$ & $>4.9$ (1.5) & 2.6/3 \\
\underline{Mrk 766} & 0.50 & $0.94^{+0.02}_{-0.01}$ & $>9.5$ (5.2) & 6.3/13 & \underline{Mrk 335} & 0.50 & $0.97\pm0.02$ & $>10.3$ (3.2) & 4.2/3 \\
& 1.50 & $1.00^{+\text{---}}_{-0.02}$ & $1.7^{+1.7}_{-0.5}$ & 3.4/4 & & 1.50 & $1.00^{+\text{---}}_{-0.04}$ & $>2.4$ (1.1) & 0.1/2 \\
& 2.00 & $0.97^{+0.02}_{-0.06}$ & $>0.8$ (0.6) & 2.9/1 & & 2.00 & $1.00^{+\text{---}}_{-0.05}$ & $>2.4$ (1.0) & 1.9/1 \\
& 2.83 & --- & --- & --- & & 2.83 & --- & --- & --- \\
\hline
\end{tabular}
\vspace{1cm}
\end{table*}

%%%%%%%%%%%%%%%%%%%% TABLE 6 %%%%%%%%%%%%%%%%%
\begin{table*}
 \caption{Best-fitting results for the $C(\overline{E},3\,\mathrm{keV})$ vs. $\overline{E}/(3\,\mathrm{keV})$, and $\nu_{\mathrm{b}}(\overline{E},3\,\mathrm{keV})$ vs. $\overline{E}/(3\,\mathrm{keV})$ data. The models are defined by equations \ref{eq8} and \ref{eq9}, respectively, and the best-fitting models are shown as red dashed lines in Figs. \ref{fig4} and \ref{fig5}.}
 \label{table6}
 \begin{tabular}{ccccccc}
  \hline
  (1) & (2) & (3) & (4) & (5) & (6) & (7) \\
  Source & $a$ & $b$ & $\log{(E_{*}/3\,\mathrm{keV})}$ & $\chi^2_{\mathrm{min}}/\mathrm{dof}$ &  $\nu_{\mathrm{b},0}$ & $\chi^2_{\mathrm{min}}/\mathrm{dof}$ \\
  &  &  &  &  & ($10^{-3}\,\mathrm{Hz}$) \\
  \hline
  \hline
  1H0707--495 & $1.036\pm0.009$ & $0.15\pm0.02$ & $-0.24\pm0.07$ & $0.6/2$ & $0.57\pm0.03$ & $0.9/3$ \\
  MCG--6-30-15 & $1.009\pm0.005$ & $0.11\pm0.01$ & $-0.09\pm0.05$ & $0.2/2$ & $1.03\pm0.09$ & $1.8/3$ \\
  NGC 4051 & $1$ & $0.13\pm0.02$ & $0$ & $9.7/3$ & $3.7\pm0.3$ & $0.8/3$ \\
  PKS 0558--504 & $1.09\pm0.04$ & $0.36\pm0.09$ & $-0.24\pm0.13$ & $1.3/2$ & --- & --- \\
  Mrk 766 & $1.09\pm0.03$ & $0.49\pm0.05$ & $-0.18\pm0.06$ & $1.4/2$ & --- & --- \\
  Ark 564 & $1.01\pm0.02$ & $0.11\pm0.04$ & $-0.11\pm0.14$ & $0.5/2$ & $0.34\pm0.04$ & $3.9/3$ \\
  IRAS 13224--3809 & --- & --- & --- & --- & $0.162\pm0.008$ & $0.05/2$ \\
  MCG--5-23-16 & --- & --- & --- & --- & --- & --- \\
  NGC 7314 & --- & --- & --- & --- & --- & --- \\
  Mrk 335 & $1.09\pm0.02$ & $0.37\pm0.04$ & $-0.24\pm0.05$ & $0.4/2$ & --- & --- \\
  \hline
 \end{tabular}
\end{table*}

Based on the shape of the sample intrinsic coherence of most sources, we fitted the data with the following model:
\noindent
\begin{equation}
\hat{\gamma}^2_{\mathrm{int}}(\nu;\overline{E},3\,\mathrm{keV})=C(\overline{E},3\,\mathrm{keV})\,\mathrm{exp}[-\nu/\nu_{\mathrm{b}}(\overline{E},3\,\mathrm{keV})].
\label{eq7}
\end{equation}
\noindent
Equation \ref{eq7} describes a function that is constant at low frequencies (equal to $C(\overline{E},3\,\mathrm{keV})$), and then decreases exponentially at frequencies above a `break' frequency, $\nu_{\mathrm{b}}(\overline{E},3\,\mathrm{keV})$. We determined the best-fitting model parameters using standard $\chi^2$ minimisation techniques (similar to the modelling of the sample time-lag spectra). The best-fitting results are listed in Table \ref{table5}, and the best-fitting models are shown as red dashed lines in the relevant Appendix \ref{appb} figures. In some cases we did not detect a significant break frequency, and we list the 68 per cent lower limit on the corresponding best-fitting $\nu_{\mathrm{b}}(\overline{E},3\,\mathrm{keV})$ values in column 4 of Table \ref{table5} (we also show the 95 per cent lower limits in parentheses). Furthermore, in some cases, the best-fitting $C(\overline{E},3\,\mathrm{keV})$ value was equal to one, and we list the respective 68 per cent lower limit in the same column.

In general, the model fits the data well in almost all cases. In most sources, $C(\overline{E},3\,\mathrm{keV})$ decreases with increasing energy separation between the light curves. The loss of coherence is reinforced by the simultaneous decrease of $\nu_{\mathrm{b}}(\overline{E},3\,\mathrm{keV})$ with increasing energy separation (e.g. MGC--6-30-15, and NGC 4051). In some cases (e.g. IRAS 13224--3809) the intrinsic coherence is equal to one at low frequencies, for all energy separation values we considered. The loss of coherence in this case is because $\nu_{\rm b}(\overline{E},3\,\mathrm{keV})$ decreases strongly with increasing energy separation between the light curves. We investigate below these issues in more detail.

%%%%%%%%%%%%SECTION 4.2%%%%%%%%%%%%%%%%%
\subsection{The energy dependence of the intrinsic coherence} \label{sec42}

%%%%%%%%%%%% Figure 4 %%%%%%%%%%%%%%%%
\begin{figure}
 \includegraphics[width=\hsize]{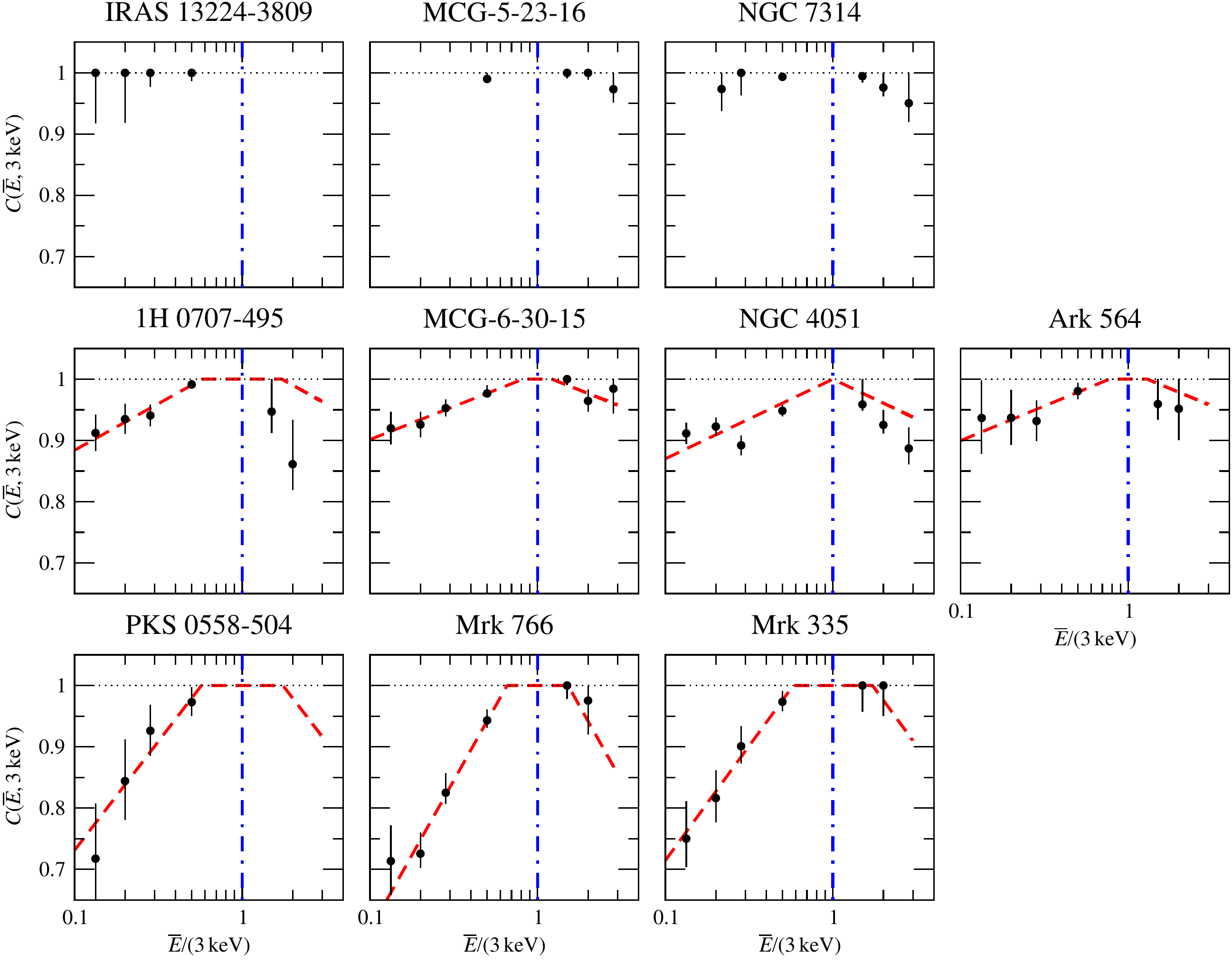}
 \caption{Plots of the best-fitting low-frequency constant intrinsic-coherence value, $C(\overline{E},3\,\mathrm{keV})$ (black points), as a function of $\overline{E}/(3\,\mathrm{keV})$. The vertical, blue dotted-dashed line indicates $\overline{E}/(3\,\mathrm{keV})=1$, while the red dashed lines show the best-fitting model for the $C(\overline{E},3\,\mathrm{keV})$ vs. $\overline{E}/(3\,\mathrm{keV})$ data (see the text for more details).}
\label{fig4}
\end{figure}

%%%%%%%%%%%% Figure 5 %%%%%%%%%%%%%%%%
\begin{figure}
 \includegraphics[width=\hsize]{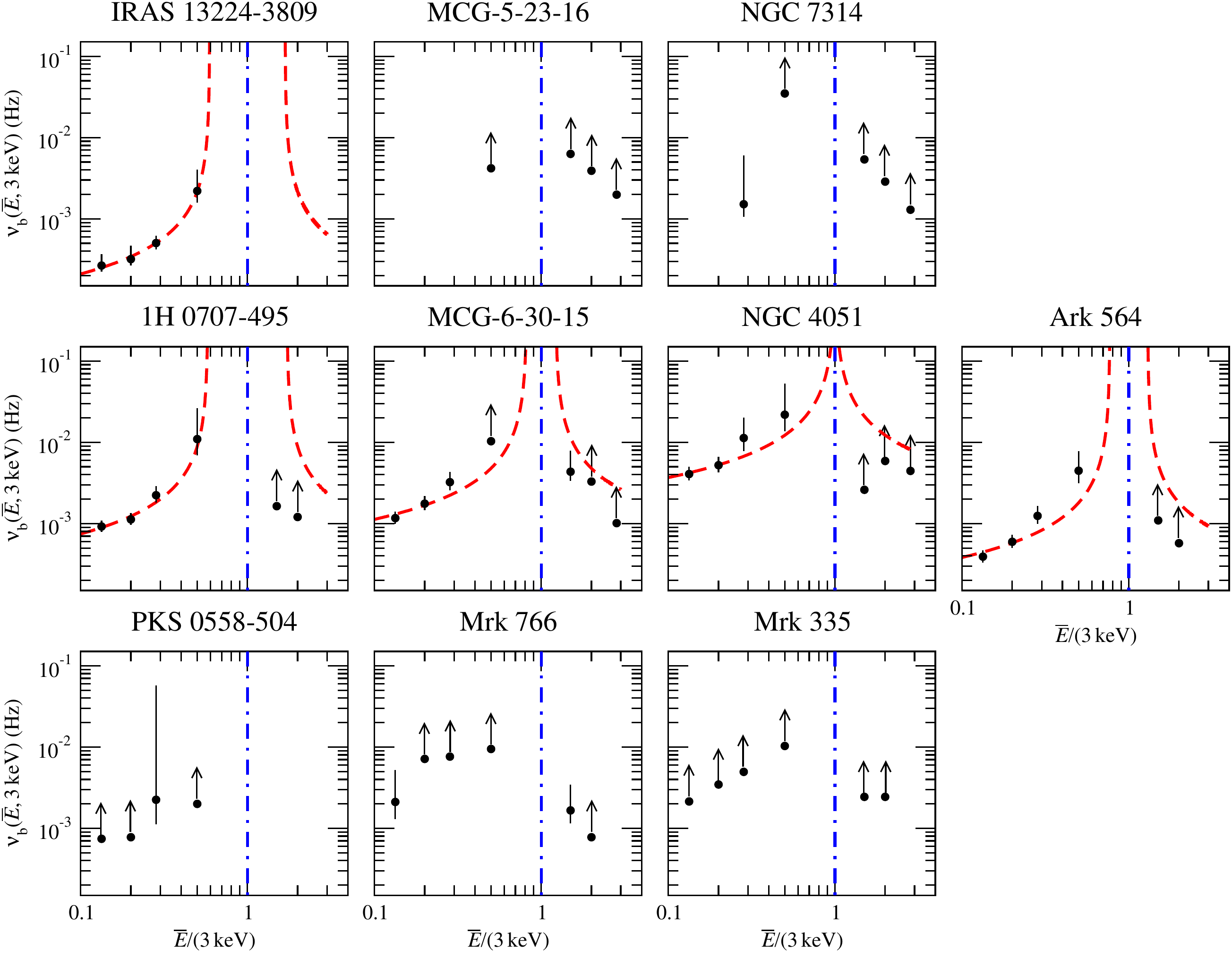}
 \caption{Plots of the best-fitting intrinsic coherence break-frequency, $\nu_{\mathrm{b}}(\overline{E},3\,\mathrm{keV})$ (black points), as a function of $\overline{E}/(3\,\mathrm{keV})$. The vertical, blue dotted-dashed lines indicate $\overline{E}/(3\,\mathrm{keV})=1$, while the red dashed lines show the best-fitting model for the $\nu_{\mathrm{b}}(\overline{E},3\,\mathrm{keV})$ vs. $\overline{E}/(3\,\mathrm{keV})$ data (see the text for more details).}
 \label{fig5}
\end{figure}

Figures \ref{fig4} and \ref{fig5} show the best-fitting $C(\overline{E},3\,\mathrm{keV})$ and $\nu_{\mathrm{b}}(\overline{E},3\,\mathrm{keV})$ values as a function $\overline{E}/(3\,\mathrm{keV})$. Each panel in these figures corresponds to a different source. The sources are divided into three groups (corresponding to the three rows in each figure) according to a common phenomenological behaviour of $C(\overline{E},3\,\mathrm{keV})$ as a function of energy.

The first group (first row in Figs. \ref{fig4} and \ref{fig5}) consists of IRAS 13224--3809, MCG--5-23-16, and NGC 7314 (henceforth, Group A). The best-fitting $C(\overline{E},3\,\mathrm{keV})$ values of the Group A sources are consistent with one at all energies. The second group (second row in the same figures) consists of 1H 0707--495, MCG--6-30-15, NGC 4051, and Ark 564 (henceforth, Group B). The best-fitting $C(\overline{E},3\,\mathrm{keV})$ values of Group B show a moderate (up to 10 per cent) decrease from the value of one as the energy separation increases. Arguably, the uncertainty of the best-fitting $C(\overline{E},3\,\mathrm{keV})$ values of the Group A sources is larger than that those of the Group B sources, hence a meaningful quantitative comparison these two Groups cannot be determined very accurately. The third group (third row in the same figures) consists of PKS 0558--504, Mrk 766, and Mrk 335 (henceforth, Group C). The best-fitting $C(\overline{E},3\,\mathrm{keV})$ values of Group C show a stronger (up to 30 per cent) decrease from the value of one as the energy separation increases.

To further investigate the dependence of $C(\overline{E},3\,\mathrm{keV})$ on energy separation, we fitted the data plotted in the panels of Fig. \ref{fig4} to a function of the form
\begin{equation}
C(\overline{E},3\,\mathrm{keV})=a+b\log(\overline{E}/3\,\mathrm{keV}).
\label{eq8}
\end{equation}
We did not fit the Group A data because there either are few $C(\overline{E},3\,\mathrm{keV})$ estimates, or the they are consistent with one. We only fitted the model to the $C(\overline{E},3\,\mathrm{keV})$ values at soft energies ($\overline{E}<3\,\mathrm{keV}$), as their error is smaller than at hard energies ($\overline{E}>3\,\mathrm{keV}$). The best-fitting results are listed in Table \ref{table6}. The model provides a statistically acceptable fit to the data of all sources. The Group B and Group C sources are characterised by significantly different best-fitting $b$ values. The weighted-mean $b$ value of the Group B and C sources is $0.120\pm0.008$ and $0.41\pm0.03$, respectively. The $b$ values of the individual sources within the two Groups are consistent, within the errors, with the Group's weighted-mean value.

Column 4 of Table \ref{table6} lists $\log(E_*/3\,\mathrm{keV})$, where $E_*$ is the energy at which $C(\overline{E},3\,\mathrm{keV})$ becomes equal to one. According to equation \ref{eq8}, $\log(E_*/3\,\mathrm{keV})=(1-a)/b$. The value of $E_*$ cannot exceed $3\,\mathrm{keV}$, since this is the mean energy of the reference band\footnote{The best-fitting $a$ and $b$ values of NGC 4051 were such that $E_*>3\,\mathrm{keV}$; for that reason we fitted the NGC 4051 data by setting $a=1$ during the fitting procedure, to force an amplitude of 1 for $\overline{E}=3\,\mathrm{keV}$.}. The best-fitting models are shown as red dashed lines in Fig. \ref{fig4}. The extension of the best-fitting lines at energies $>E_*$ was done assuming that $C(\overline{E},3\,\mathrm{keV})=1$ at energies between $E_*$ and $3\,\mathrm{keV}$, and that the $C(\overline{E},3\,\mathrm{keV})$ vs. $\overline{E}/(3\,\mathrm{keV})$ model is symmetric around $3\,\mathrm{keV}$, whereby $\overline{E}/(3\,\mathrm{keV})=1$ (indicated by the vertical, blue dotted-dashed lines in the same figure). This assumption appears to be consistent with the MCG--6-30-15 and NGC 4051 data, where the hard-energy $C(\overline{E},3\,\mathrm{keV})$ values are as accurately determined as the corresponding soft-energy values. The weighted-mean $\log(E_*/3\,\mathrm{keV})$ value is $-0.18\pm0.03$ (which corresponds to a weighted-mean $E_*$ value of $1.98\pm0.14\,\mathrm{keV}$). The results indicate that, with the exception of the Group A sources and NGC 4051, the low-frequency constant intrinsic-coherence value is consistent with one when $|\log(\overline{E}/3\,\mathrm{keV})|\lesssim0.2$.

Owing to the fact that the frequency range $[\nu_{\rm low},\nu_{\rm max}]$ is relatively narrow, we could only obtain lower limits on $\nu_{\mathrm{b}}(\overline{E},3\,\mathrm{keV})$ in most cases. IRAS 13224--3809 and the Group B sources stand as exceptions; for these sources $\nu_{\mathrm{b}}(\overline{E},3\,\mathrm{keV})$ increases as the light-curve energy separation decreases. To investigate the energy dependence of $\nu_{\mathrm{b}}(\overline{E},3\,\mathrm{keV})$, we fitted the $\nu_{\mathrm{b}}(\overline{E},3\,\mathrm{keV})$ vs. $\overline{E}/(3\,\mathrm{keV})$ data of these sources to the model
\noindent
\begin{equation}
\nu_{\mathrm{b}}(\overline{E},3\,\mathrm{keV})=\nu_{\mathrm{b},0}/|\log[(\overline{E}/3\,\mathrm{keV})-\log(E_0/3\,\mathrm{keV})]|.
\label{eq9}
\end{equation}
\noindent
The above function increases rapidly towards $+\infty$ as $\overline{E}\rightarrow E_0$, while its normalization is set by $\nu_{\mathrm{b},0}$. When we originally left $E_0$ as a free parameter during the fitting procedure, its best-fitting value for the Group B sources was always consistent with the respective best-fitting $E_*$ values. We therefore set $E_0=E_*$ for these sources, to determine $\nu_{\mathrm{b},0}$ more accurately. In the case of IRAS 13224--3809 we left $E_0$ as a free parameter, and obtained a best-fitting value of $\log(E_0/3\,\mathrm{keV})=-0.23\pm0.01$. The results indicate that the sample intrinsic coherence function in the Group B sources is flat (at least over the sampled frequency range) when $|\log(\overline{E}/3\,\mathrm{keV})|\lesssim0.2$. This holds true for IRAS 13224--3809 as well.

The model defined by equation \ref{eq9} provided statistically acceptable fits for all aforementioned sources. The best-fitting results are listed in Table \ref{table6}, and the red dashed lines in Fig. \ref{fig5} show the best-fitting models. Just like with the best-fitting models plotted in Fig. \ref{fig4}, in plotting the best-fitting models at hard energies we assumed that $\nu_{\mathrm{b},0}$ tends to infinity when $\overline{E}$ is between $E_*$ and $3\,\mathrm{keV}$, and that the models are symmetric around $\overline{E}=3\,\mathrm{keV}$.

\section{Discussion and conclusions} \label{sec5}

We performed a systematic analysis of the X-ray continuum time-lags and intrinsic coherence in ten AGN, using all available \textit{XMM-Newton} observations. The AGN we studied are X-ray bright, highly variable, and have a large amount of \textit{XMM-Newton} archival data ($\gtrsim0.3\,\mathrm{Ms}$). The BH mass estimates for most sources in our sample are clustered around $\sim1-5\times10^6\,\mathrm{M}_{\odot}$, with the exception of  MCG--5-23-16, Mrk 335, and PKS 0558--504, whose BH mass estimates are $\sim8$, $\sim26$ and $\sim250\times 10^6\,\mathrm{M}_{\odot}$, respectively. Their X-ray Eddington ratio estimates, $\lambda_\mathrm{X}$, are relatively uniformly distributed over the range $\sim0.002-0.08$.

We considered light curves in seven energy bands ($0.3-0.5$, $0.3-0.7$, $0.7-1$, $1-2$, $4-5$, $5-7$, and $7-10\,\mathrm{keV}$). We kept the width of the energy bands as narrow as possible to increase the energy resolution, and, at the same time, maintain a reasonably high count rate for the resulting light curves. This is necessary for the meaningful estimation of the time-lags and intrinsic coherence over the broadest possible frequency range (which depends on the intrinsic variability amplitude, and the mean count rate of the light curves). We chose $2-4\,\mathrm{keV}$ as the reference band. The observed variations in this band should be representative of the X-ray continuum variations, as it is expected to be relatively free of warm absorber effects, as well as contributions from relativistically smeared X-ray reflection from the inner disc. In addition, the mean count rate of $2-4\,\mathrm{keV}$ light curves is reasonably large in most sources, and is located (roughly) in the middle of the energy range of \textit{XMM-Newton}'s EPIC-pn detector. As a result, the energy separation between the reference band and the lowest/highest energy bands we considered is balanced.

We used the mean of each energy band to study the energy dependence of the observed time-lag spectra and intrinsic coherence functions. In principle, the mean energy of the photons detected in each band should depend on the slope of the X-ray spectrum, and on the response of the detector. However, given the narrow width of the energy bands we considered, the mean energy of each band should be a reasonable approximation of the mean photon energy. In any case, the uncertainty introduced by this choice should not be significant, given the magnitude of the errors from the statistical analysis of the data.

%%%%%%%%%%%%%%% SECTION 5.1: TIME LAGS SUMMARY %%%%%%%%%%%%%%%%%%%
\subsection{Summary of the time-lag analysis} \label{sec51}

\begin{figure}
\centering
\includegraphics[width=\hsize]{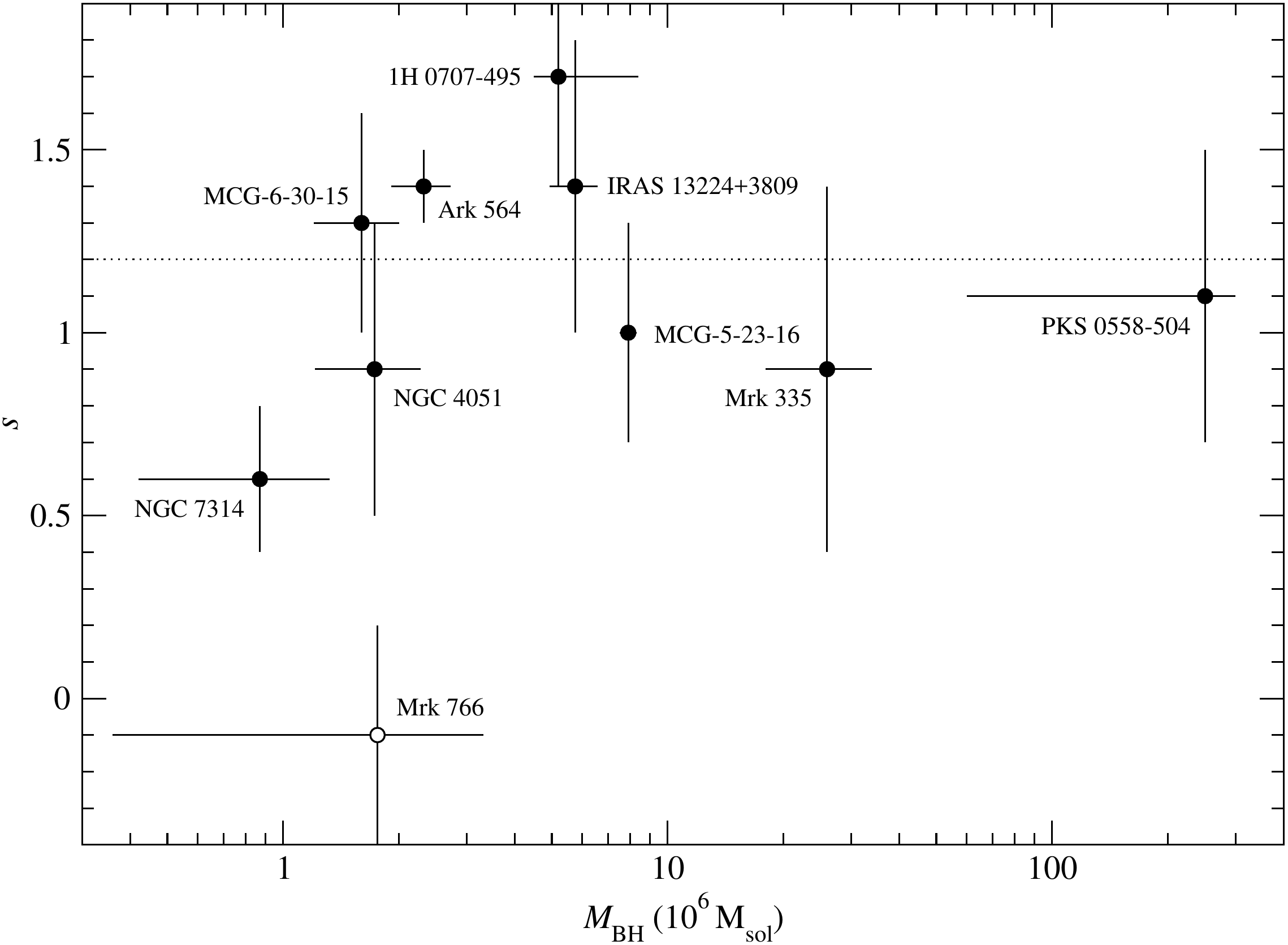}
\caption{Best-fitting power-law time-lag model slope vs. BH mass. The horizontal dotted line indicates the weighted-mean slope value.}
\label{fig6}
\end{figure}

The observed time-lags at low frequencies show a power-law-like dependence on frequency, at all energies and for all sources (see the relevant figures in Appendix \ref{appb}). The time-lags are either positive or negative, depending on whether the energy band is below or above the reference band, respectively. This is a well-known result; this behaviour is commonly referred to as hard time-lags: variations in hard energy-bands are delayed with respect to variations in softer energy-bands. We defined a frequency range where the sample time-lag spectra are dominated by the X-ray continuum time-lags (see Section \ref{sec31}), and fitted the data with a power-law model. Our results are summarised below:
\begin{itemize}
\item[1.] A power-law model fits the continuum time-lags well, at all energies, and for all sources.
\item[2.] The power-law slope is energy independent. Figure \ref{fig6} shows a plot of the best-fitting power-law slopes as a function of the BH mass. The weighted-mean slope value is $1.2\pm0.1$ (indicated by the horizontal dotted line in the same figure). The mean slope, as well as the individual  best-fitting slopes, are consistent with a value of $-1$, except for Mrk 766. The best-fitting slope in this case is consistent with zero: the time-lags have approximately the same value at all (sampled) frequencies.
\item[3.] At a given frequency, the time-lag amplitude increases logarithmically with the light-curve mean-energy ratio (see Fig. \ref{fig3}).
\end{itemize}

The above results are broadly consistent with previous works \citep[e.g.][]{2001ApJ...554L.133P,2004MNRAS.348..783M,2006MNRAS.372..401A,2008MNRAS.388..211A,2009ApJ...700.1042S}, in that the X-ray continuum time-lags between two light curves at energies $E_1$ and $E_2$ $(E_1<E_2$), $\tau(\nu;E_1, E_2)$, follow a relation of the form $\tau(\nu;E_1, E_2)\propto \log(E_2/E_1)\nu^{-1}$.

%%%%%%%%%%%%%%% SUB-SECTION 5.1.1 %%%%%%%%%%%%%%%%%%%
\subsubsection{Low-frequency turn-over} \label{sec511}

In the case of 1H 0707--495 and MCG--6-30-15, the time-lags show a turn-over at the lowest sampled frequencies, at all energies (see Figs. \ref{figb1} and \ref{figb3} in Appendix \ref{appb}). This turn-over could be the result of an additional time-lag component that has the opposite sign to the hard lag component (i.e. a `soft lag' component), which becomes more significant at low frequencies. This could be due to X-ray reverberation soft lags, but only at energies lower than the reference band (see Section \ref{sec513} below for a more detailed discussion on this topic). Recently, \citet{2016A&A...596A..79S} showed that a warm absorber can also produce soft lags in AGN, up to tens of seconds on time-scales of hours. In this case, the time delays are associated with the response of the absorbing gas to changes in the ionising source. Therefore, such a soft lag component could be expected in sources where ionised material is located close to their X-ray emitting region, and is responding to to changes in the ionizing continuum (like MCG--6-30-15, for example).

Such a component might also explain the (peculiar) time-lag spectra of Mrk 766, which remain almost constant at low frequencies. On the other hand, we do not detect a noticeable low frequency turn-over in the time-lag spectra of NGC 4051 \citep[the source studied by][]{2016A&A...596A..79S}. The time-lag amplitude in this source is low, but this could be explained by its low X-ray luminosity (see the discussion in the section below). Time-lag spectra properly determined over a wider frequency range are necessary to investigate the presence of low frequency turn-overs in the time-lag spectra of AGN.

%%%%%%%%%%%%%%% SUB-SECTION 5.1.2 %%%%%%%%%%%%%%%%%%
\subsubsection{Low-frequency time-lag amplitude} \label{sec512}

\begin{figure}
\centering
\includegraphics[width=\hsize]{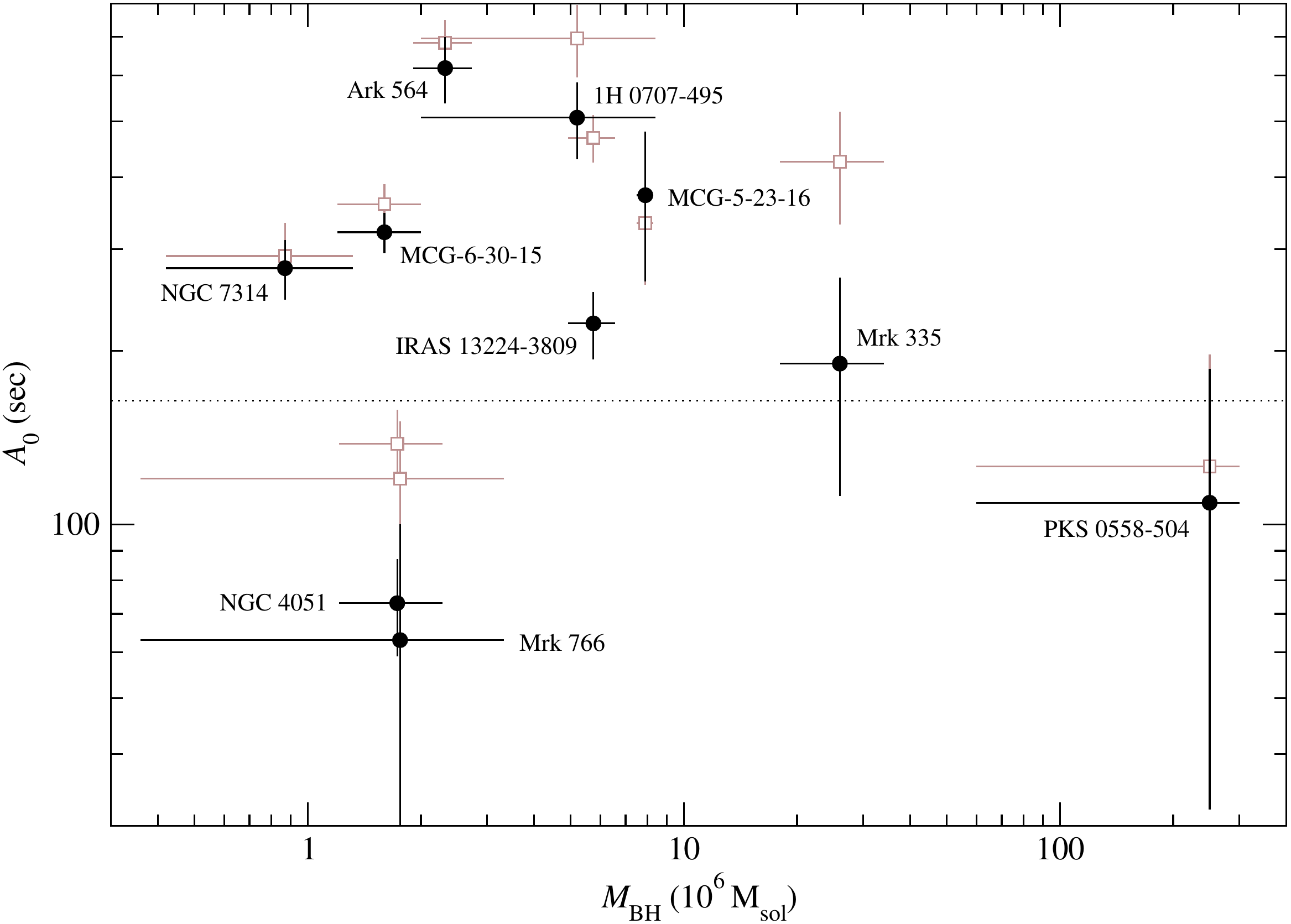}
\caption{The time-lag amplitude, $A_0$, vs. BH mass (filled black circles). The open brown squares show the corresponding amplitude when subtracting the model reverberation time-lag component (see Sect \ref{sec513} for details).}
\label{fig7}
\end{figure}

Figure \ref{fig7} shows the low frequency time-lag amplitude plotted as a function of BH mass. As a measure of the time-lag amplitude we used the best-fitting $A_0$ values listed in Table \ref{fig4}. The horizontal dashed line indicates the weighted-mean $A_0$ value, which is equal to $164\pm10\,\mathrm{sec}$. The points are scattered around the mean, and the scatter is significant, as the individual points are not consistent with the mean (we find $\chi^2_{\mathrm{min}}/\mathrm{dof}=157.6/9$ when we fit the data with the dotted line shown in the same figure). The scatter of the points around the mean appears to be random, i.e. we do not observe a systematic trend which would indicate that $A_0$ depends on $M_{\mathrm{BH}}$. Indeed, the correlation coefficient for the $A_0$ vs. $M_{\mathrm{BH}}$ data is $r=-0.33$, with $p_{\mathrm{null}}=0.35$. On the other hand, we notice that sources with high $\lambda_\mathrm{X}$ values, such as Ark 564, have systematically higher $A_0$ values than sources with low $\lambda_\mathrm{X}$ values, such as NGC 4051.

\begin{figure}
\centering
\includegraphics[width=\hsize]{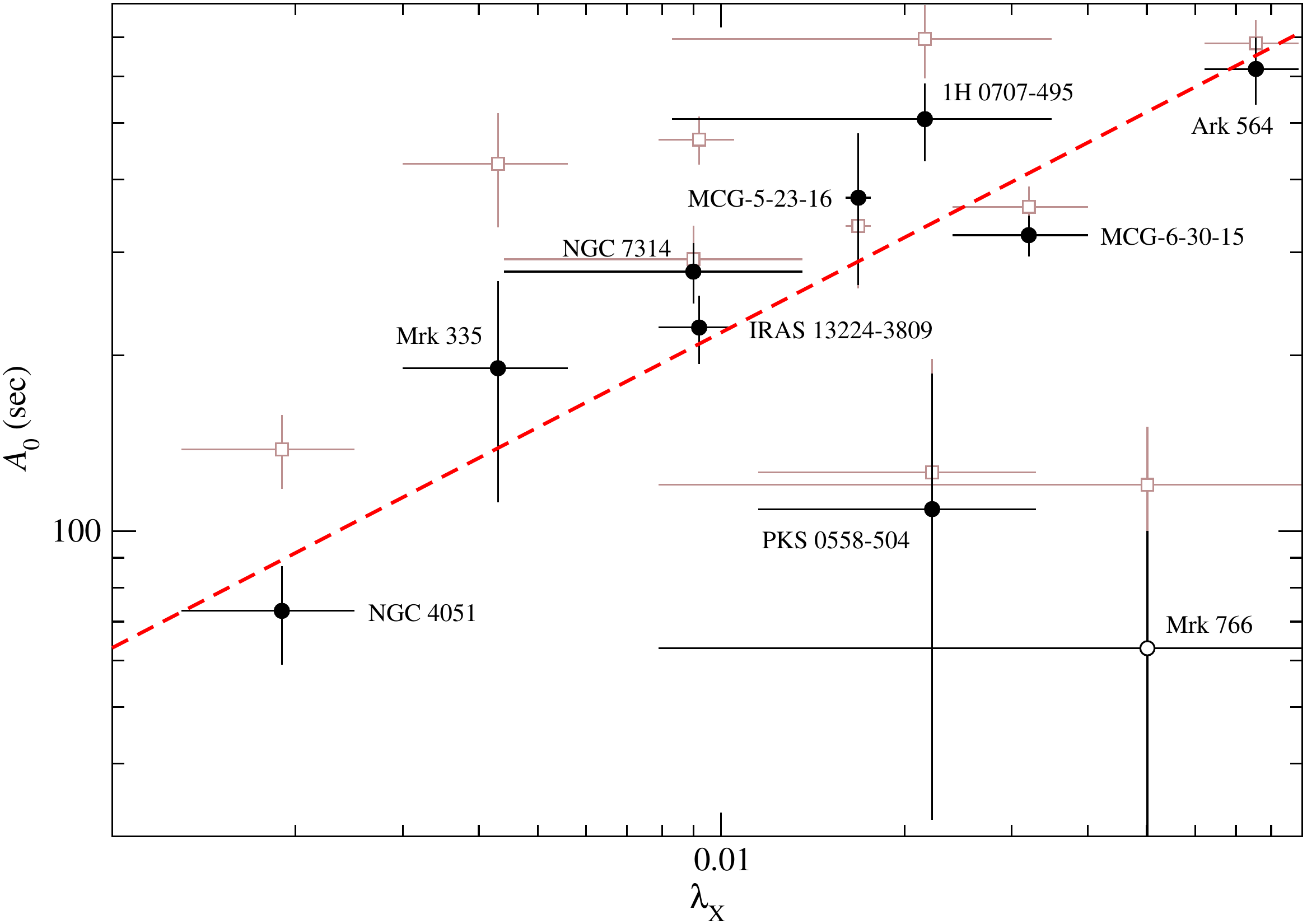}
\caption{The time-lags amplitude, $A_0$, as a function of the X-ray Eddington ration, $\lambda_\mathrm{X}$ (filled black circles). The open brown squares show the corresponding amplitude when subtracting the model reverberation time-lag component (see Section \ref{sec513} for details). The red dashed line indicates the best-fitting power-law model to the data.}
\label{fig8}
\end{figure}

Figure \ref{fig8} shows a plot of $A_0$ as a function of $\lambda_{\mathrm{X}}$. The plot shows that $A_0$ and $\lambda_\mathrm{X}$ are positively correlated, except perhaps in the case of PKS 0558--504 and Mrk 766 (in particular). We fitted the data with a power-law model of the form $A_0=B\lambda_\mathrm{X}^\beta$ (excluding Mrk 766 at first). The best-fitting model is indicated by the red dashed line in the same figure. The model fits the data well ($\chi^2_{\mathrm{min}}/\mathrm{dof}=9.1/7$). The best-fitting model parameter values are $B=3.42\pm0.13$ and $\beta=0.55\pm0.07$. These values do not change significantly even when we include the Mrk 766 data in the fit; though the statistical quality of the fit worsens ($\chi^2_{\mathrm{min}}/\mathrm{dof}=16.9/8$), it remains acceptable ($p_{\mathrm{null}}>0.01$). We thus conclude that the magnitude of the continuum time-lags in AGN scales approximately with the square root of the X-ray Eddington ratio.

Our final conclusion is that the X-ray continuum time-lags in AGN follow a relation of the form:
\noindent
\begin{equation} \label{eq10}
\tau(\nu;E_1,E_2)\sim\left[2600\sqrt{\lambda_\mathrm{X}}\log\left(\frac{E_2}{E_1}\right)\right]\times\left(\frac{\nu}{10^{-4}\,\mathrm{Hz}}\right)^{-1}\,\mathrm{sec}.
\end{equation}
\noindent
For a given light-curve energy separation the continuum time-lags are inversely proportional to frequency, and, at a given frequency, their amplitude increases logarithmically with the light-curve mean-energy ratio, ($E_2/E_1$). For a given light-curve energy separation and at a given frequency, the continuum time-lags increase with the square root of the X-ray Eddington ratio of an AGN.

\subsubsection{Effects of X-ray reverberation} \label{sec513}

Depending on the X-ray source and disc geometry, a significant amount of X-rays may illuminate disc and be reflected. Due to the different light travel paths between photons arriving directly from the source and those reflected off the surface of the disc, variations in the reprocessed disc emission are expected to be delayed with respect to variations in the X-ray continuum. The  magnitude of these delays will depend on the size and location (with respect to the disc) of the X-ray source, the viewing angle, the mass and spin of the BH, as well as the ionization state of the disc.

Since we chose $2-4\,\mathrm{keV}$ as the reference band, the sign of the X-ray reverberation time-lags should be opposite to the sign of the continuum time-lags at soft energies $\lesssim2\,\mathrm{keV}$ (soft lags). At harder energies, both the reverberation and the continuum time-lags have the same sign. \citet{2016A&A...594A..71E} showed that, under general assumptions, the observed time-lag spectra at each frequency should be equal to the sum of the continuum plus the reverberation time-lag component. Therefore, the amplitude of the continuum time-lags when the energy is below (above) $3\,\mathrm{keV}$, which is the mean energy of our reference band, may be underestimated (overestimated). To disentangle the two time-lag components we must model the reverberation time-lags as well. This was performed by E16, who modelled the $2-4$ vs. $5-7\,\mathrm{keV}$ time-lags in the context of a simple lamp-post geometry. However, it is beyond the scope of the present work to fit the observed time-lag spectra at all energies, for all sources, in this way. We thus performed a simpler test to get an estimate of the strength of the reverberation time-lag component in each case.

To calculate the theoretical X-ray reverberation time-lag spectra, we used the model of Dov\v{c}iak et al. (2017; in prep.), which is similar to the model used by E16 (see section 4 in their paper). The most important feature of the new version of the model is that it takes disc ionization into account to determine the X-ray reflection spectrum. In this way, the model can accurately predict the reverberation time-lag spectra at energies below $\sim2\,\mathrm{keV}$ as well. As input model parameters we used the BH mass and the X-ray Eddington ratio estimates listed in column 1 of Table \ref{table1}. We set the X-ray source height to $7.5r_{\mathrm{g}}$, which is representative of the mean source height for the sources E16 considered (see their fig. 4). We set the accretion disc density (which was assumed to have a constant radial profile) to $10^{15}\,\mathrm{cm}^{-3}$, and the X-ray source photon index to $\Gamma=2$.

We then calculated, for all sources, the model reverberation time-lag amplitude at $10^{-4}\,\mathrm{Hz}$ as a function of energy, using $2-4\,\mathrm{keV}$ as the reference band. Our results are shown as open brown squares in Fig. \ref{fig3}. We then subtracted these values from the amplitudes $A(\overline{E},3\,\mathrm{keV})$ determined from the observed time-lag spectra (represented by the filled black circles in the same figure), to determine the amplitude of the hard lags only, $A_{{\rm hard}}(\overline{E},3\,\mathrm{keV})$. Then, we fitted the $A_{{\rm hard}}({\overline E},3\,{\rm keV})$ vs. $\overline{E}/(3\,{\rm keV})$ data, using the same model that we used to fit the original data (defined by equation \ref{eq6}). The resulting best-fitting $A_{0,\mathrm{hard}}$ values are plotted as open brown squares in Figs. \ref{fig7} and \ref{fig8}, respectively.
 
These points suggest that X-ray reverberation is unlikely to explain our results. For example, even the $A_{0,{\rm hard}}$ values show a significant scatter around their mean, without an indication of a correlation with $M_{\mathrm{BH}}$. Furthermore, Fig. \ref{fig8} shows that $A_{0, {\rm hard}}$ and $\lambda_{\mathrm{X}}$ are still positively correlated. We fitted the $A_{0,{\rm hard}}$ vs. $\lambda_{\mathrm{X}}$ data with a power-law model; the best-fitting slope value is $0.41\pm0.06$, which is still consistent with 0.5 at the $\sim1.5\sigma$ level. We believe that this result demonstrates that the dependence of the continuum time-lag amplitude on the square root of the X-ray Eddington ratio still holds in all likelihood. We therefore conclude that, on average, our results are not significantly affected by the (possible) dilution of the hard lags by a soft-lag component (like the one expected in the case of X-ray reverberation).

\subsection{Summary of the intrinsic-coherence analysis} \label{sec52}

We presented the results from a detailed investigation of the the statistical properties of standard Fourier-based intrinsic coherence estimates. We provide practical `guidelines' (see Section \ref{seca5}) for constructing an intrinsic coherence estimator that is minimally biased, and has known, reliable errors. Our results indicate that the distribution of the intrinsic coherence estimates at frequencies lower than $\nu_{\rm max}$ (defined by equation \ref{eqa6}) is similar to a Gaussian. Consequently, they can be used to model the intrinsic coherence using traditional $\chi^2$ minimisation techniques. We stress that this is an approximate result. Strictly speaking, the distribution of the intrinsic coherence estimates, especially at frequencies close to $\nu_{\rm max}$, is almost certainly not a Gaussian. If a model fails to fit the the observed intrinsic coherence, the results should be treated with caution. At the very least, the data should be fitted up to frequencies $\sim \nu_{\rm max}/2$, as the hypothesis of Gaussianity should be more appropriate at these frequencies. Perhaps the most interesting result for practical applications is that the range `$\pm1\times{\rm (corrected) error}$' (`$\pm2\times{\rm (corrected) error}$') corresponds to the $\sim68$ per cent ($\sim95$ per cent) confidence interval of the intrinsic coherence estimates.

Using the available \textit{XMM-Newton} data for the sourves in our sample, we managed to estimate their intrinsic coherence at frequencies between $\sim5\times10^{-5}\,\mathrm{Hz}$ and $\sim1.5\times10^{-3}\,\mathrm{Hz}$. Our results are summarised below:
\begin{itemize}
\item[1.] For a given light-curve energy separation, the intrinsic coherence is approximately constant at low frequencies. This constant level depends logarithmically on the light-curve mean-energy ratio (see Fig. \ref{fig4}).
\item[2.] For half the sources in our sample (IRAS 13224--3809, 1H 0707--495, MGC--6-30-15, NGC 4051, and Ark 564) the intrinsic coherence decreases exponentially with increasing frequency above a certain break-frequency (see the relevant figures in Appendix \ref{appb}). The break frequency depends logarithmically on the light-curve mean-energy ratio (see Fig. \ref{fig5}).
\end{itemize}

\subsubsection{The low-frequency constant intrinsic-coherence value} \label{sec521}

In some cases, the low-frequency constant intrinsic-coherence value is consistent with one (perfect coherence), at all energies (e.g. IRAS 13224--3809, MCG--5-23-16, and NGC 7314; the Group A sources). For most sources, this constant level is smaller than one and decreases with increasing light-curve energy separation (see Fig. \ref{fig5}, and equation \ref{eq8}). Its energy dependence is not the same in all sources; in some cases it decreases rapidly as the energy separation increases (e.g. PKS 0558--504, Mrk 766, and Mrk 335; the Group C sources), while in the remaining sources (1H 0707--495, MCG--6-30-15, NGC 4051, and Ark 564; the Group B sources), the dependance is less steep.

We found no universal scaling of the constant level (for a given energy separation) with either the BH mass or the X-ray Eddington ratio for the AGN in our sample. Its value is, however, consistent with one when the energy separation, parametrised by $|\log(E_2/E_1)|$, is smaller than $\sim0.2$ for all sources.

\subsubsection{The high-frequency break} \label{sec522}

\begin{figure}
\centering
\includegraphics[width=\hsize]{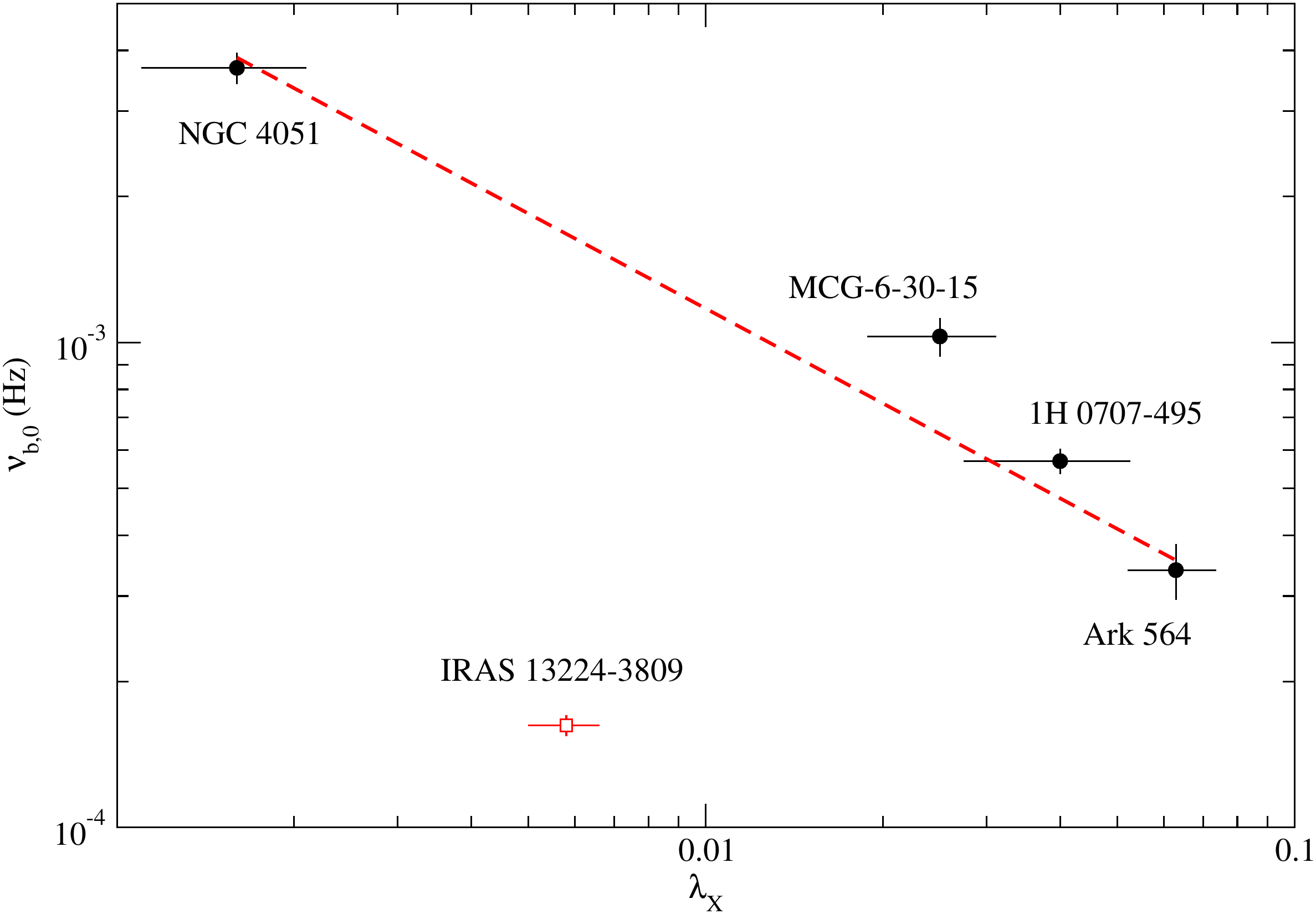}
\caption{Best-fitting intrinsic coherence break-frequency, $\nu_{\mathrm{b},0}$, as a function of the X-ray Eddington ratio, $\lambda_{\mathrm{X}}$.}
\label{fig9}
\end{figure}

Figure \ref{fig9} shows the break frequency for a given energy separation, $\nu_{\mathrm{b},0}$ (as defined by equation \ref{eq9}), as a function of $\lambda_{\mathrm{X}}$. We observe a strong anti-correlation between $\nu_{\mathrm{b},0}$ and $\lambda_\mathrm{X}$, but only for the Group B sources. The IRAS 13224--3809 data (open red square in the same figure) are not consistent with the other sources. For the Group B sources, we fitted the $\nu_{\mathrm{b},0}$ vs. $\lambda_\mathrm{X}$ data with a power-law model of the form $\nu_{\mathrm{b},0}\propto\lambda_{\mathrm{X}}^\gamma$. The best-fitting slope value is $\gamma=-0.65\pm0.08$, and the best-fitting model is indicated with the red dashed line in the same figure. This trend is very similar to the trend of the continuum time-lag amplitude with $\lambda_{\mathrm{X}}$.

However, it is not clear whether the intrinsic coherence functions of all AGN exhibit high-frequency breaks, or whether the corresponding break frequencies have the same dependence on $\lambda_\mathrm{X}$ as those exhibited by the Group B sources (for IRAS 13224--3809, we know that this is not the case). Our inability to better constrain the break frequencies in half the sources of our sample is not exactly due to the lack of good quality data. For example, NGC 7314 hosts a BH with a low mass, but its $1-2$ vs. $2-4\,\mathrm{keV}$ intrinsic coherence estimates (Fig. \ref{figb18}) do not indicate a high-frequency break, even though they are reliably estimated over a frequency range comparable to the corresponding data of e.g. NGC 4051 (Fig. \ref{figb6}), which has a similarly low BH mass estimate. The same remark holds true for other sources as well. It is therefore possible that the phenomenological differences regarding the energy dependence of the intrinsic coherence for the Group A, B, and C sources are real, although we cannot identify the common physical parameter that characterises the AGN of each Group.

\subsection{Implications of our results} \label{sec53}

The results of our work, which are based on a quantitative (rather than a qualitative) analysis of the observed low-frequency time-lags and intrinsic coherence, can, in principle, be used to constrain theoretical models of AGN X-ray variability. For example, we discuss below a few implications of our results in the context of the so-called propagating fluctuations model \citep[e.g.][]{1997MNRAS.292..679L,2001MNRAS.327..799K,2006MNRAS.367..801A}, which can explain many X-ray variability properties of compact accreting systems. According to this model, fluctuations in the mass accretion rate of the disc are produced at different radii, and then propagate to the centre. The fluctuations are coupled, in the sense that low-frequency fluctuations produced at large radii modulate higher-frequency fluctuations produced further in.

\subsubsection{Continuum time-lags} \label{sec531}

In the propagating fluctuations model, the characteristic fluctuation time-scale at a radius $R$ is assumed to correspond to the viscous time-scale at that radius, $t_{\mathrm{visc}}(R)$. In the context of standard thin-disc theory \citep{1973A&A....24..337S}, this time-scale is given by $t_{\mathrm{visc}}(R)=[(R/R_{\mathrm{g}})^{3/2}(H/R)^{-2}\alpha^{-1}]\,t_{\mathrm{g}}$, where $(H/R)$ is the disc scale-height to radius ratio, $\alpha$ is the viscosity parameter, and $t_{\mathrm{g}}\equiv R_{\mathrm{g}}/c\sim5\,(M_{\mathrm{BH}}/M_6)\,\mathrm{sec}$ ($R_{\mathrm{g}}\equiv GM_{\mathrm{BH}}/c^2$ is the gravitational radius). Assuming that a) the fluctuations move inwards with a speed corresponding to the so-called drift velocity, $v(R)=R/t_{\mathrm{visc}}(R)=[(R/R_{\mathrm{g}})^{-1/2}(H/R)^2\alpha]\,c$, and b) the emissivity profile of the disc is energy-dependent, with harder photons being produced closer to the centre, the model predicts that $\tau(\nu;E_1,E_2)\propto\nu^{-1}\log(E_2/E_1)$. This is entirely consistent with our results (see Section \ref{sec512}).

While the model was initially developed for XRBs, in the case of AGN the X-ray emission cannot be produced by the disc. For the model to be applicable to AGN the disc fluctuations must therefore propagate to an extended X-ray source, which should have an emissivity profile that hardens closer to the centre. We will henceforth assume that the time-scales of the fluctuations that propagate through the X-ray source, as well as their inward-propagation speed, is identical to what is assumed in standard thin-disc theory.

The continuum time-lags should flatten below a certain characteristic frequency, $\nu_{\mathrm{flat}}$, which corresponds to the viscous time-scale at the outer radius of the X-ray source. This flattening could explain the turn-over in the observed continuum time-lags at frequencies $\sim10^{-4}\,\mathrm{Hz}$ we detected in 1H 0707--495 and MCG--6-30-15. Assuming that $\alpha\sim0.1$ and $(H/R)\sim0.1$ (the standard values adopted in the thin-disc approximation), and that the X-ray source has a size $\sim10R_{\mathrm{g}}$ \citep[as suggested by quasar microlensing studies; e.g.][]{2016AN....337..356C}, we get $\nu_{\mathrm{flat}}\sim 1/t_{\mathrm{visc}}(10R_{\mathrm{g}})\sim6\times10^{-6}(M_{\mathrm{BH}}/M_6)^{-1}\,\mathrm{Hz}$. For $M_{\mathrm{BH}}=1.7M_6$ (the weighted-mean value for 1H 0707--495 and MCG--6-30-15), this gives $\nu_{\mathrm{flat}}\sim4\times10^{-6}\,\mathrm{Hz}$, which is below what we observe. A very small X-ray emitting region of size $\sim1R_{\mathrm{g}}$ (as inferred from X-ray reverberation studies of AGN) is required to explain this discrepancy. It might therefore be possible that the low frequency turn-over in the observed time-lag spectra is caused by the above effect.

As shown by \citet{2006MNRAS.367..801A}, the typical time-lag magnitudes predicted by the model are $\sim1-10$ per cent of the variability time-scale; i.e. $\sim10^{2-3}\,\mathrm{sec}$ at $10^{-4}\,\mathrm{Hz}$. According to equation \ref{eq10}, for a typical AGN in our sample with $\lambda_{\mathrm{X}}\sim0.04$ and for energy separation values $\log(E_2/E_1)\sim0.3-0.9$ (the total range we considered), the corresponding magnitudes are $\sim100-500\,\mathrm{sec}$, i.e. $1-5$ per cent of the variability time-scale. This is consistent with the model prediction. However, the observed scaling of the time-lag magnitudes with the square root of the X-ray Eddington ratio appears difficult to explain. Assuming that $\alpha$ remains the same for all sources, the aforementioned scaling implies that $(H/R)$ should increase with increasing X-ray Eddington ratio. This is contrary to what one would expect if AGN are simply scaled-up versions of XRBs, as in the latter it is generally believed that an increase in the accretion rate (of which the X-ray Eddington ratio is a proxy) results a `thinner' disc (and vice-versa). This discrepancy is therefore perhaps one of the most interesting results of our work, which could constrain AGN X-ray variability models.

\subsubsection{Intrinsic coherence} \label{sec532}

As discussed by VN97, a (near-)unity intrinsic coherence between X-ray emission in any two energy bands is generally expected to be the exception rather than the rule. This is because unity coherence would imply that the corresponding fluxes are related by a linear transformation. Our results are thus broadly consistent with this expectation, as we find evidence for near-unity intrinsic coherence values only for three out of the ten sources we studied (IRAS 13224--3809, MCG--5-23-16, and NGC 7314).

In the context of the propagating fluctuations model, the intrinsic coherence depends on the (unknown) power-spectral density function (PSD) of the accretion rate fluctuations. For example, \citet{2006MNRAS.367..801A} considered the case whereby the intrinsic PSD of the accretion rate fluctuations generated at each radius has the shape of a Lorentzian function centred at the local viscous frequency; they showed that, the narrower the Lorentzian is, the closer the intrinsic coherence between any two energy bands is to unity. This is because the observed variability a given frequency will have contributions from incoherent fluctuations originating from several radii, which will, in fact, increase in number as the energy separation increases. Moreover, the loss of coherence becomes more severe at higher frequencies, as there is increasingly less variability at time-scales shorter than the viscous time-scale of the inner-most X-ray source radius.  Our results regarding the shape of the observed X-ray intrinsic coherence in AGN are thus in broad agreement with the aforementioned theoretical expectations.

Another mechanism that can lead to a loss of coherence is the presence of a variable warm absorber. In the case of NGC 4051, the presence of a warm absorber has been shown to lead to a smaller loss of coherence that what is observed. Moreover, this warm absorber should cause an almost uniform loss of coherence over all frequencies, contrary to the observed exponential decrease at high frequencies \citep[compare the $0.3-1$ vs. $2-4\,\mathrm{keV}$ intrinsic coherence panel in Fig. \ref{figb6} with the bottom panel of fig. 10 in][]{2016A&A...596A..79S}. Therefore, it appears unlikely that a variable warm absorber alone can explain the observed loss of coherence in NGC 4051.

As discussed in Section \ref{sec522}, our results indicate that (contrary to the continuum time-lags) the intrinsic coherence does not appear to have a universal energy- and frequency-dependence that scales with either the BH mass, or the accretion rate in the sources we studied. This argues for the existence of an additional physical parameter, whose determination poses an interesting challenge to AGN X-ray variability models.

\section*{Acknowledgements} 
We thank the referee for their helpful comments and suggestions. This work was supported in part by the AGNQUEST project, which was implemented under the Aristeia II Action of the Education and Lifelong Learning operational programme of the GSRT, Greece. It has also received funding from the European Research Council under the European Union's Seventh Framework Programme (FP/2007-2013) / ERC Grant Agreement n. 617001. This work has made use of: a) the NASA/IPAC Extragalactic Database (NED) which is operated by the Jet Propulsion Laboratory, California Institute of Technology, under contract with the National Aeronautics and Space Administration, and b) data provided by the University of California, San Diego Center for Astrophysics and Space Sciences, X-ray Group (R.E. Rothschild, A.G. Markowitz, E.S. Rivers, and B.A. McKim), obtained at http://cass.ucsd.edu/$\sim$rxteagn/.

\bibliographystyle{mnras}
\bibliography{refs}

\newpage

%%%%%%%%%%%%%%%%%%%%%%%%%%%%%%%%%%%%%%%%%%%%%%%%%%

%%%%%%%%%%%%%%%%% APPENDICES %%%%%%%%%%%%%%%%%%%%%

\appendix

%%%%%%%%%%%%%%%%%%%%%%%%%%%%%%%%%%%%%%%%%%
%%%%%%%%%%%%%%%%%%%%%%%%%%%%%%%%%%%%%%%%%%
\section{The intrinsic coherence estimate} \label{appa}
%%%%%%%%%%%%%%%%%%%%%%%%%%%%%%%%%%%%%%%%%%
%%%%%%%%%%%%%%%%%%%%%%%%%%%%%%%%%%%%%%%%%%

EP16 discussed the effects of the measurement errors in the coherence of two processes in their appendix C. Following their notation, we denote with $\gamma^2_{XY}(\nu)$ the intrinsic coherence of the discrete version of two continuous random processes (discretisation is almost unavoidable in every observation of a continuous signal), and with $\gamma^2_{XY,\mathrm{n}}(\nu)$ the coherence of the discrete processes in the presence of measurement noise. EP16 demonstrated that $\gamma^2_{XY,\mathrm{n}}(\nu)$ is always smaller than $\gamma^2_{XY}(\nu)$, at all frequencies. In fact, $\gamma^2_{XY,\mathrm{n}}$ will tend to zero (irrespective of the true value of $\gamma^2_{XY}$) at frequencies where the amplitude of the noise variations is significantly larger than the amplitude of the intrinsic variations. EP16 also showed that the coherence estimate, $\hat{\gamma}^2_{xy}(\nu_p)$ (equation \ref{eq3}), is a biased estimate even of $\gamma^2_{XY,\mathrm{n}}(\nu)$ (let alone $\gamma^2_{XY}(\nu)$): at frequencies where $\gamma^2_{XY,\mathrm{n}}(\nu)$ tends to zero, the mean of $\hat{\gamma}^2_{xy}(\nu_p)$ will tend to $\sim 1/m$, where $m$ is the number of light curve segments. VN97 proposed the following estimator of the intrinsic coherence (i.e. $\gamma^2_{XY}(\nu)$):
\noindent
\begin{equation} \label{eqa1}
\hat{\gamma}^2_{\mathrm{int},xy}(\nu_p)=\frac{|\hat{C}_{xy}(\nu_p)|^2-|\hat{\varsigma}(\nu_p)|^2}{[\hat{P}_x(\nu_p)-P_{\epsilon_x}][\hat{P}_y(\nu_p)-P_{\epsilon_y}]},
\end{equation}
\noindent
where
\noindent
\begin{equation} \label{eqa2}
|\hat{\varsigma}(\nu_p)|^2=\frac{1}{m}[\hat{P}_x(\nu_p)P_{\epsilon_y}+\hat{P}_y(\nu_p)P_{\epsilon_x}-P_{\epsilon_x}P_{\epsilon_y}],
\end{equation}
\noindent
and $\{P_{\epsilon_x},P_{\epsilon_y}\}$ are the power spectra of the experimental noise components in the observed light curves (which are usually constant at all frequencies).

VN97 described various recipes for estimating the error of $\hat{\gamma}^2_{\mathrm{int},xy}(\nu_p)$ in different frequency regimes, depending on the relative strength of the experimental noise over the intrinsic variations. When the latter dominate over the former, VN97 suggested the following analytic estimate for the error of $\hat{\gamma}^2_{\mathrm{int},xy}(\nu_p)$:
\noindent
\begin{align} \label{eqa3}
\nonumber
\hat{\sigma}_{\hat{\gamma}^2_{\mathrm{int}}}(\nu_p)=\frac{\hat{\gamma}^2_{\mathrm{int},xy}(\nu_p)}{\sqrt{m}}\Big\{\frac{2|\hat{\varsigma}(\nu_p)|^4m}{[|\hat{C}_{xy}(\nu_p)|^2-|\hat{\varsigma}(\nu_p)|^2]^2} \\
+\left[\frac{P_{\epsilon_x}}{\hat{P}_x(\nu_p)-P_{\epsilon_x}}\right]^2+\left[\frac{P_{\epsilon_y}}{\hat{P}_y(\nu_p)-P_{\epsilon_y}}\right]^2+m\left[\frac{\sigma_{\hat{\gamma}^2}(\nu_p)}{\hat{\gamma}^2_{xy}(\nu_p)}\right]^2\Big\}^{1/2}.
\end{align}
\noindent
Equations \ref{eqa1} and \ref{eqa3} are often used to estimate the intrinsic coherence between light curves in different energy bands in the context of both AGN and XRB X-ray variability studies.

One of the aims of this work is to study the statistical properties of $\hat{\gamma}^2_{\mathrm{int},xy}(\nu_p)$, namely: a) its bias (i.e. the difference between its mean value and $\gamma^2_{XY}(\nu)$), b) how well equation \ref{eqa3} approximates the true scatter of $\hat{\gamma}^2_{\mathrm{int},xy}(\nu_p)$ around its mean, and c) its probability distribution. To our knowledge, the results from such a study have not been reported in the literature so far. We used the same simulated light curves that EP16 used in their study. For completness, we summarise below the way EP16 constructed these light curves.

%%%%%%%%%%%%%%%%%%%%%%%%%%%%%%%%%%%%%%%%%%
\subsection{Simulation setup} \label{seca1}
%%%%%%%%%%%%%%%%%%%%%%%%%%%%%%%%%%%%%%%%%%

We considered three different numerical experiments, each corresponding to a different prescribed time-lag spectrum: a) a constant time-lag spectrum of $10\,\mathrm{sec}$ at each frequency (henceforth, experiment CD), b) a power-law time-lag spectrum of the form $100(\nu/10^{-4}\,\mathrm{Hz})^{-1}\,\mathrm{sec}$ (henceforth, experiment PLD), and c) a time-lag spectrum expected when the two random processes are related by a constant response function equal to $0.2/(200\,\mathrm{sec})$ for $|t-200\,\mathrm{sec}|\le100\,\mathrm{sec}$, and zero otherwise (henceforth, experiment THRF). As discussed in EP16, these functions are frequently used to model the observed X-ray time-lag spectra in AGN. In all cases, we assumed unity intrinsic coherence at all frequencies.

For each numerical experiment we generated 100 light-curve pairs with a duration of $10.24\,\mathrm{Ms}$, and a sampling rate of $1\,\mathrm{sec}$. We followed \citet{1995A&A...300..707T} to generate the light curves, assuming a `smoothly-bending' power-law PSD with low-frequency slope $-2$, high frequeny slope $-1$, and `bend-frequency' $2\times10^{-4}\,\mathrm{Hz}$. The original light curves were subsequently binned at $100\,\mathrm{sec}$ and chopped into 500 $20\,\mathrm{ks}$-segments, to simulate the effects of finite binning and light-curve duration. For each numerical experiment we thus ended up with $500\times100=5\times10^4$ light-curve segments of $20\,\mathrm{ks}$ duration (LS20 light curves, hereafter). To simulate the effects of measurement errors, we created five copies of each LS20 light-curve pair corresponding to a different S/N combination, $\{(\mathrm{S/N})_x,(\mathrm{S/N})_y\}$: $\{3,3\}$, $\{9,3\}$, $\{18,3\}$, $\{9,9,\}$, and $\{18,9\}$. We then added a Gaussian random number of zero mean and appropriate variance to each point of the LS20 light curves with a given S/N combination.

We calculated the $m=10$, $20$, $30$, and $40$ averaged cross-periodogram and periodograms to calculate the intrinsic coherence estimate, along with its analytic error, according to equations \ref{eqa1} and \ref{eqa3}, respectively, and did not consider frequencies above $\nu_{\mathrm{crit}}$. The number of intrinsic coherence estimates in each experiment and every S/N combination were thus 5000, 2500, 1666, and 1250. Figures \ref{figa4}, \ref{figa6}, \ref{figa8}, \ref{figa10}, and \ref{figa12} at the end of this appendix show our results. Each column in the these figures corresponds to a different S/N combination. Black circles, green squares, and blue diamonds correspond to experiment PLD, CD, and THRF, respectively.

The mean sample intrinsic coherence, $\langle\hat{\gamma}^2_{\mathrm{int},xy}(\nu_p)\rangle$, is plotted in the top rows. The horizontal dotted lines indicate the intrinsic coherence, which is equal to one (by the way we constructed the simulated light curves). We plot the mean `error ratio' in the middle row panels. This is defined as the ratio of the mean analytic error, $\langle\hat{\sigma}_{\hat{\gamma}^2_{\mathrm{int}}}(\nu_p)\rangle$, over the standard deviation of the intrinsic coherence estimates, $\sigma_{\hat{\gamma}^2_{\mathrm{int}}}(\nu_p)$. In the bottom panels we plot the  probability $p_{1\hat{\sigma}}$ and $p_{2\hat{\sigma}}$, that the intrinsic coherence estimates lie within 1 and $2\langle\hat{\sigma}_{\hat{\gamma}^2_{\mathrm{int}}}\rangle$, respectively. The horizontal dotted lines indicate the values of 0.68 and 0.95, which correspond to the percentage of points that lie within 1 and 2$\sigma$ around the mean for a Gaussian random variable.

%%%%%%%%%%%%%%%%%%%%%%%%%%%%%%%%%%%%%%%%%%
\subsection{Bias of the intrinsic coherence estimate} \label{seca2}
%%%%%%%%%%%%%%%%%%%%%%%%%%%%%%%%%%%%%%%%%%

The top row panels in Figs. \ref{figa4}, \ref{figa6}, \ref{figa8}, \ref{figa10}, and \ref{figa12} show that the mean sample intrinsic coherence is close to unity (i.e. it is equal to the intrinsic coherence) at low frequencies. At higher frequencies it increases (in most cases), and then decreases (the scatter increases steadily with increasing frequency). The pattern is similar for all three numerical experiments (within the scatter of the points), which suggests that our results are probably independent of the intrinsic CS of the time series.

To investigate the bias of  $\hat{\gamma}^2_{\mathrm{int},xy}(\nu_p)$ in more detail, we first averaged the mean sample intrinsic coherence obtained from each experiment at every frequency, and then binned the resulting values over neighbouring frequencies with a logarithmic step of 1.2 (except from the two lowest frequency points). In this way we reduced the scatter of the mean sample intrinsic coherence, which increases substantially at high frequencies.\footnote{We note that, unlike the case with binned CS estimates, averaging of the intrinsic coherence estimates does not introduce any significant bias. The reason is that the mean sample intrinsic coherence does not appear to be a steep function of frequency, contrary to the case of the mean sample real and imaginary parts of the CS (see EP16 for a detailed discussion regarding the bias of CS estimates).} The resulting intrinsic coherence values are shown as filled brown up-triangles in the first row panels of Figs. \ref{figa4}, \ref{figa6}, \ref{figa8}, \ref{figa10}, and \ref{figa12}. The binned mean sample intrinsic coherence shows an increase after a certain maximum frequency, $\nu_{\mathrm{max}}$, followed by a rather steep decrease in many cases. We conclude that the intrinsic coherence estimates defined by equation \ref{eqa1} are biased estimates of the intrinsic coherence at frequencies higher than $\nu_{\rm max}$.

It would be desirable to predict analytically the bias of $\hat{\gamma}^2_{\mathrm{int},xy}(\nu)$ or, equivalently, to obtain an analytical prescription to calculate $\nu_{\mathrm{max}}$, however this is a difficult task. According to equation \ref{eqa1},
\noindent
\begin{align} \label{eqa4}
\nonumber
\mathrm{E}[\hat{\gamma}^2_{\mathrm{int},xy}(\nu_p)] &=\mathrm{E}\left\{\frac{|\hat{C}_{xy}(\nu_p)|^2-|\hat{\varsigma}(\nu_p)|^2}{[\hat{P}_x(\nu_p)-P_{\epsilon_x}][\hat{P}_y(\nu_p)-P_{\epsilon_y}]}\right\} \\
&= \frac{\mathrm{E}[|\hat{C}_{xy}(\nu_p)|^2]-\mathrm{E}[|\hat{\varsigma}(\nu_p)|^2]}{\mathrm{E}[\hat{P}_x(\nu_p)-P_{\epsilon_x}]\mathrm{E}[\hat{P}_y(\nu_p)-P_{\epsilon_y}]}+\ldots,
\end{align}
\noindent
where $\mathrm{E}$ denotes the expectation operator, and the dots indicate higher-order terms. If we assume that $\mathrm{E}[|\hat{C}_{xy}(\nu_p)|^2]\sim|C_{xy}(\nu_p)|^2+|\varsigma(\nu_p)|^2$, $\mathrm{E}[|\hat{\varsigma}(\nu_p)|^2]\sim|\varsigma(\nu_p)|^2\sim(1/m)(P_xP_{\epsilon_y}+P_yP_{\epsilon_x}+P_{\epsilon_x}P_{\epsilon_y})$, $\mathrm{E}[\hat{P}_x(\nu_p)]\sim P_x+P_{\epsilon_x}$, and $\mathrm{E}[\hat{P}_y(\nu_p)]\sim P_y+P_{\epsilon_y}$, where $C_{xy}$, $P_x$, and $P_y$ are the intrinsic CS and PSDs of the measured process (i.e. in the absence of experimental noise), equation \ref{eqa4} becomes
\noindent
\begin{align} \label{eqa5}
\nonumber
\mathrm{E}[\hat{\gamma}^2_{\mathrm{int},xy}(\nu_p)] &=\frac{|C_{xy}(\nu_p)|^2}{P_xP_y}+\ldots \\
&= \gamma^2_{XY}(\nu_p)+\ldots,
\end{align}
\noindent
where the dots again denote higher-order terms. They are usually assumed to be small, however, our results indicate that these terms become increasingly important at high frequencies. An analytical estimation of the bias requires the calculation of the next-order terms in equation \ref{eqa5}, which are not known in closed form by us. We thus proceeded to investigate how to obtain an empirical recipe for estimating $\nu_{\mathrm{max}}$.

This frequency appears to depend mainly on the S/N ratio of the light curves: it increases with increasing signal-to-noise ratio. It also depends on $m$ (for a fixed S/N combination it increases with increasing $m$), but to a lesser degree. These results indicate that $\nu_{\mathrm{max}}$ should correspond to a characteristic time-scale where the experimental noise fluctuations start dominating over the fluctuations of the intrinsic, underlying signal. Given that $\nu_{\mathrm{crit}}$, as defined in Section \ref{sec3}, is a proxy of such a time-scale, we expect the two frequencies to be correlated. To test this, we defined $\nu_{\mathrm{max}}$ as the frequency where the mean sample intrinsic coherence becomes equal to 1.05 (which corresponds to a bias of 5 per cent), and computed it by interpolating the binned sample coherence values and equating the interpolated functions to 1.05.

\begin{figure}
   \centering
   \includegraphics[width=\hsize]{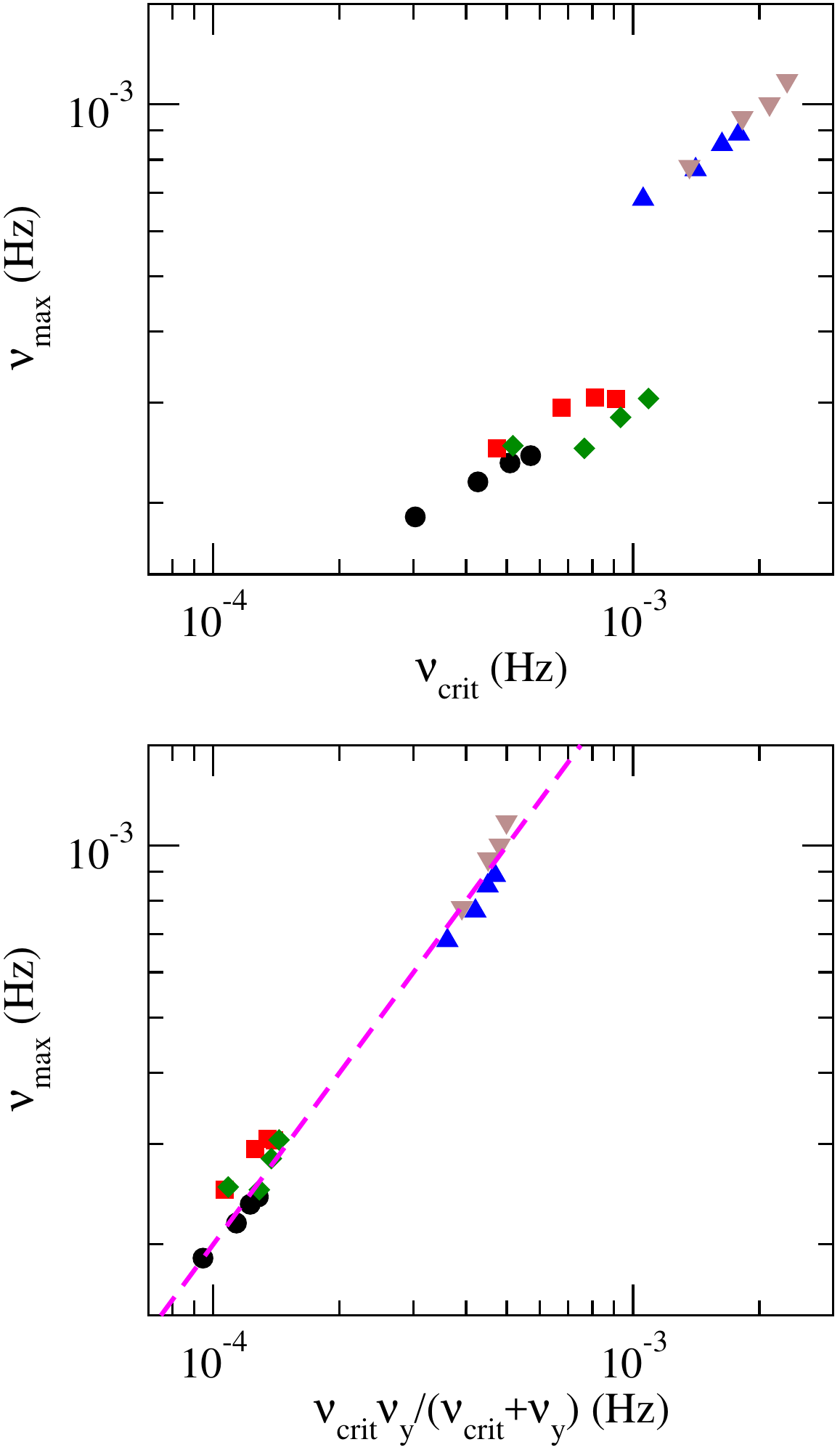}
      \caption{The maximum frequency, $\nu_{\rm max}$, which is the frequency where the bias of the intrinsic coherence estimates is equal to 1.05, as a function of $\nu_{\rm crit}$ (top panel) and $\nu_{\rm crit}\nu_{y}/(\nu_{\rm crit}+\nu_{y})$ (bottom panel; see text in Section \ref{seca2} for details).}
\label{figa1}
   \end{figure}

The top panel in Fig. \ref{figa1} shows $\nu_{\mathrm{max}}$ as a function of $\nu_{\mathrm{crit}}$. Black circles, red squares, green diamonds, blue up-triangles, and brown down-triangles show the points (for all the four different $m$ values) when $\{(\mathrm{S/N})_x,(\mathrm{S/N})_y\}=\{3,3\}$, $\{9,3\}$, $\{18,3\}$, $\{9,9\}$, and $\{18,9\}$, respectively. This plot confirms that $\nu_{\mathrm{max}}$ and $\nu_{\mathrm{crit}}$ are indeed positively correlated. On average, $\nu_{\mathrm{max}}$ increases with increasing $\nu_{\mathrm{crit}}$. It also shows that the correlation is different between the cases of low (black circles, red squares, and green diamonds) and high S/N (blue up-triangles and brown down-triangles in the same panel). This suggests that $\nu_{\mathrm{max}}$ also depends on an additional parameter that is related to the S/N of the light curve with the smallest mean count rate.

By trial and error, we discovered that the frequency $\nu_{\mathrm{crit}}\nu_y/(\nu_{\mathrm{crit}}+\nu_y)$, where $\nu_y$ is the frequency where the sample power spectrum of the light curve with the lowest S/N becomes equal to $4P_{\epsilon_y}/(1-m^{-1/2})$, is a better proxy of $\nu_{\mathrm{max}}$. We illustrate this fact in the bottom panel of Fig. \ref{figa1}, which shows $\nu_{\mathrm{max}}$ as a function of $\nu_{\mathrm{crit}}\nu_y/(\nu_{\mathrm{crit}}+\nu_y)$. The best-fitting relation in this case is $\nu_{\mathrm{max}}=2\nu_{\mathrm{crit}}\nu_y/(\nu_{\mathrm{crit}}+\nu_y)$. The dashed magenta line in the same panel shows this relation. We therefore suggest the following formula for estimating $\nu_{\mathrm{max}}$:
\noindent
\begin{equation} \label{eqa6}
\nu_{\mathrm{max}}=2\frac{\nu_{\mathrm{crit}}\nu_y}{\nu_{\mathrm{crit}}+\nu_y},
\end{equation}
\noindent
where $\nu_y$ is the frequency where the sample PSD of the light curve with the lowest S/N, $\hat{P}_y$, becomes equal to $4P_{\epsilon_y}/(1-m^{-1/2})$. The sample intrinsic coherence estimates defined by equation \ref{eqa1} are reliable estimates of the intrinsic coherence (in the sense that their bias should be $\lesssim5$ per cent) at frequencies lower than $\nu_{\mathrm{max}}$. The vertical (red) dashed lines in Figs. \ref{figa4}--\ref{figa13} indicate $\nu_{\rm max}$, which were computed using equation \ref{eqa6} in each case.

%%%%%%%%%%%%%%%%%% FIG. A1 %%%%%%%%%%%%%%

%%%%%%%%%%%%%%%%%%%%%%%%%%%%%%%%%%%%%%%%%%
\subsection{The error of the intrinsic coherence estimate} \label{seca3}
%%%%%%%%%%%%%%%%%%%%%%%%%%%%%%%%%%%%%%%%%%

The second row plots in Figs. \ref{figa4}, \ref{figa6}, \ref{figa8}, \ref{figa10}, and \ref{figa12} show that the error ratio remains approximately constant at frequencies below $\sim \nu_{\rm max}$, and then increases rapidly with increasing frequency. The frequency range over which the error ratio remains constant is roughly equal to the range between the lowest frequency and $\nu_{\rm max}$. On average, the error ratio at low frequencies is greater than unity when $m\le20$ while the ratio becomes roughly equal (or slightly smaller than) unity when $m\ge 30$. The fact that the error ratio becomes significantly larger than unity at frequencies higher than $\nu_{\rm max}$ indicates that the analytic error defined by equation \ref{eqa3} overestimates the standard deviation of the intrinsic coherence estimates in these cases. We suspect that the reason for this significant discrepancy is that equation \ref{eqa3} provides an error estimate based on the assumption that the distribution of the sample intrinsic coherence is Gaussian, which is far from true at high frequencies (see Section \ref{seca4} for a more detailed discussion on this issue).

The bottom panels in Figs. \ref{figa4}, \ref{figa6}, \ref{figa8}, \ref{figa10}, and \ref{figa12} show the percentage of the sample intrinsic coherence estimates that are within 1 and $2\langle\hat{\sigma}_{\hat{\gamma}^2_{\mathrm{int}}}\rangle$ of the sample mean, $p_{\rm 1\hat{\sigma}}$ and $p_{\rm 2\hat{\sigma}}$, respectively. The former is larger than 68 per cent, in most cases. Most of the sample intrinsic coherence estimates are closer to the sample mean than the analytic error predicts. The $p_{\rm 2\hat{\sigma}}$ values are also larger than 95 per cent, but only when $m=10$.

\begin{figure}
   \centering
   \includegraphics[width=\hsize]{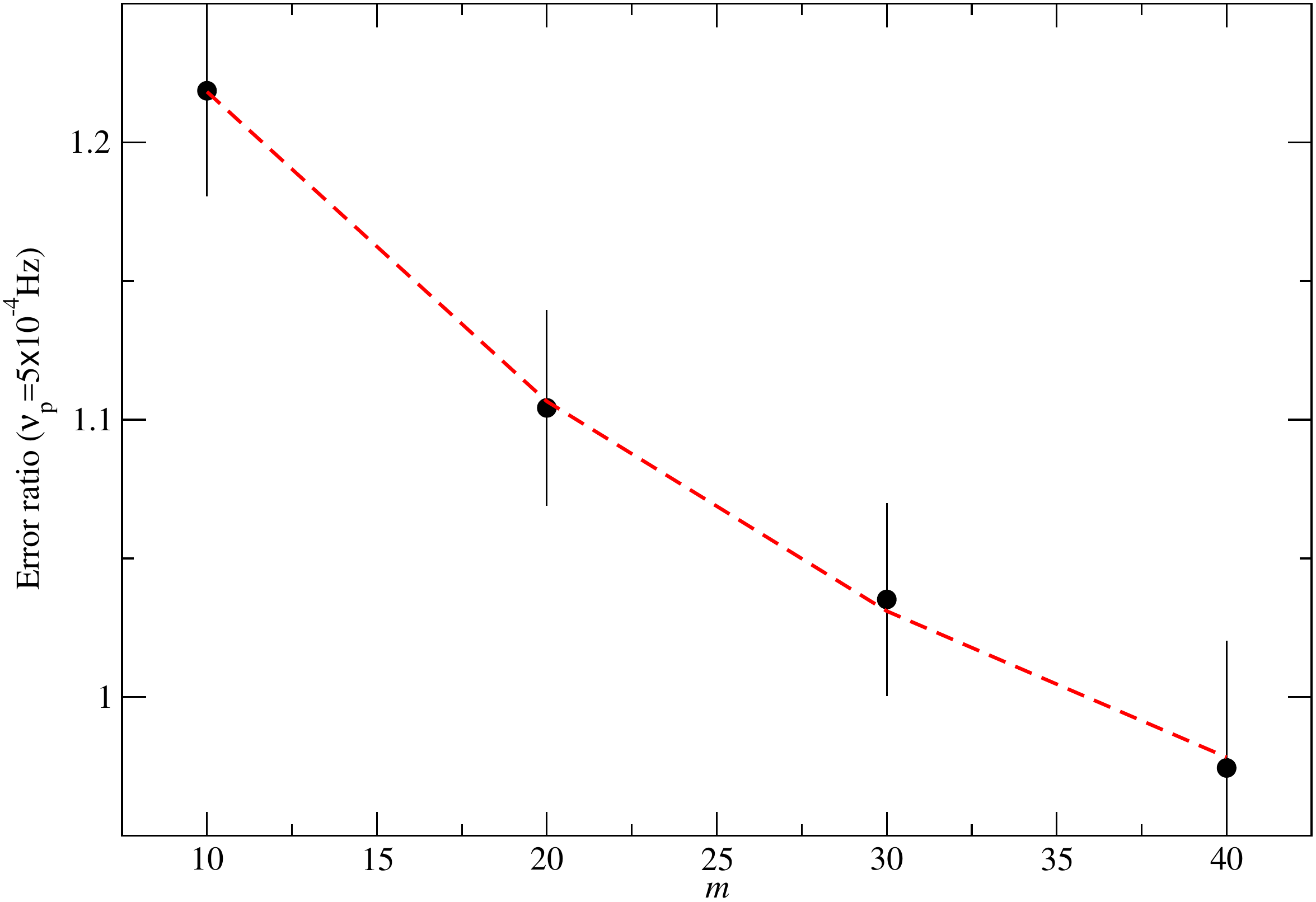}
      \caption{The mean error ratio at the lowest sample frequency, for all experiments, plotted as a function of $m$. The dashed line indicates the best-fitting line to the data (see equation \ref{eqa6}).}
\label{figa2}
   \end{figure}

We computed the weighted mean of the error ratio at the lowest frequency ($5\times10^{-4}\,\mathrm{Hz}$) for each $m$, using the results from the three different numerical experiments, for all S/N combinations. Figure \ref{figa2} shows the resulting mean error ratio plotted as a function of $m$. We found that the following relation describes the data  well:
\noindent
\begin{equation} \label{eqa7}
\frac{\langle\hat{\sigma}_{\hat{\gamma}^2_{\mathrm{int}}}(\nu_p)\rangle}{\sigma_{\hat{\gamma}^2_{\mathrm{int}}}(\nu_p)}=0.5+(3.8/\sqrt{m})-(4.8/m).
\end{equation}
\noindent
The red dashed line in the same figure shows this relation. This result suggests that, if we divide the analytic error estimate by $0.5+(3.8/\sqrt{m})-(4.8/m)$, then the `corrected' error estimates will better approximate the true scatter of the sample intrinsic coherent estimates around their mean.

In the top panels of Figs. \ref{figa5}, \ref{figa7}, \ref{figa9}, \ref{figa11}, and \ref{figa13} we plot the corrected error ratio. As long as $m\ge20$, the ratio is constant and $\sim0.8-1$ at frequencies lower than $\nu_{\rm max}$. The middle panels in the same figures show the corrected $p_{\rm 1\hat{\sigma}}$ and $p_{\rm 2\hat{\sigma}}$ values in each case. When $m\ge 20$, $p_{\rm 1\hat{\sigma}}$ and $p_{\rm 2\hat{\sigma}}$ values are $\sim0.68-0.75$ and almost identical to $0.95$, respectively, at frequencies lower than $\nu_{\rm max}$. These results indicate that, as long as $m\gtrsim20$, the corrected analytic error approximates well the true scatter of the sample intrinsic coherence around the mean, and, at the same time, the standard deviation of these estimates corresponds to the standard deviation of a Gaussian variable at frequencies lower than $\nu_{\rm max}$. This suggests that the distribution of the intrinsic coherence estimates may approximate a Gaussian in the same frequency range.

%%%%%%%%%%%%%%%%%% FIG. A2 %%%%%%%%%%%%%%

%%%%%%%%%%%%%%%%%%%%%%%%%%%%%%%%%%%%%%%%%%
\subsection{Probability distribution of the intrinsic coherence estimates} \label{seca4}
%%%%%%%%%%%%%%%%%%%%%%%%%%%%%%%%%%%%%%%%%%

%%%%%%%%%%%%%%%%%% FIG. A3 %%%%%%%%%%%%%%
   \begin{figure}
   \centering
   \includegraphics[width=\hsize]{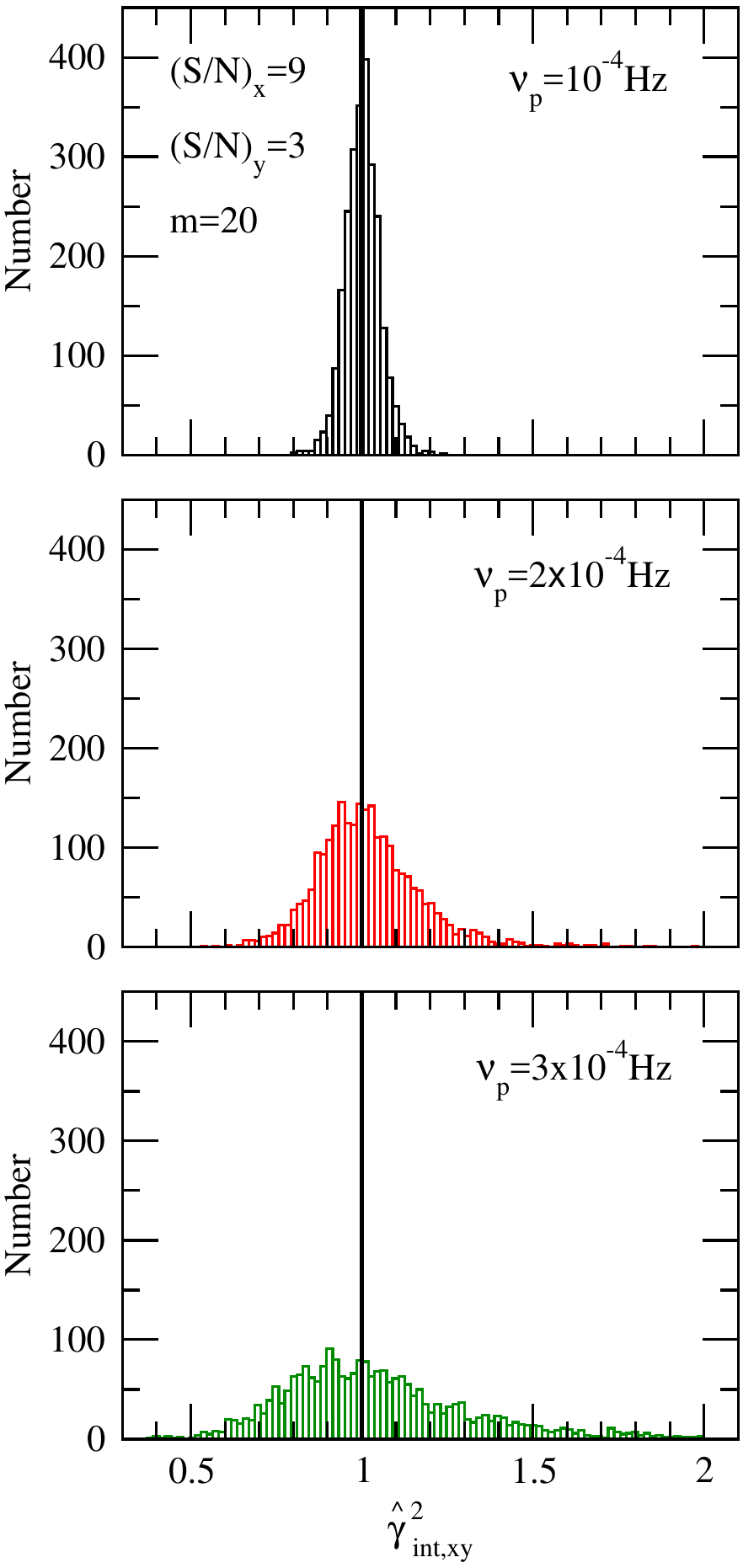}
      \caption{The probability distribution of the sample intrinsic estimates, using the results from experiment PLD, in two frequencies that are lower, and one that is higher than $\nu_{\rm max}$ (top and middle, and bottom panel, respectively). The solid vertical line indicates the unity intrinsic coherence value in each case.}
\label{figa3}
   \end{figure}

Figure \ref{figa3} shows the probability distribution of the intrinsic coherence estimates from experiment PLD, at three different frequencies: $10^{-4}$ (top panel), $2\times10^{-4}$ (middle panel), and $3\times10^{-4}\,\mathrm{Hz}$ (bottom panel). The first two frequencies, which are lower than $\nu_{\mathrm{max}}$, lie within the range where the bias is less than 5 per cent, and the corrected error ratio is close to unity. The third frequency is higher than $\nu_{\mathrm{max}}$, and both the bias as well as the corrected error ratio have increased. The plots show that, as the frequency increases, the width of the probability distribution increases as well, and the distribution becomes more skewed towards values greater than unity. The presence of this tail in the probability distribution results in a mean value larger than the intrinsic value of one, although the most probable value (i.e. the peak of the probability distribution, shown as a solid vertical line in the same figure) is almost equal to unity at all frequencies. This effect explains why the intrinsic coherence estimates become increasingly biased at higher frequencies. The effect reverses at even higher frequencies, where the distribution shows a shift towards values smaller than one. In the first two frequencies, the corrected analytic error approximates well the standard deviation of the distributions, and it indicates correctly the range which includes 68 per cent and 95 per cent of the sample intrinsic coherence values. In the third case, (even the corrected) analytic error overestimates significantly the (already large) standard deviation of the distribution.

The panels in the bottom row of Figs. \ref{figa5}, \ref{figa7}, \ref{figa9}, \ref{figa11}, and \ref{figa13} show the probability, $p_{\rm KS}(\nu_p)$, that the distribution of the intrinsic coherence estimates is Gaussian, with a mean and variance equal to the mean and variance of the sample distribution, respectively. This probability was estimated using the Kolmogorov-Smirnov (KS) test. The dotted lines in all panels indicate the value 0.01. This is the typical threshold probability that would normally be considered if one wanted to reject the hypothesis of a Gaussian distribution for the coherence estimates. The vertical lines in the panels show that, when $m\ge 20$, the hypothesis of Gaussianity for the distribution of the intrinsic coherence estimates cannot be rejected at the 0.01 significance level for most (but not all) frequencies that are lower than $\nu_{\rm max}$.

%%%%%%%%%%%%%%%%%%%%%%%%%%%%%%%%%%%%%%%%%%
\subsection{A prescription for estimating the intrinsic coherence} \label{seca5}
%%%%%%%%%%%%%%%%%%%%%%%%%%%%%%%%%%%%%%%%%%

Based on the results presented in the previous sections, we propose the following prescription for estimating the intrinsic coherence between two light curves: a) Use at least $m=20$ light curve segments and equations \ref{eqa1} and \ref{eqa3} to calculate the intrinsic coherence estimates at frequencies lower than $\nu_{\rm max}$, which is defined by equation \ref{eqa6}. b) Divide the analytic error by the quantity $0.5+(3.8/\sqrt{m})-(4.8/m)$. In this way, the intrinsic coherence estimates should be less than 5 per cent biased, the corrected error will be equal to (and, in the worse case, no more than $\sim20$ per cent smaller than) the true standard deviation of the estimates, and their distribution will be rather well approximated by a Gaussian.

We point out that the distribution is almost certainly not identical to a Gaussian. However, perhaps the most interesting result for practical applications is that the range `$\pm1\times{\rm (corrected) error}$' (`$\pm2\times{\rm (corrected) error}$') corresponds to the 68 per cent (95 per cent) confidence interval of the intrinsic coherence estimates. Furthermore, the fact that the estimates are approximately Gaussian implies that they can be used to model the intrinsic coherence estimates of two light curves, using traditional $\chi^2$ minimisation techniques. We note though that, if a model fails to fit the data well at frequencies close to $\nu_{\rm max}$, one should try to repeat the fit by only considering intrinsic coherence estimates at frequencies lower than $\sim \nu_{\rm max}/2$, as the results presented in the previous section indicate that the hypothesis of Gaussianity should be stronger at these frequencies.

\clearpage

\begin{figure*}
 \includegraphics[width=410pt]{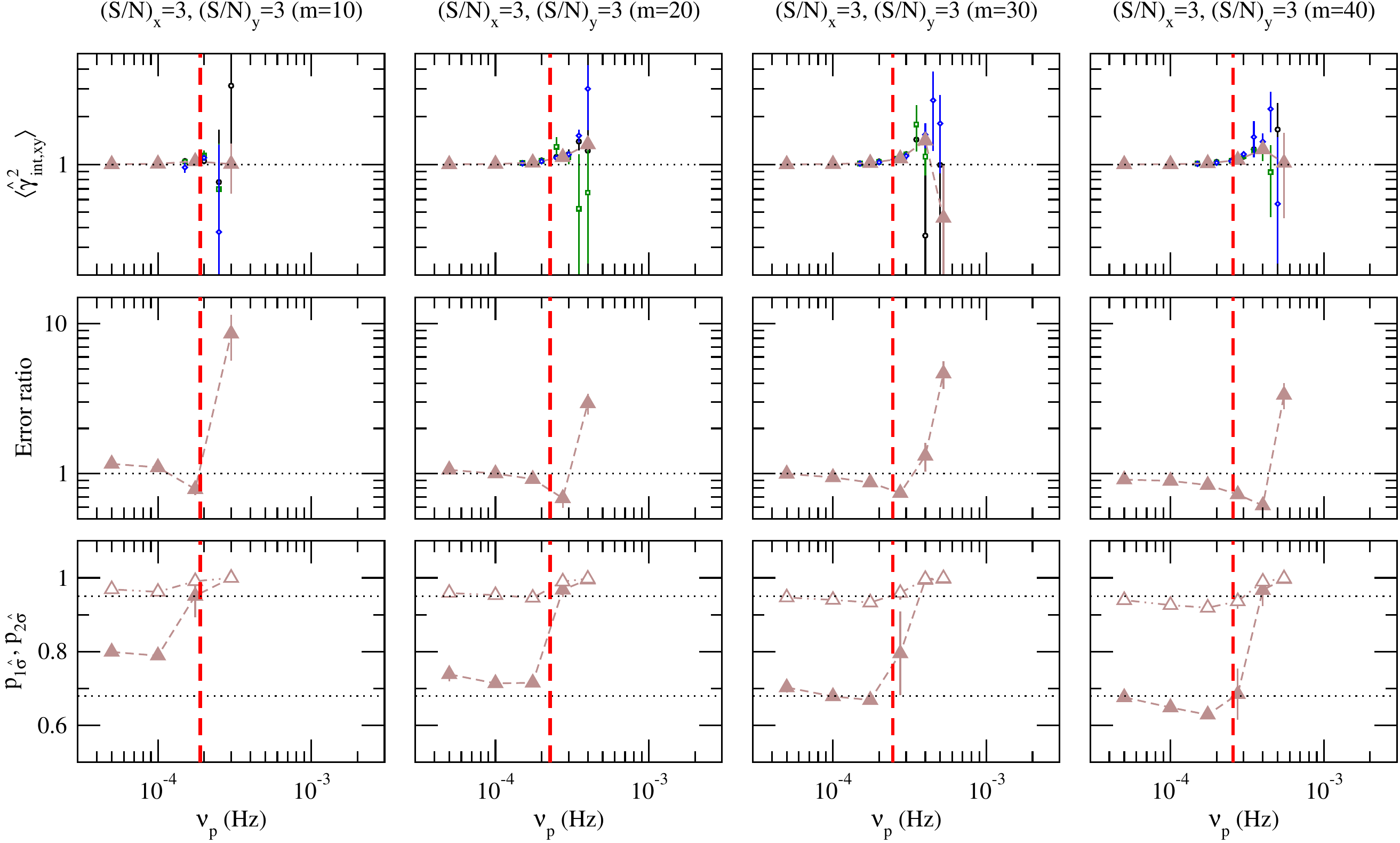}
 \caption{\textit{Top row}: Mean of the sample intrinsic coherence estimates (the horizontal dotted lines indicate the unity intrinsic coherence). \textit{Middle row}: The mean error ratio (the horizontal dotted lines indicate the unity value, in which case the mean analytic error of the sample intrinsic coherence estimates is equal to the standard deviation of the sample distributions). \textit{Bottom row}: The percentage of points in the sample intrinsic coherence distributions that are within a region equal to 1 and 2 times the mean analytic error around the mean of the distribution (the lower and upper horizontal lines indicate the 68 per cent and 95 per cent values, which hold in the case of a Gaussian distribution). In this, and all subsequent similar figures, the different columns correspond to different number of light curve segments ($m$ is indicated on the top of each column). The points in the top row panels indicate the mean sample intrinsic coherence for each experiment, while filled triangles show the binned results. For clarity reasons, in the other panels we plot only the binned results, for all experiments (see Appendix \ref{appa} for details).}.
\label{figa4}
\end{figure*}

\begin{figure*}
 \includegraphics[width=410pt]{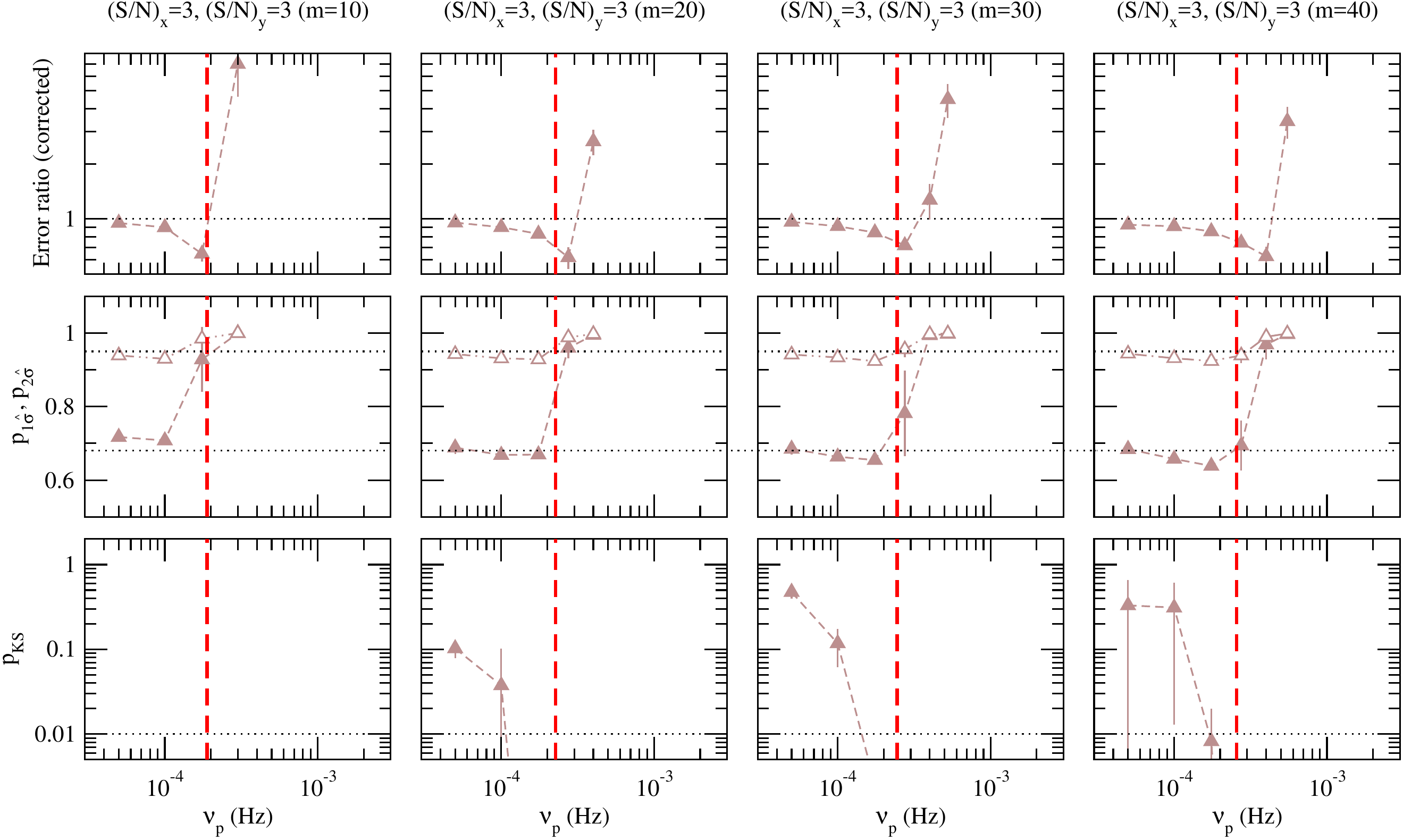}
\caption{\textit{Top row}: The mean corrected error ratio (the the horizontal dotted lines indicate the value of one, in which case the mean analytic error of the sample intrinsic coherence estimates is equal to the standard deviation of the sample distributions). \textit{Middle row}: The percentage of points in the sample intrinsic coherence distributions that are within a region equal to 1 and 2 times the mean corrected analytic error around the mean of the distribution (the horizontal dotted lines and points are as in the respective panels in Fig. \ref{figa4}). \textit{Bottom row}: Probability that the sample intrinsic coherence estimates are Gaussian-distributed (see Appendix \ref{appa} for details).}
\label{figa5}
\end{figure*}

\begin{figure*}
 \includegraphics[width=480pt]{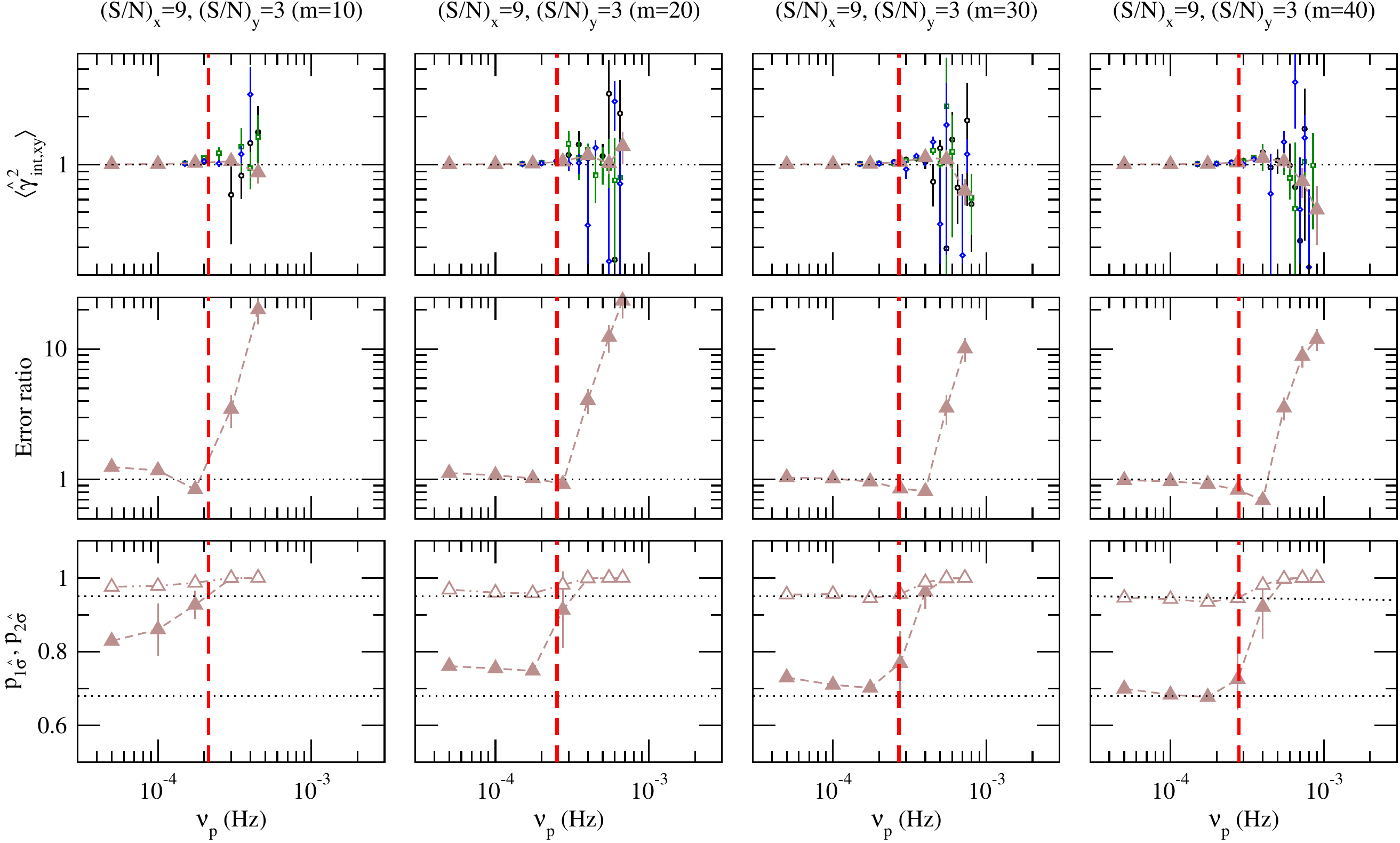}
 \caption{As in Fig. \ref{figa4}, for $(\mathrm{S/N})_x=9$ and $(\mathrm{S/N})_y=3$.}
\label{figa6}
\end{figure*}

\begin{figure*}
 \includegraphics[width=480pt]{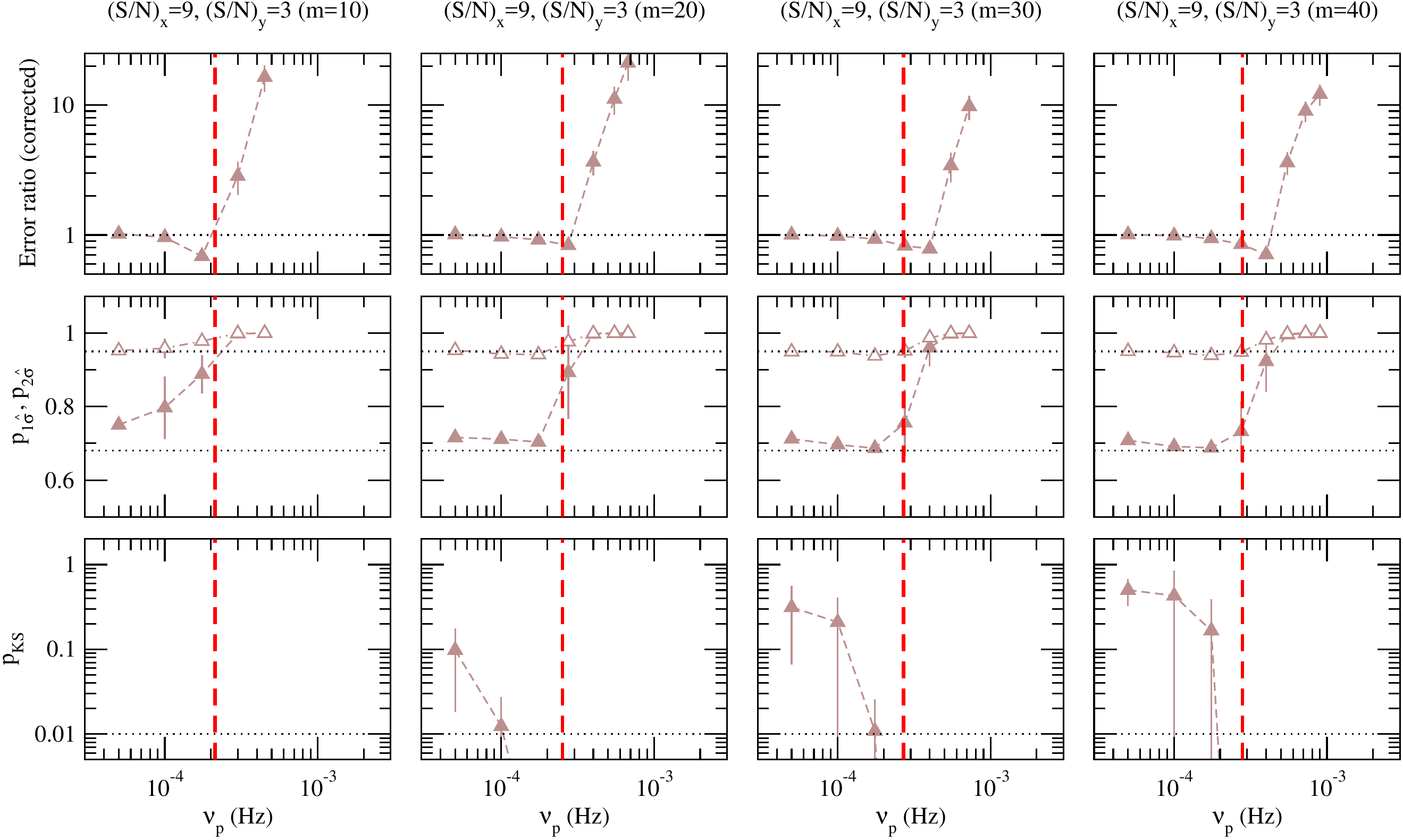}
\caption{As in Fig. \ref{figa5}, for $(\mathrm{S/N})_x=9$ and $(\mathrm{S/N})_y=3$.}
\label{figa7}
\end{figure*}

\begin{figure*}
 \includegraphics[width=480pt]{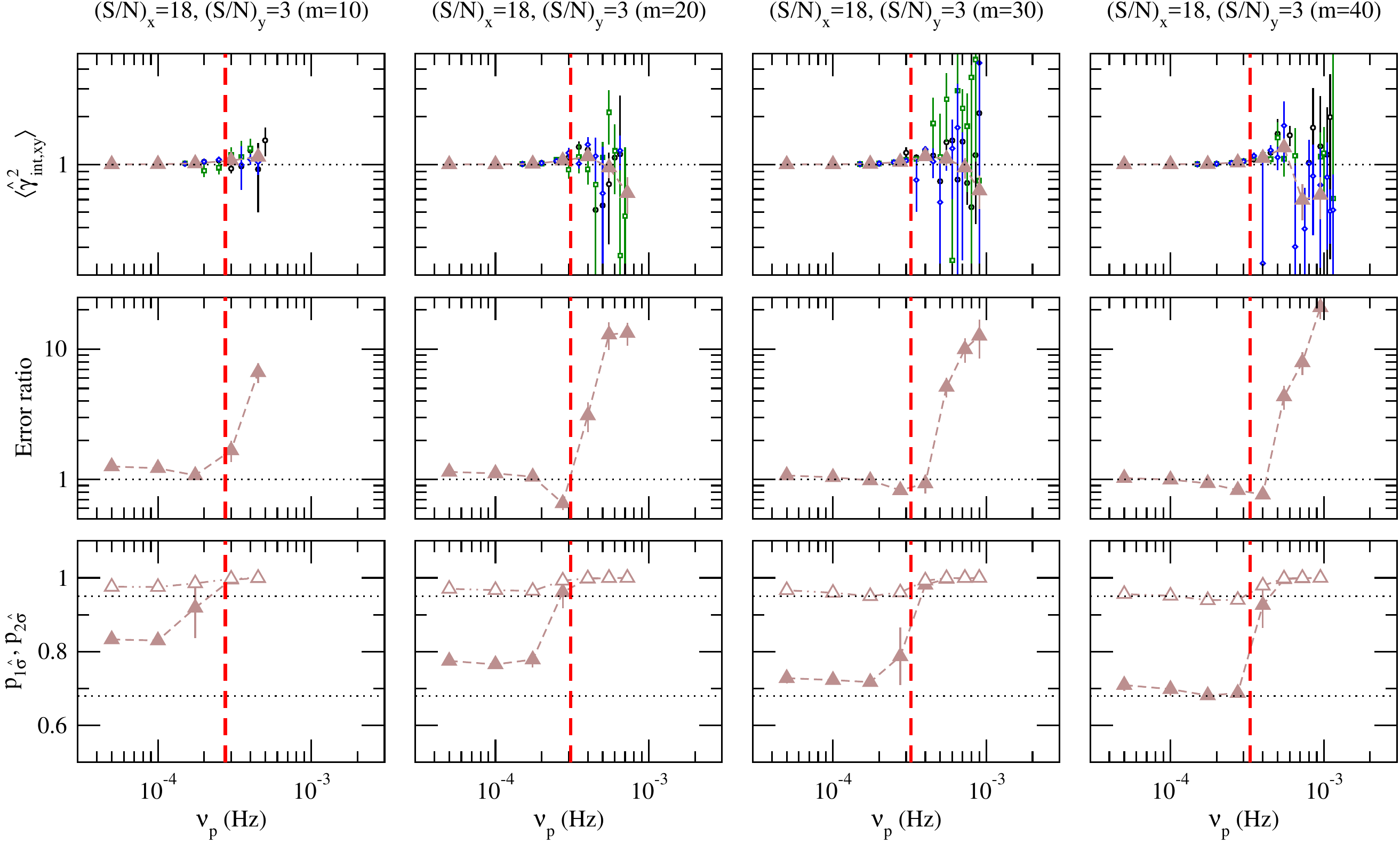}
 \caption{As in Fig. \ref{figa4}, for $(\mathrm{S/N})_x=18$ and $(\mathrm{S/N})_y=3$.}
\label{figa8}
\end{figure*}

\begin{figure*}
 \includegraphics[width=480pt]{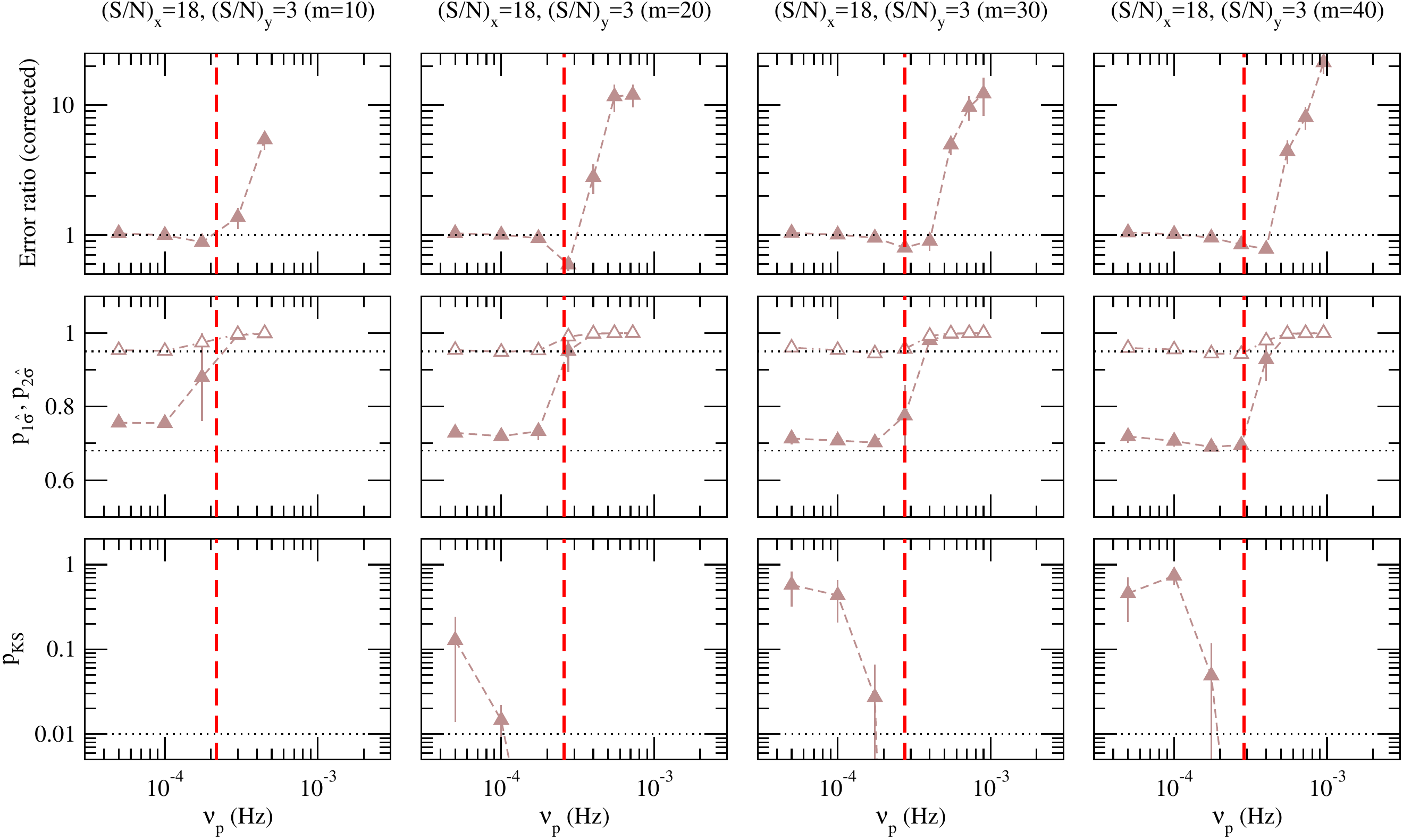}
 \caption{As in Fig. \ref{figa5}, for $(\mathrm{S/N})_x=18$ and $(\mathrm{S/N})_y=3$.}
\label{figa9}
\end{figure*}

\begin{figure*}
 \includegraphics[width=480pt]{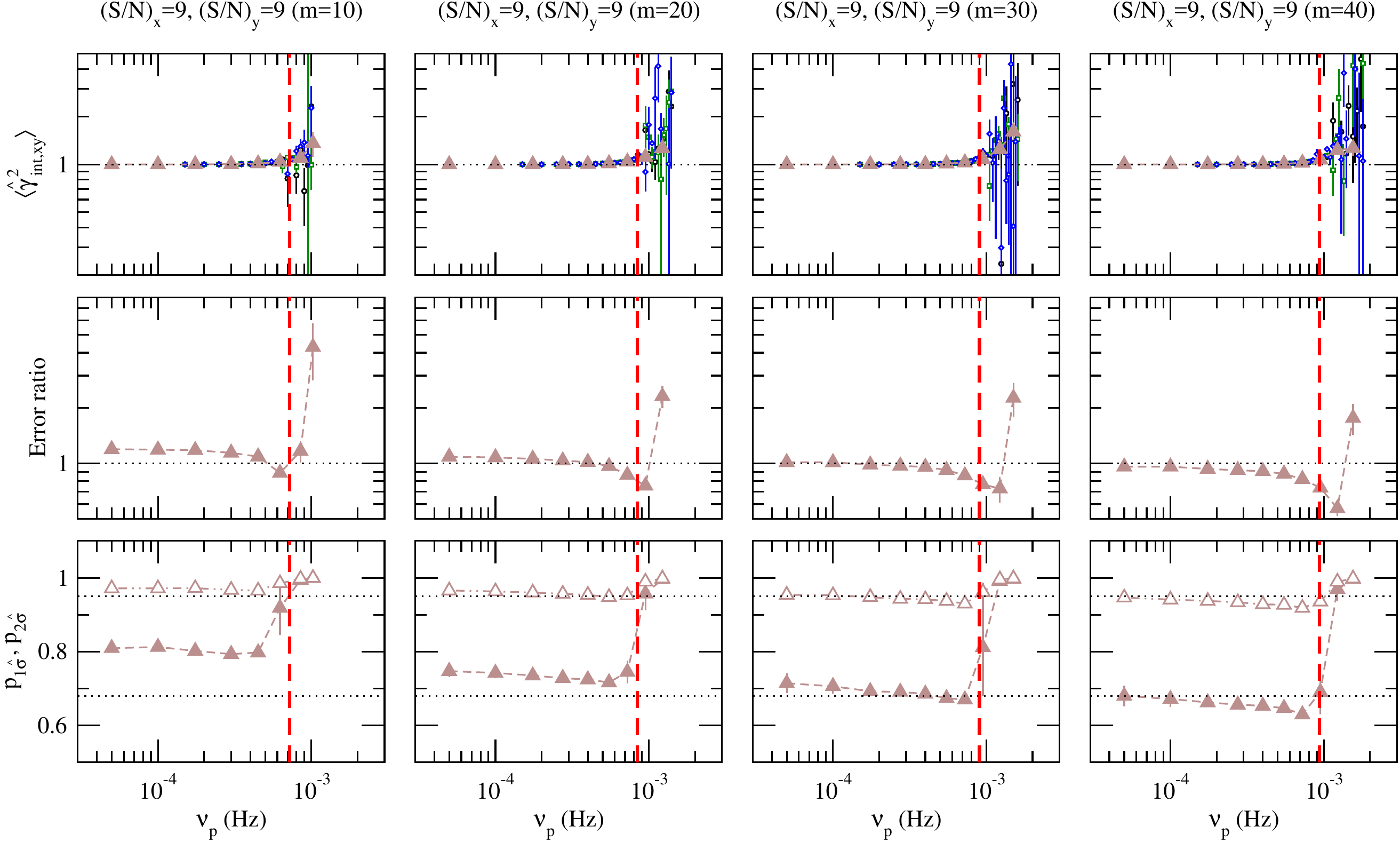}
 \caption{As in Fig. \ref{figa4}, for $(\mathrm{S/N})_x=9$ and $(\mathrm{S/N})_y=9$.}
\label{figa10}
\end{figure*}

\begin{figure*}
 \includegraphics[width=480pt]{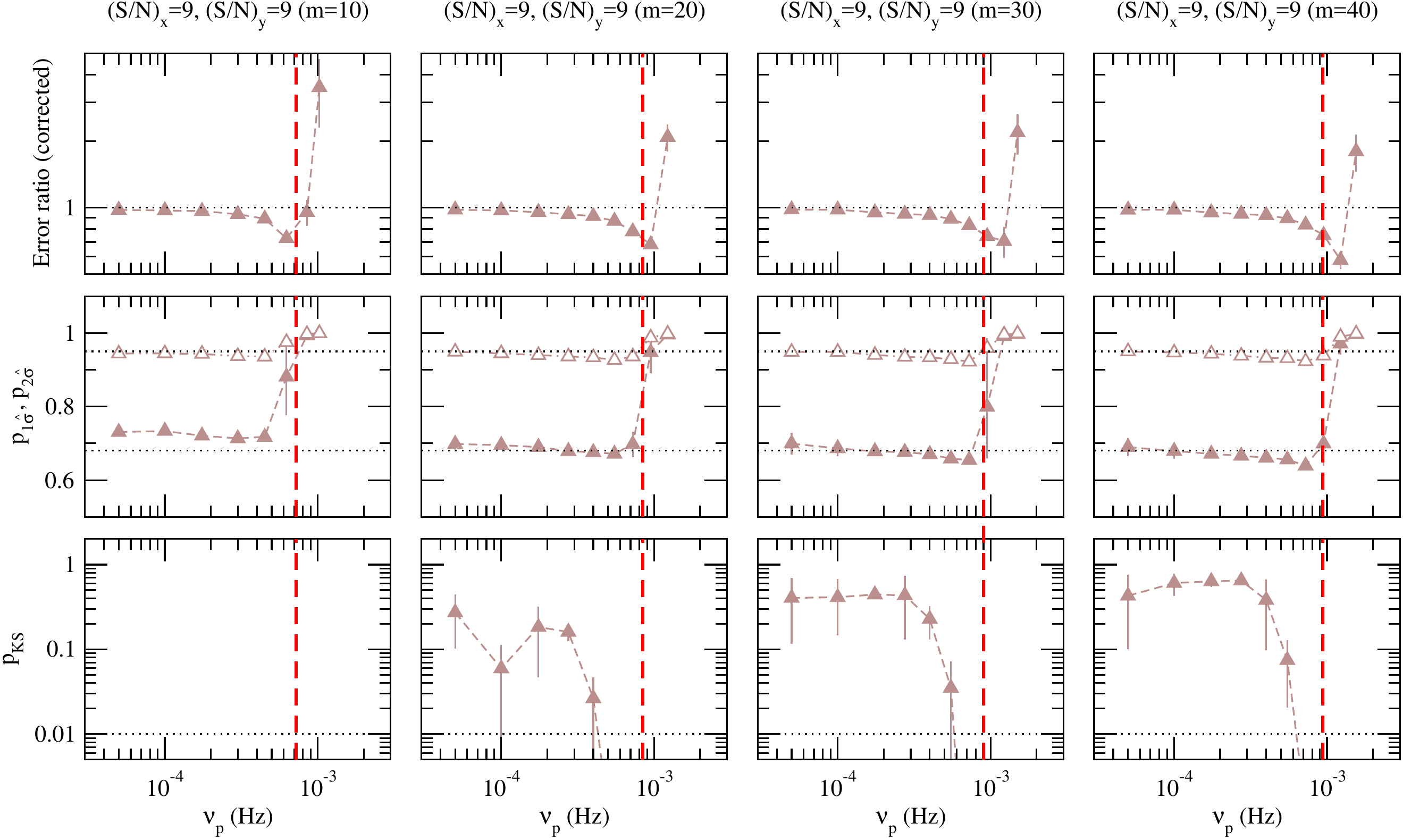}
 \caption{As in Fig. \ref{figa5}, for $(\mathrm{S/N})_x=9$ and $(\mathrm{S/N})_y=9$.}
\label{figa11}
\end{figure*}

\begin{figure*}
 \includegraphics[width=480pt]{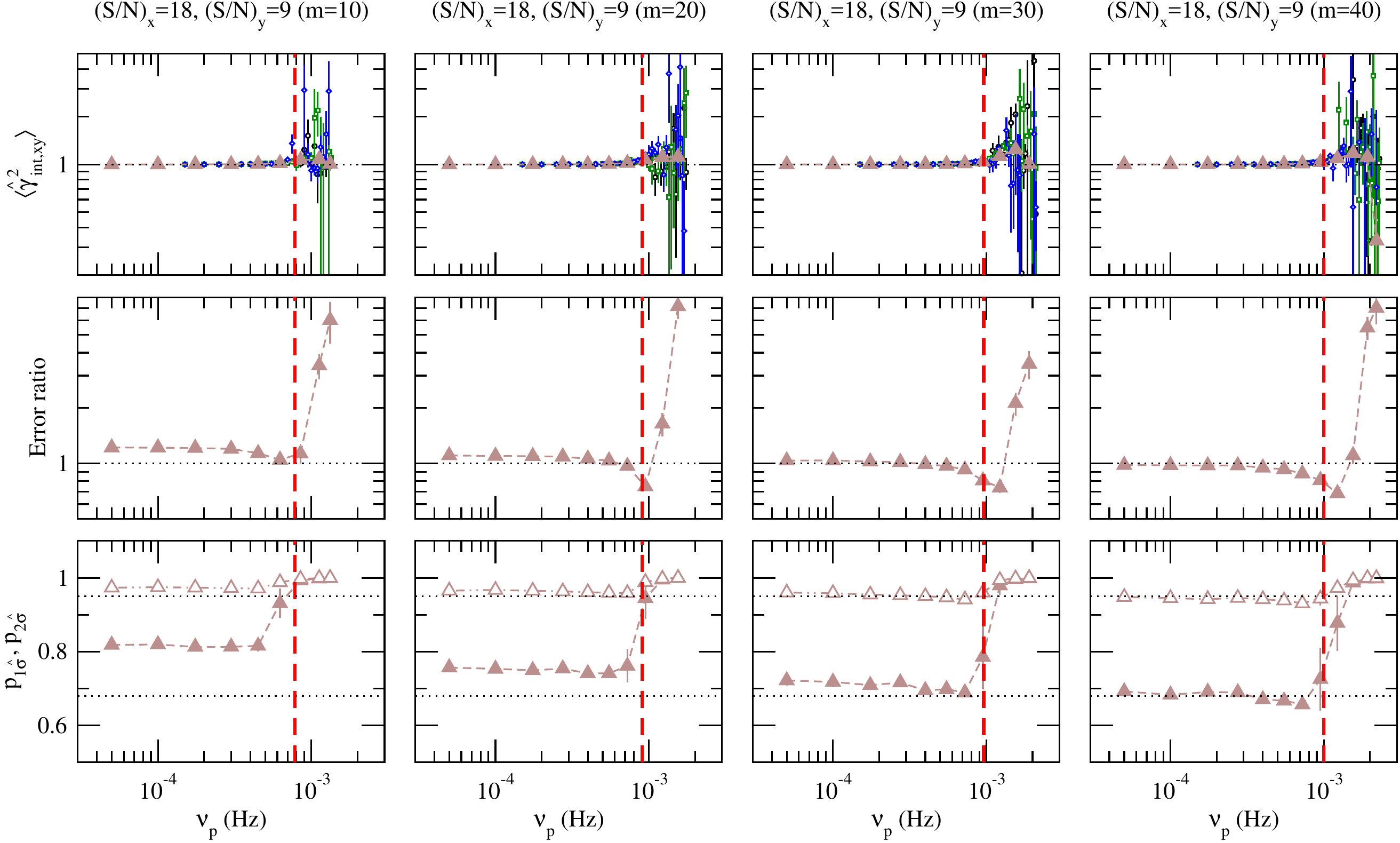}
 \caption{As in Fig. \ref{figa4}, for $(\mathrm{S/N})_x=18$ and $(\mathrm{S/N})_y=9$.}
\label{figa12}
\end{figure*}

\begin{figure*}
 \includegraphics[width=480pt]{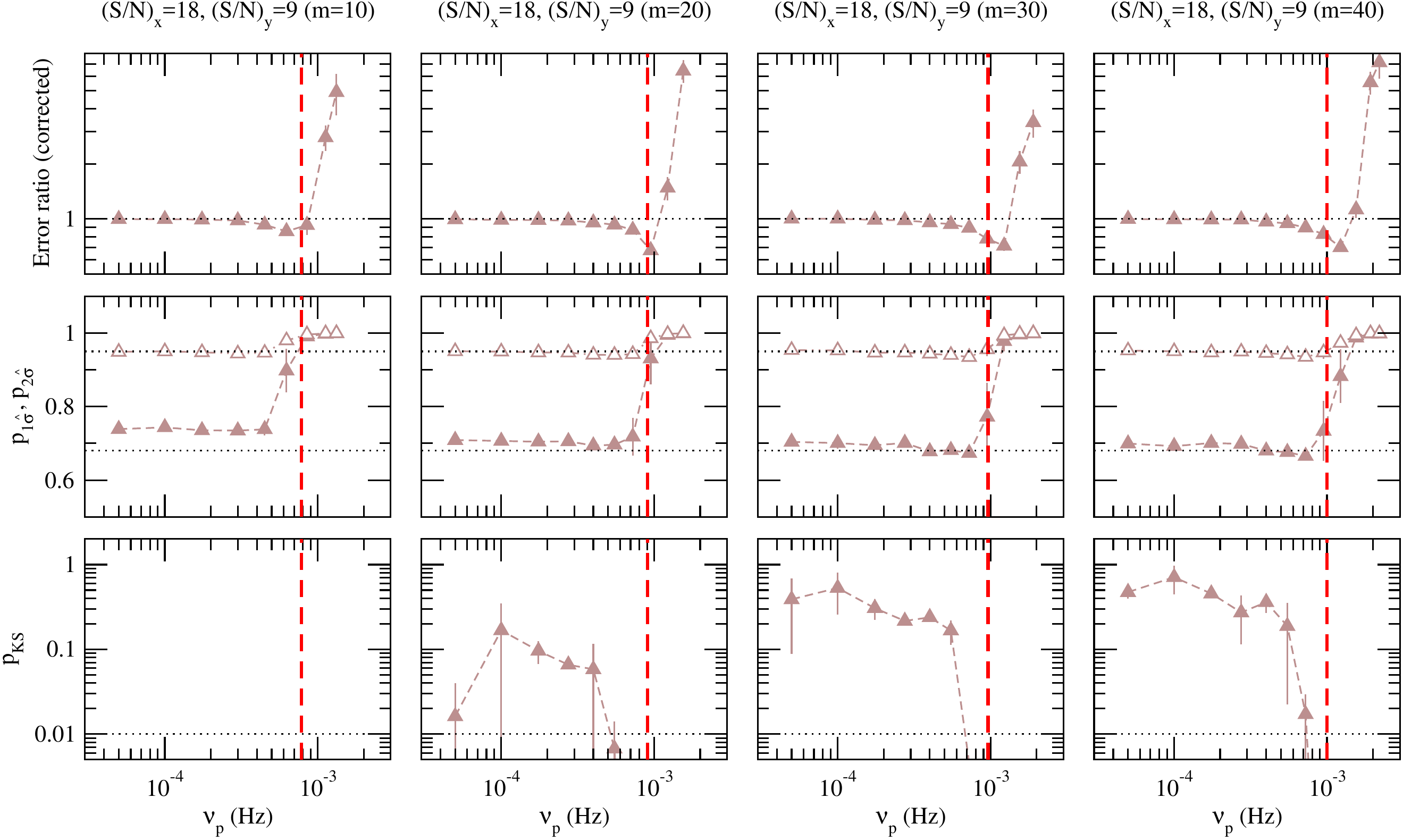}
 \caption{As in Fig. \ref{figa5}, for $(\mathrm{S/N})_x=18$ and $(\mathrm{S/N})_y=9$.}
\label{figa13}
\end{figure*}

\clearpage

\section{Observed time-lags and intrinsic coherence} \label{appb}

\newpage

\begin{figure*}
\includegraphics[width=340pt]{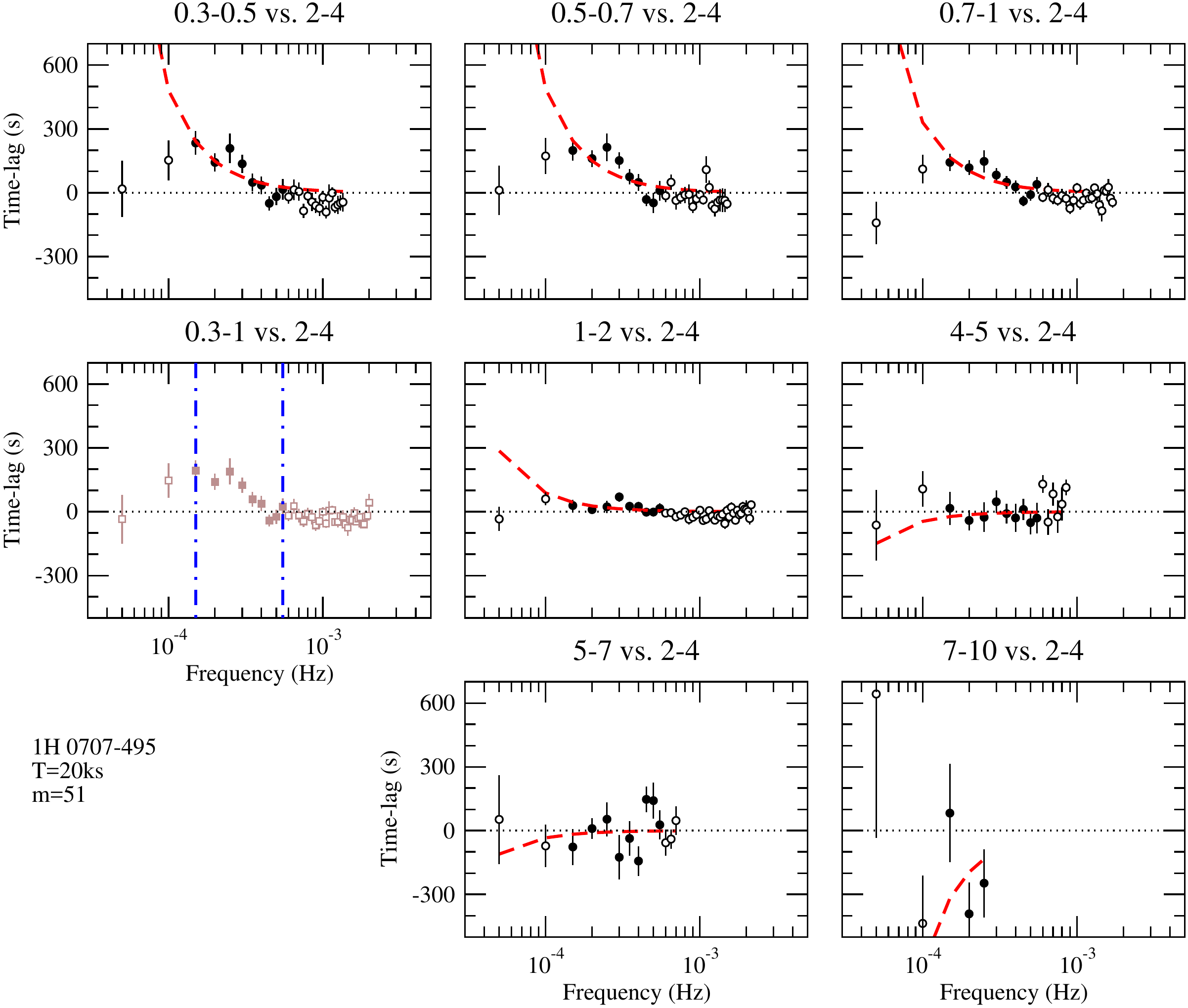}
\caption{Sample time-lag spectra of 1H 0707--495 between the $2-4\,\mathrm{keV}$ band and various energy bands. The red dashed lines indicate the best-fitting power-law model, while the vertical blue dotted-dashed lines indicate the frequency range used for the fitting procedure (see Section \ref{sec31} for more details).}
\label{figb1}
\end{figure*}

\begin{figure*}
\includegraphics[width=340pt]{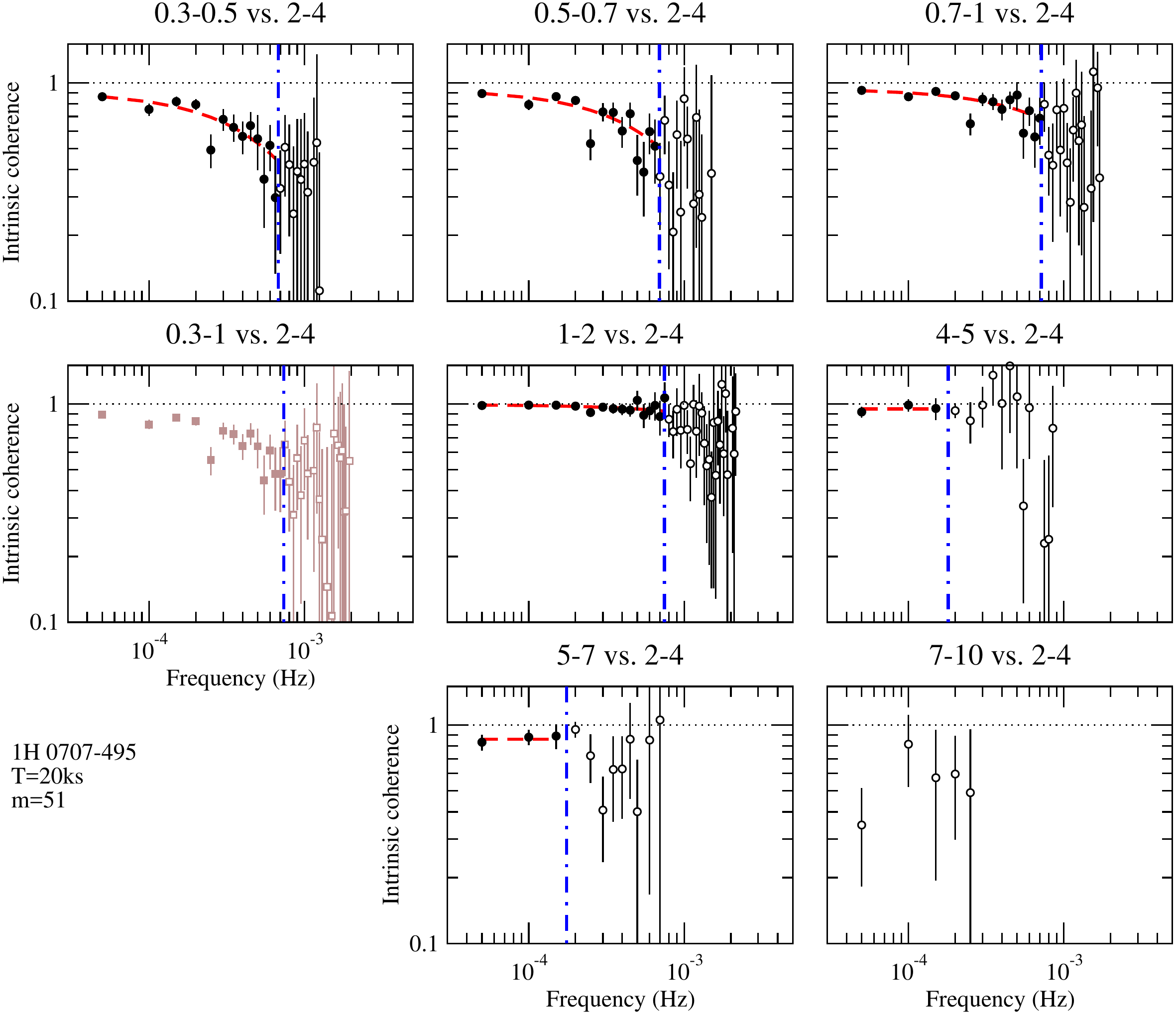}
\caption{Sample intrinsic coherence of 1H 0707--495 between the $2-4\,\mathrm{keV}$ band and various energy bands. The red dashed lines indicate the best-fitting model, while the vertical blue dotted-dashed lines indicate the maximum frequency, $\nu_{\mathrm{max}}$, below which the intrinsic coherence can be reliable estimated (see Appendix \ref{appa} for more details).}
\label{figb2}
\end{figure*}

\begin{figure*}
\includegraphics[width=350pt]{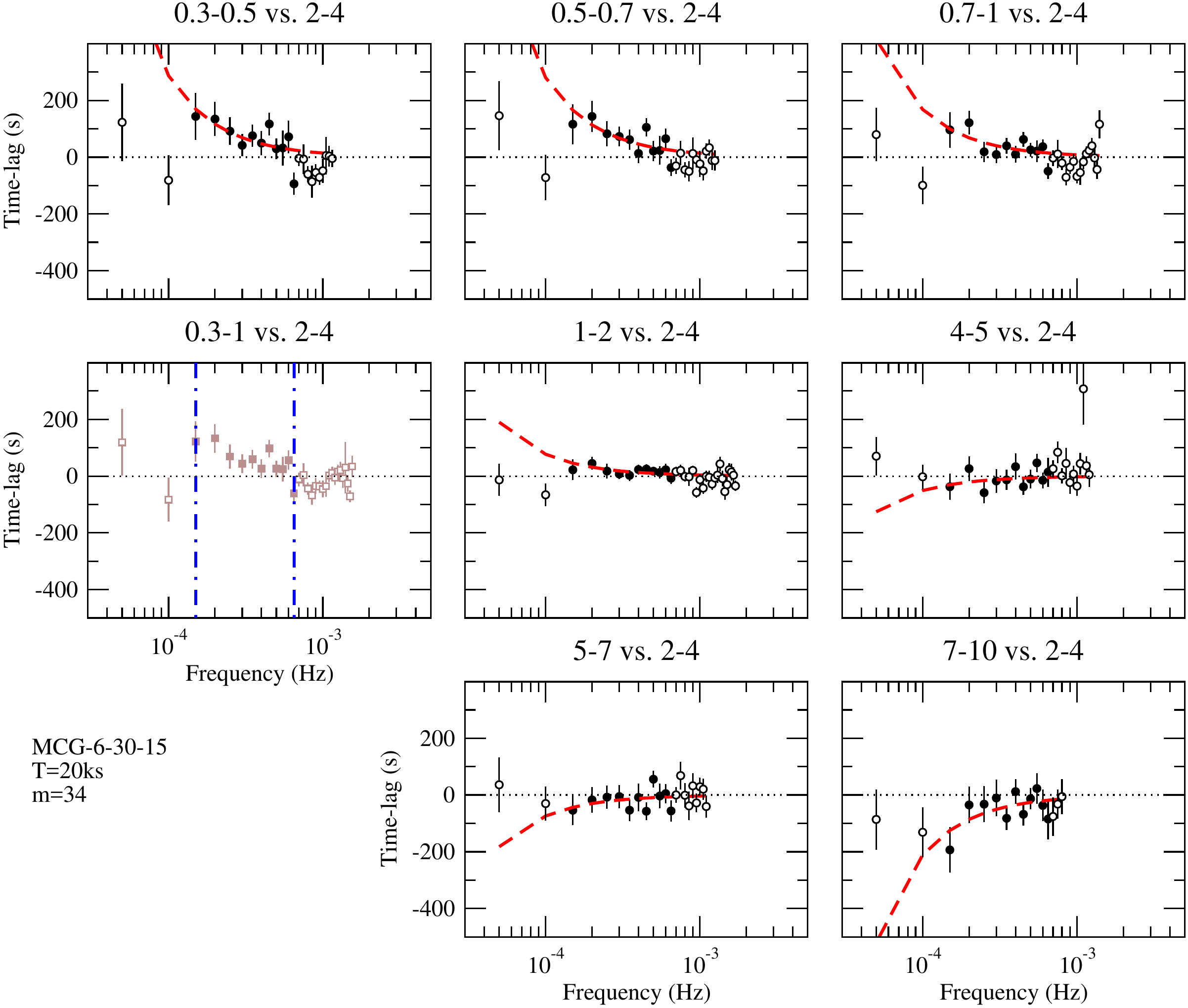}
\caption{As in Fig. \ref{figb1}, for MCG--6-30-15.}
\label{figb3}
\end{figure*}

\begin{figure*}
\includegraphics[width=350pt]{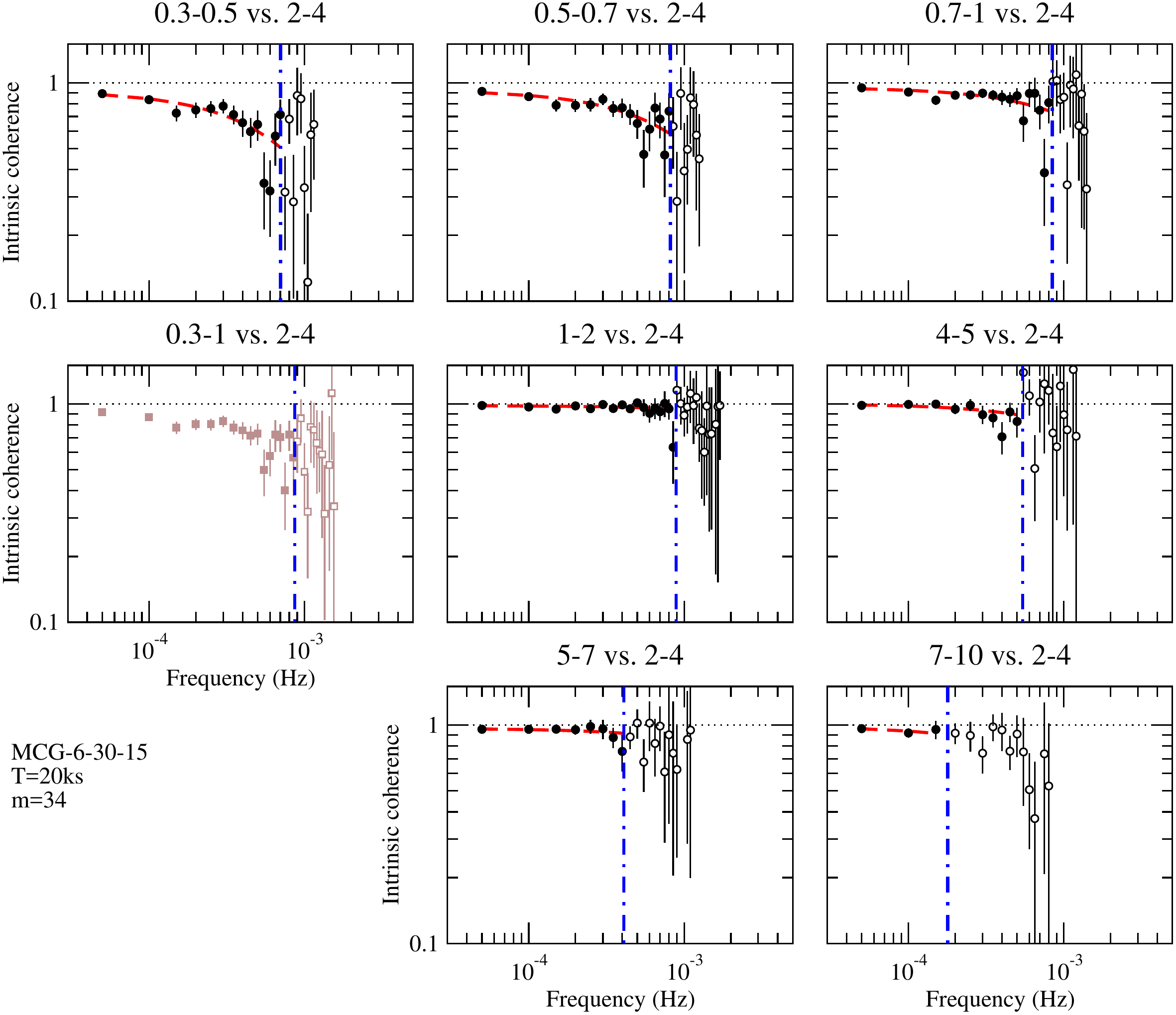}
\caption{As in Fig. \ref{figb2}, for MCG--6-30-15.}
\label{figb4}
\end{figure*}

\begin{figure*}
\includegraphics[width=350pt]{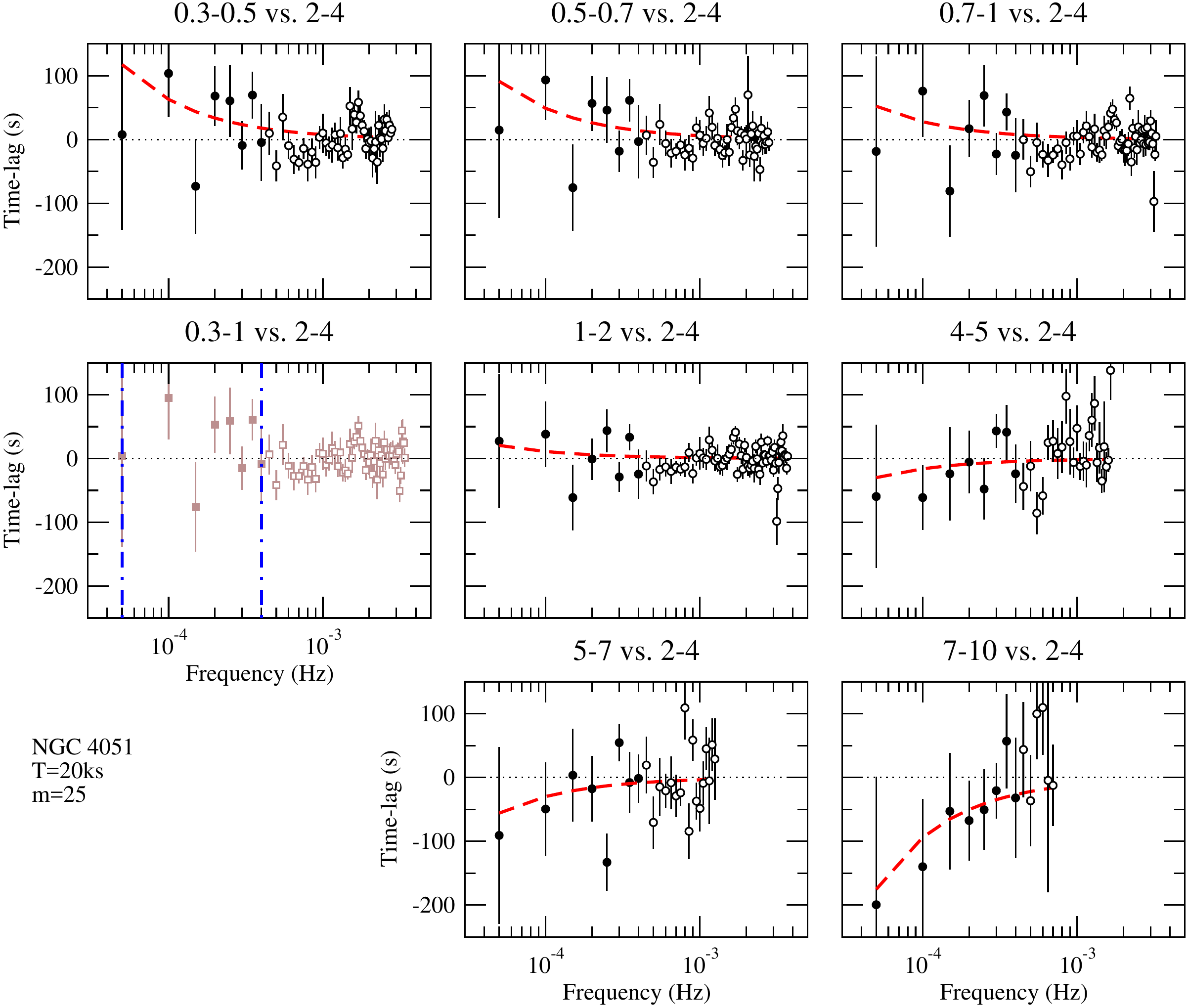}
\caption{As in Fig. \ref{figb1}, for NGC 4051.}
\label{figb5}
\end{figure*}

\begin{figure*}
\includegraphics[width=350pt]{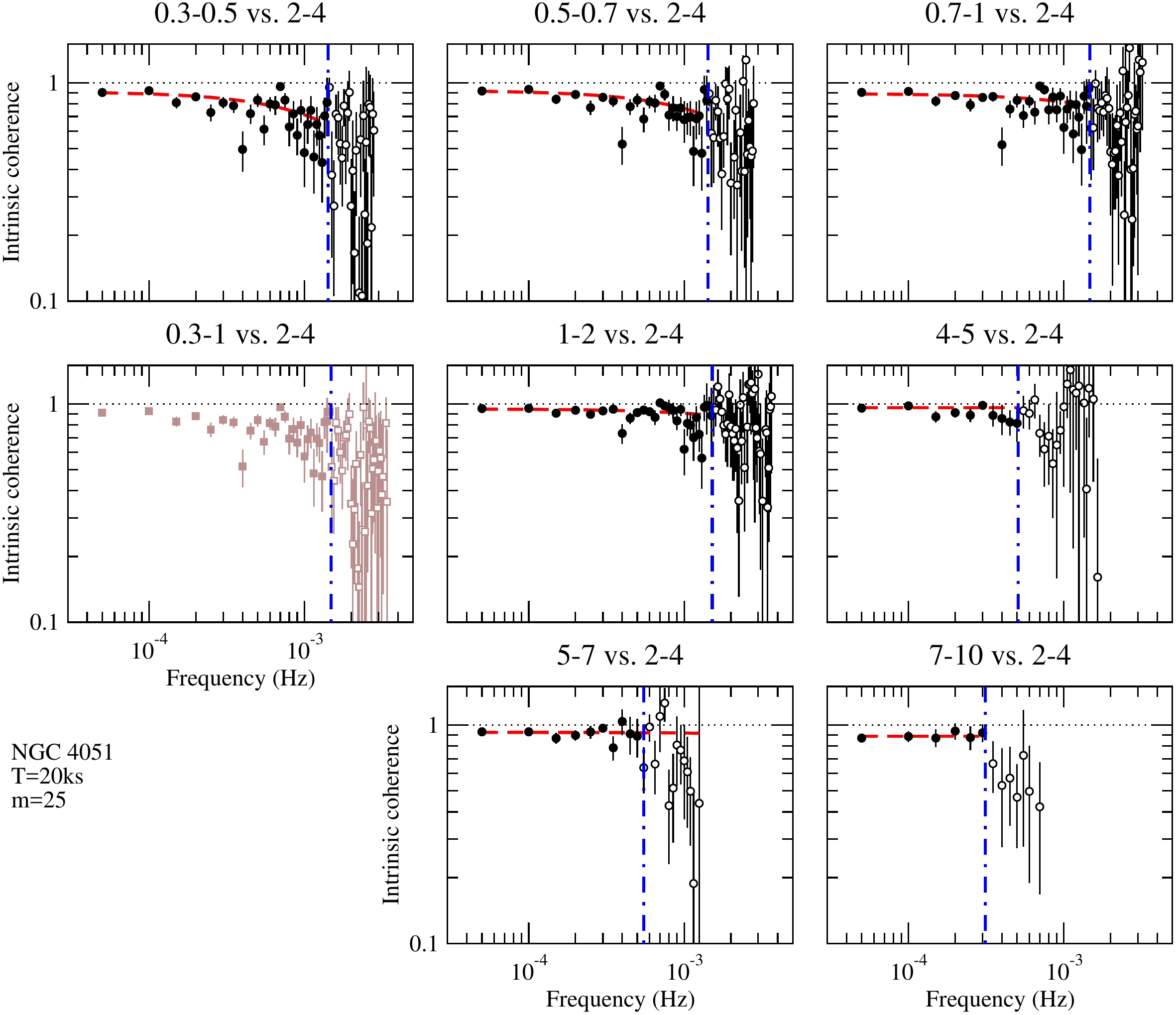}
\caption{As in Fig. \ref{figb2}, for NGC 4051.}
\label{figb6}
\end{figure*}

\begin{figure*}
\includegraphics[width=350pt]{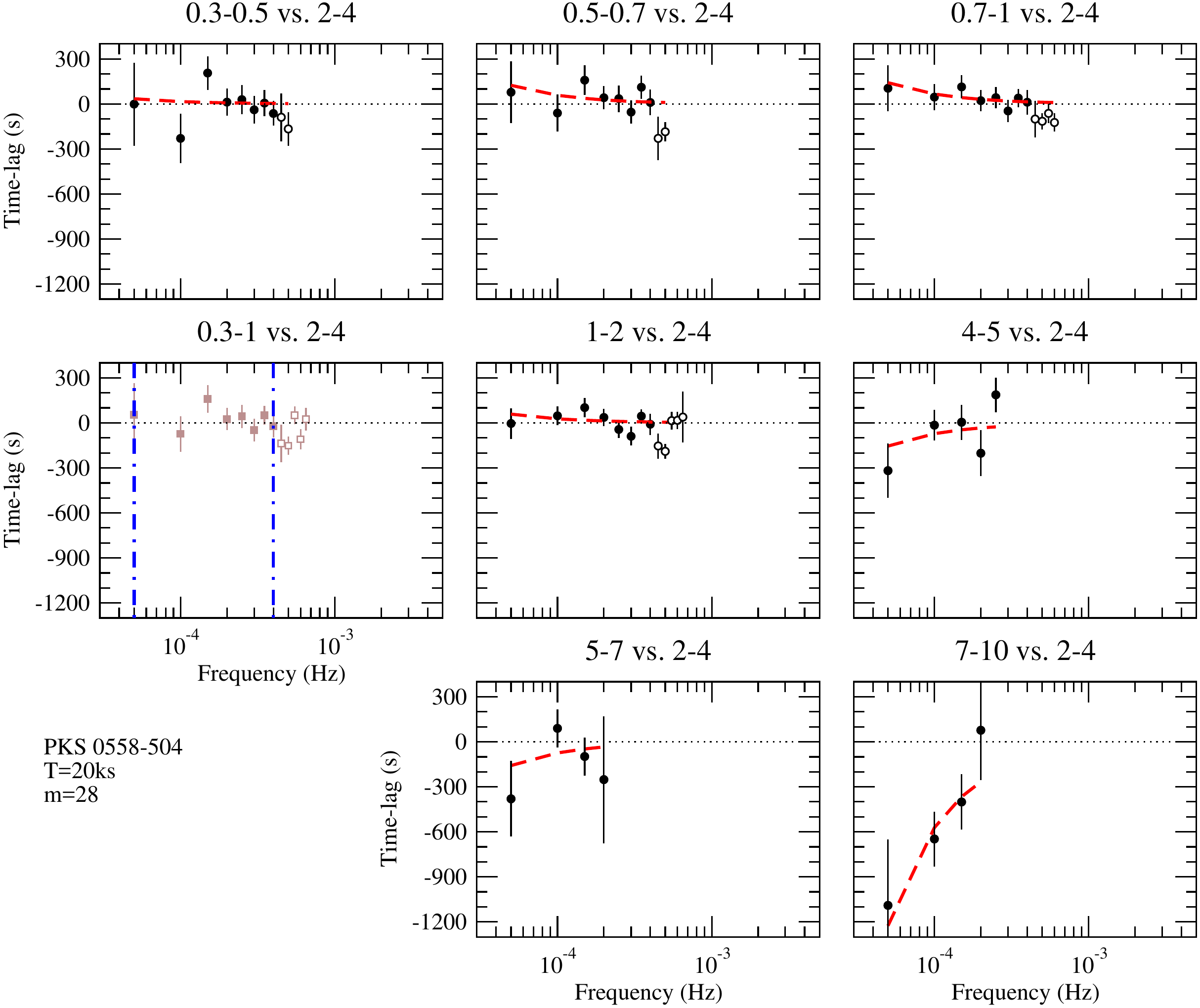}
\caption{As in Fig. \ref{figb1}, for PKS 0558--504.}
\label{figb7}
\end{figure*}

\begin{figure*}
\includegraphics[width=350pt]{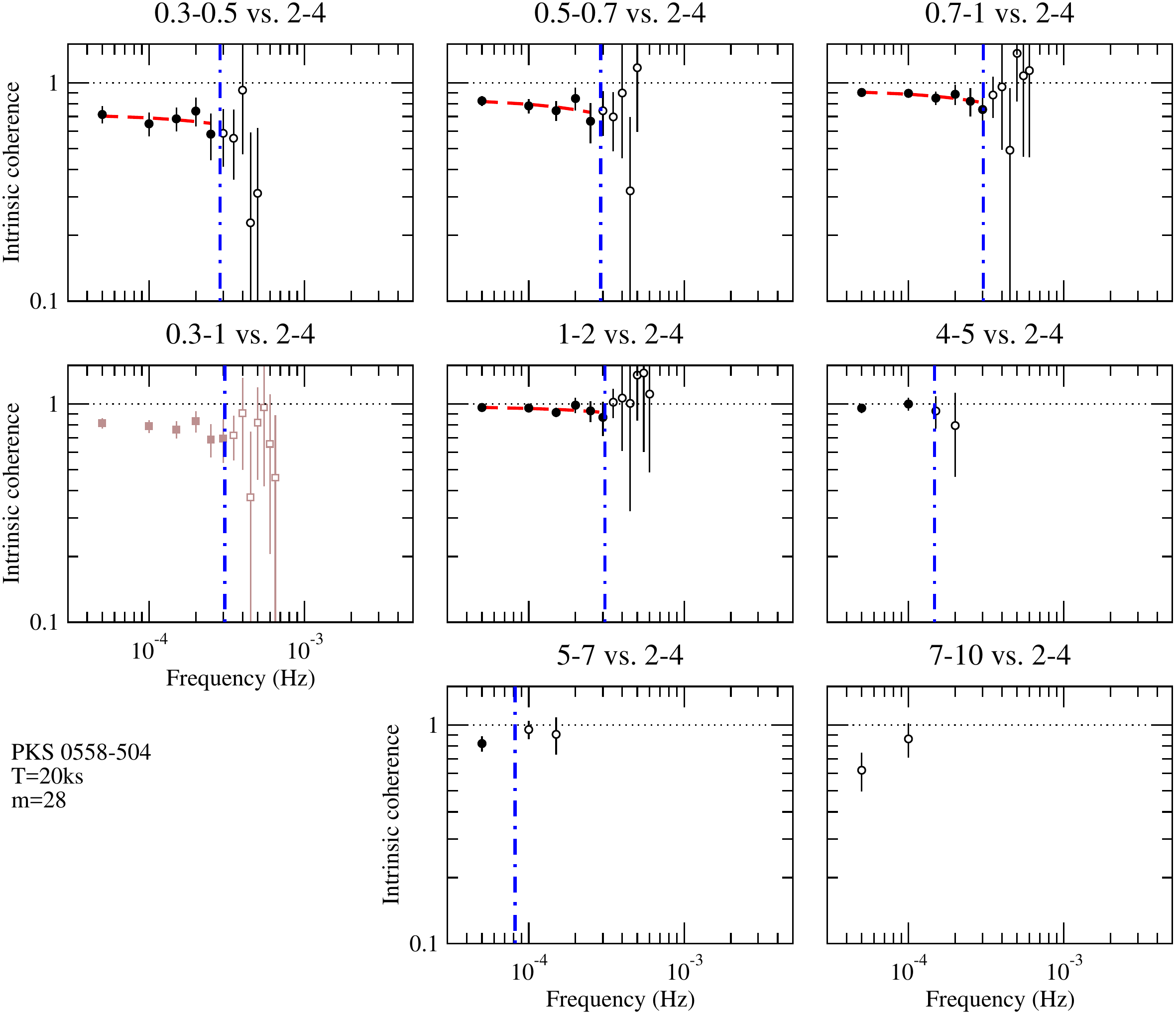}
\caption{As in Fig. \ref{figb2}, for PKS 0558--504.}
\label{figb8}
\end{figure*}

\begin{figure*}
 \includegraphics[width=350pt]{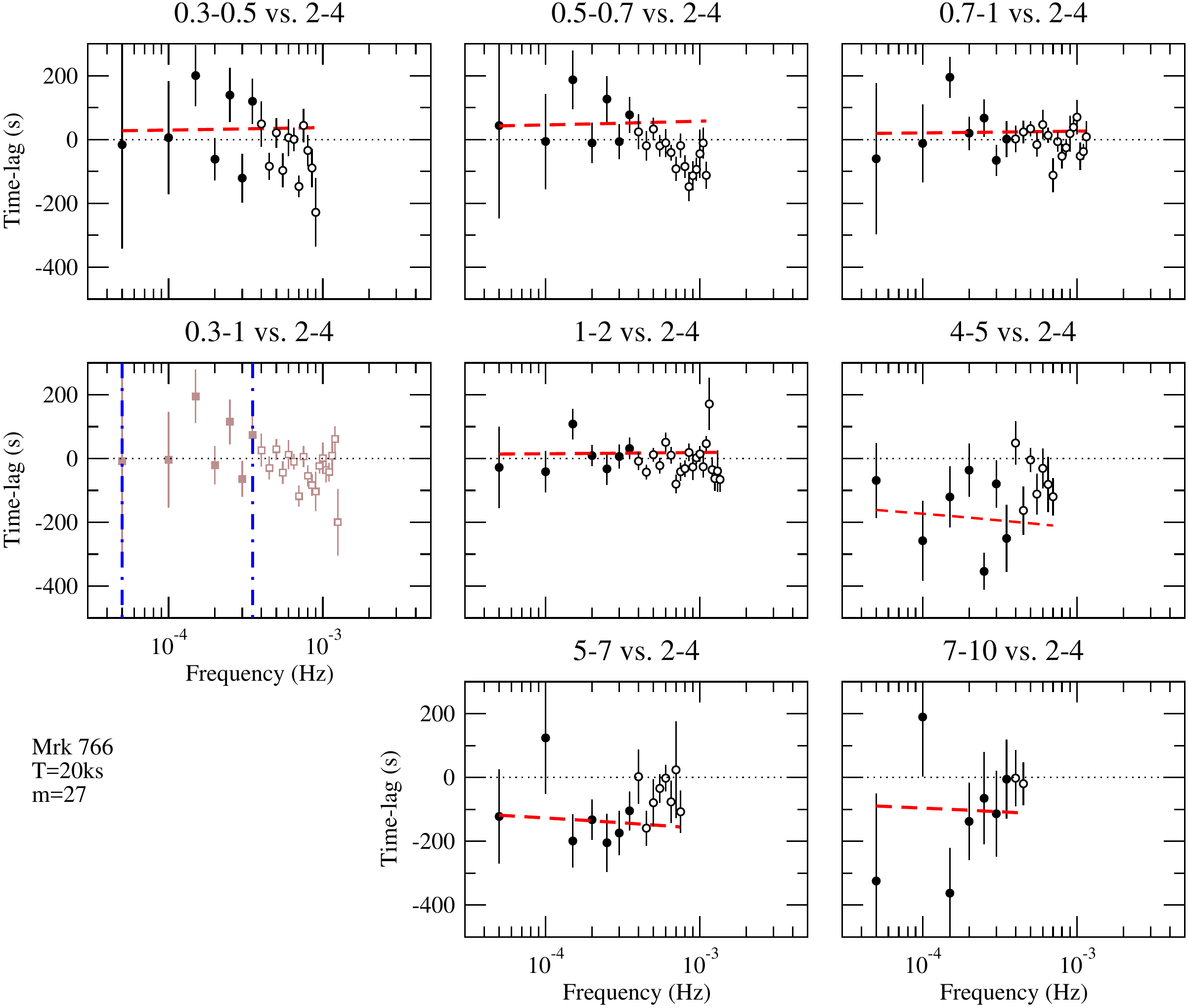}
 \caption{As in Fig. \ref{figb1}, for Mrk 766.}
\label{figb9}
\end{figure*}

\begin{figure*}
 \includegraphics[width=350pt]{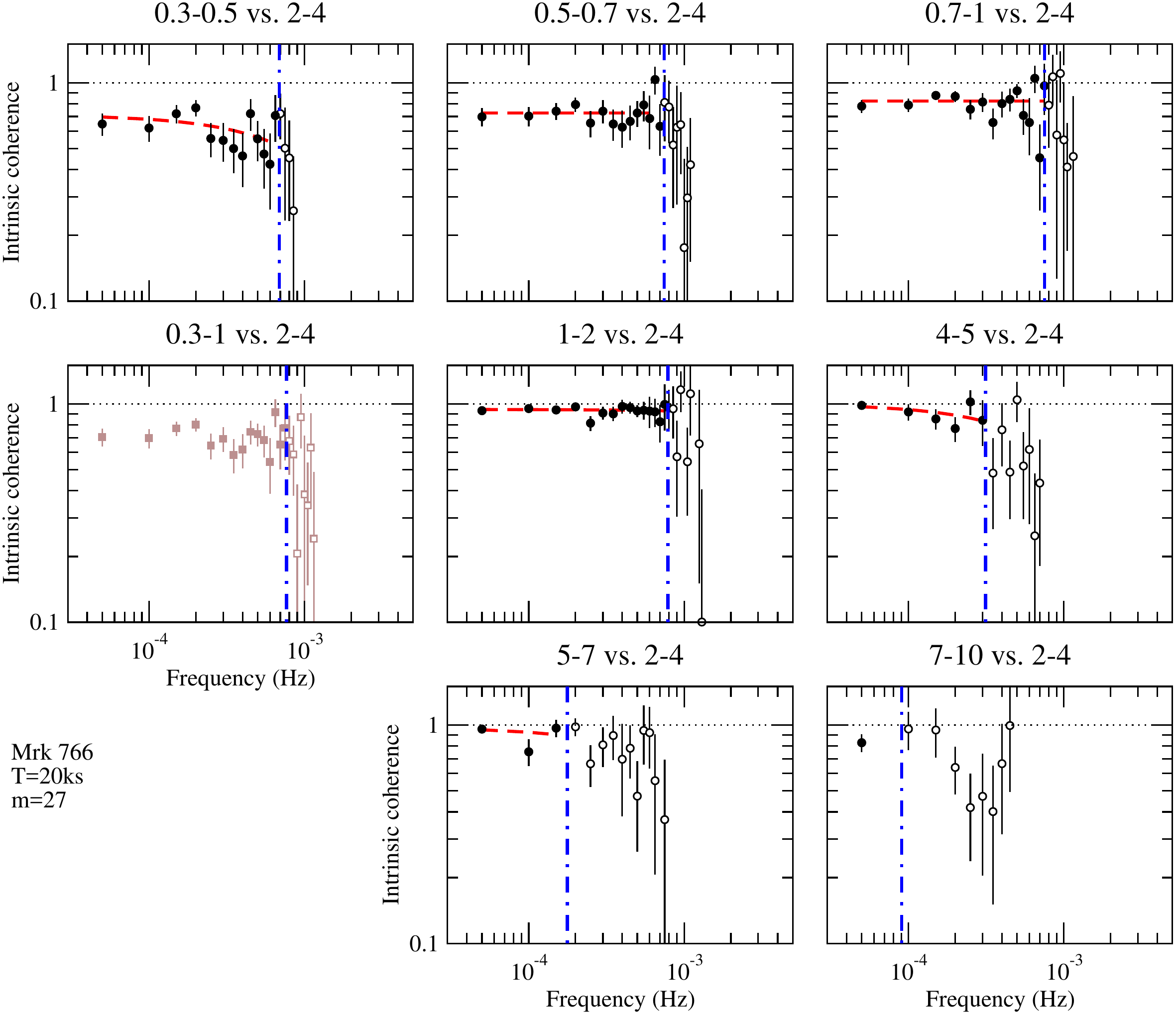}
 \caption{As in Fig. \ref{figb2}, for Mrk 766.}
\label{figb10}
\end{figure*}

\begin{figure*}
 \includegraphics[width=350pt]{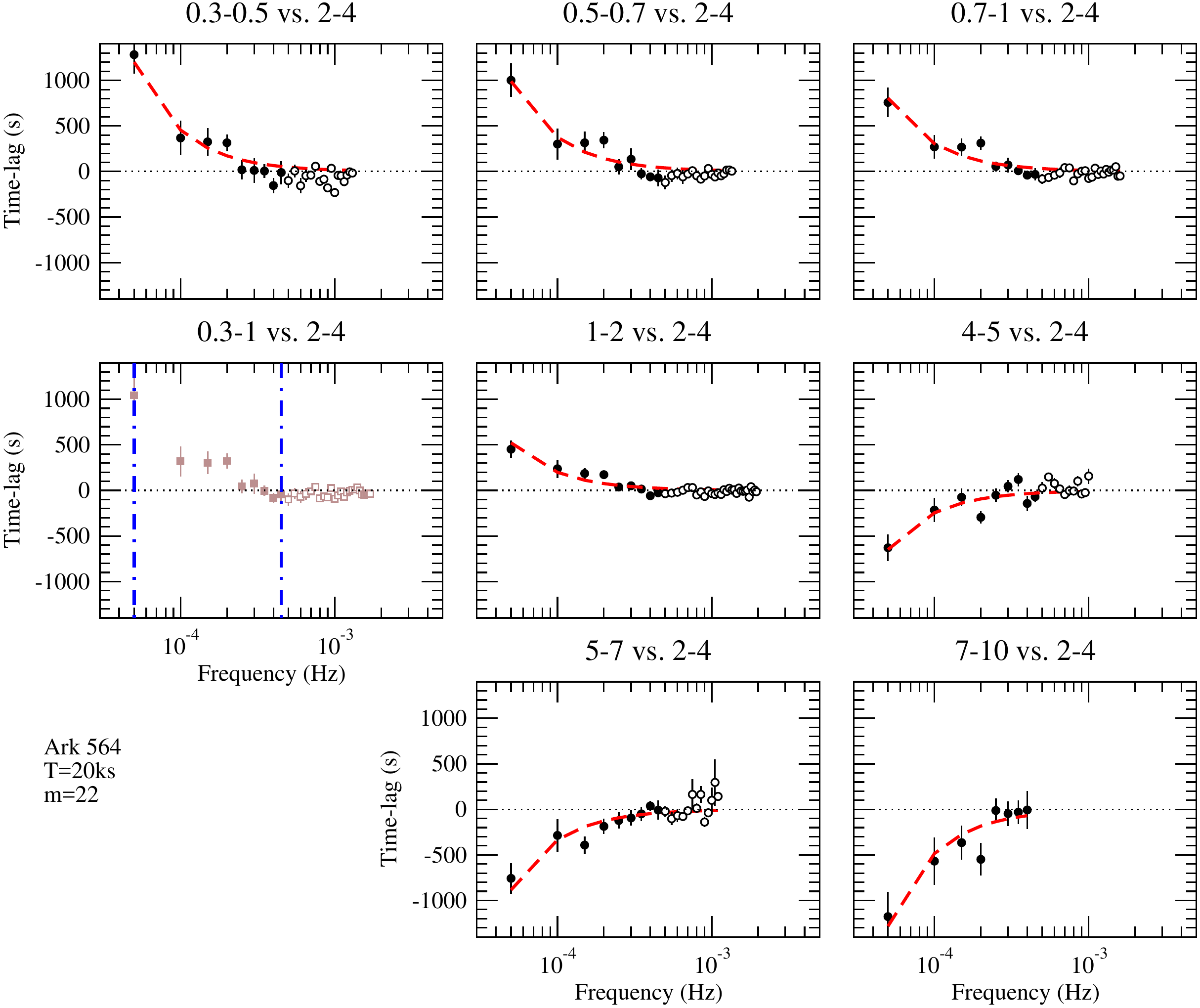}
 \caption{As in Fig. \ref{figb1}, for Ark 564.}
\label{figb11}
\end{figure*}

\begin{figure*}
 \includegraphics[width=350pt]{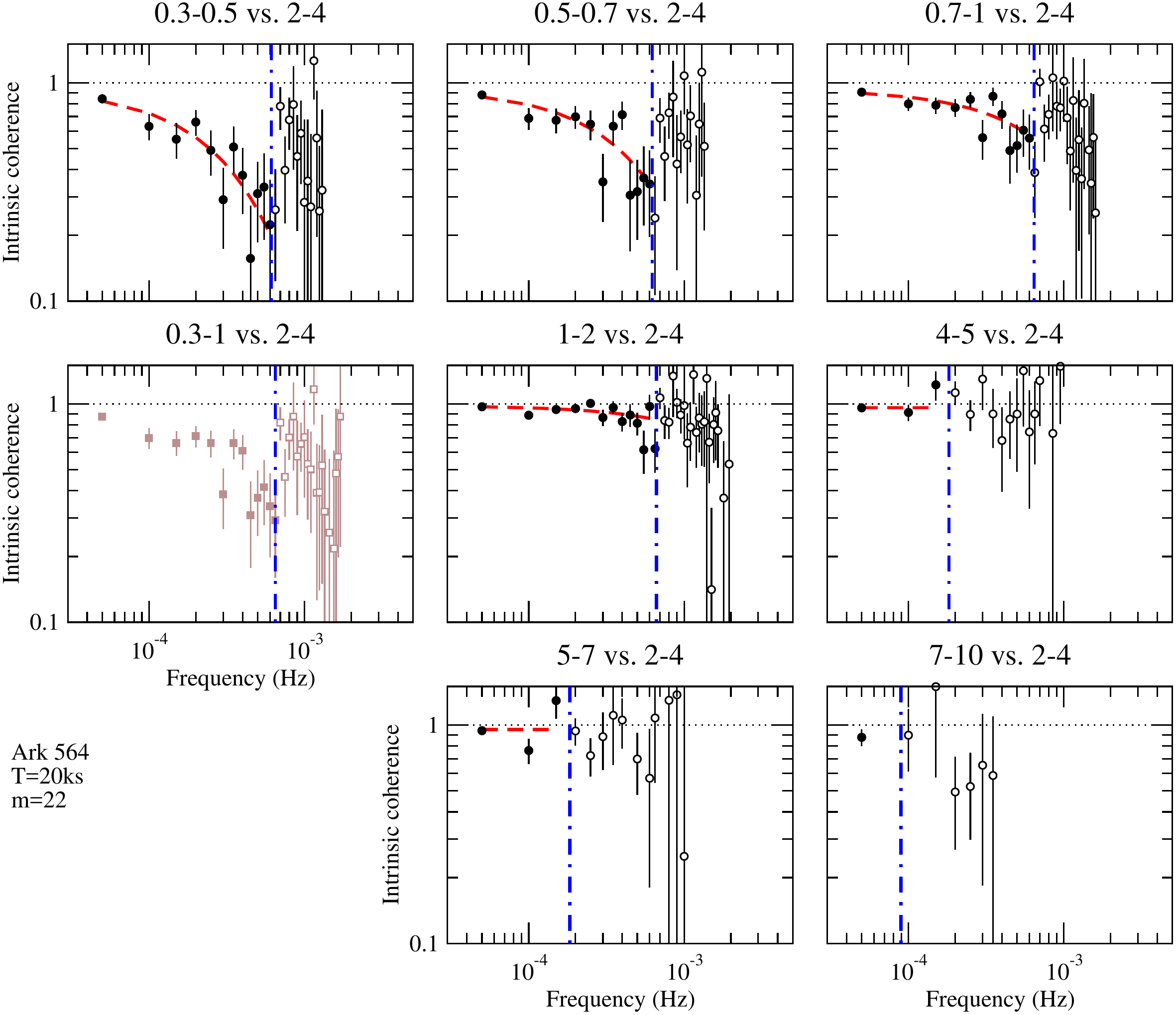}
 \caption{As in Fig. \ref{figb2}, for Ark 564.}
\label{figb12}
\end{figure*}

\begin{figure*}
 \includegraphics[width=350pt]{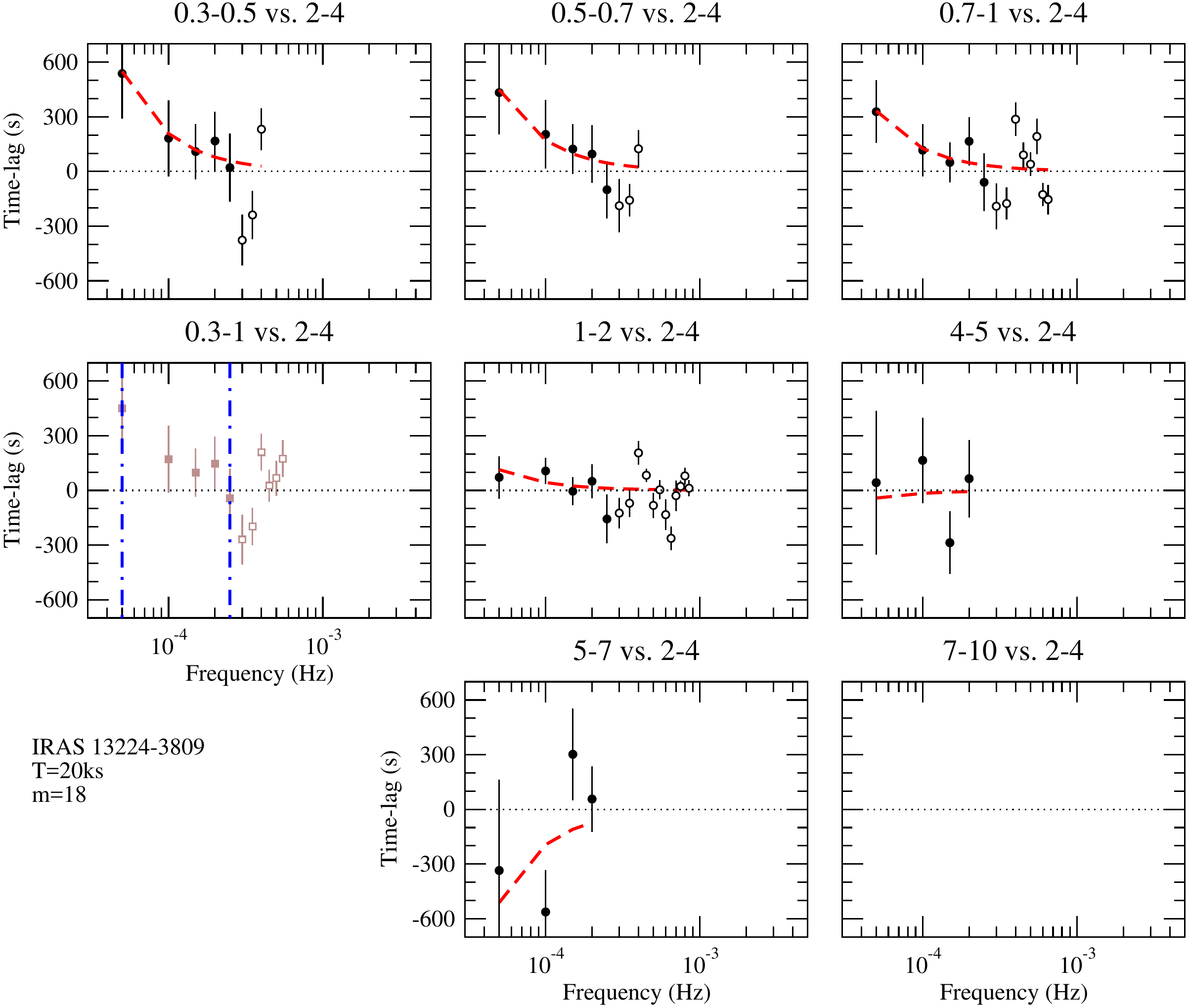}
 \caption{As in Fig. \ref{figb1}, for IRAS 13224--3809.}
\label{figb13}
\end{figure*}

\begin{figure*}
 \includegraphics[width=350pt]{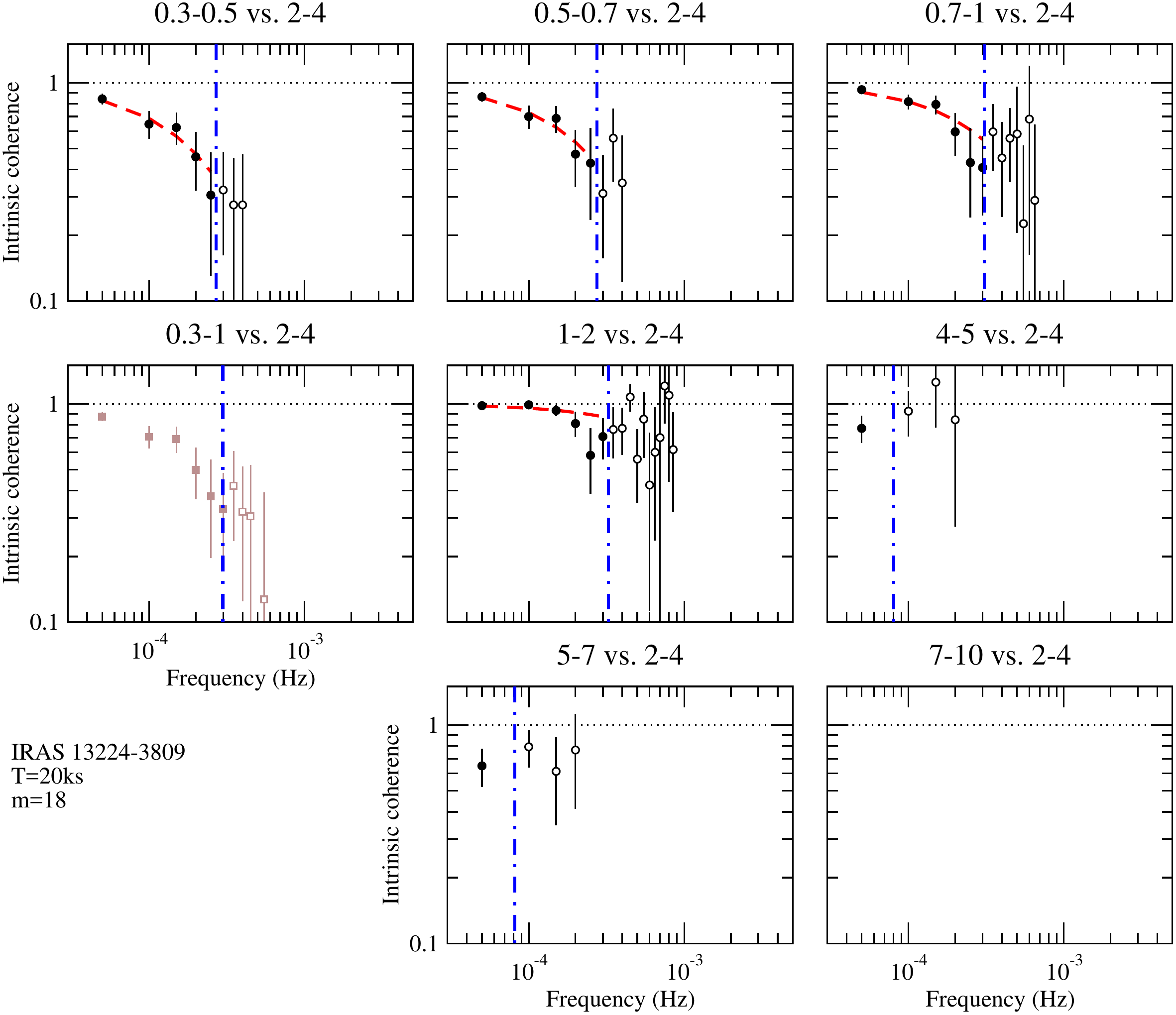}
 \caption{As in Fig. \ref{figb2}, for IRAS 13224--3809.}
\label{figb14}
\end{figure*}

\begin{figure*}
 \includegraphics[width=350pt]{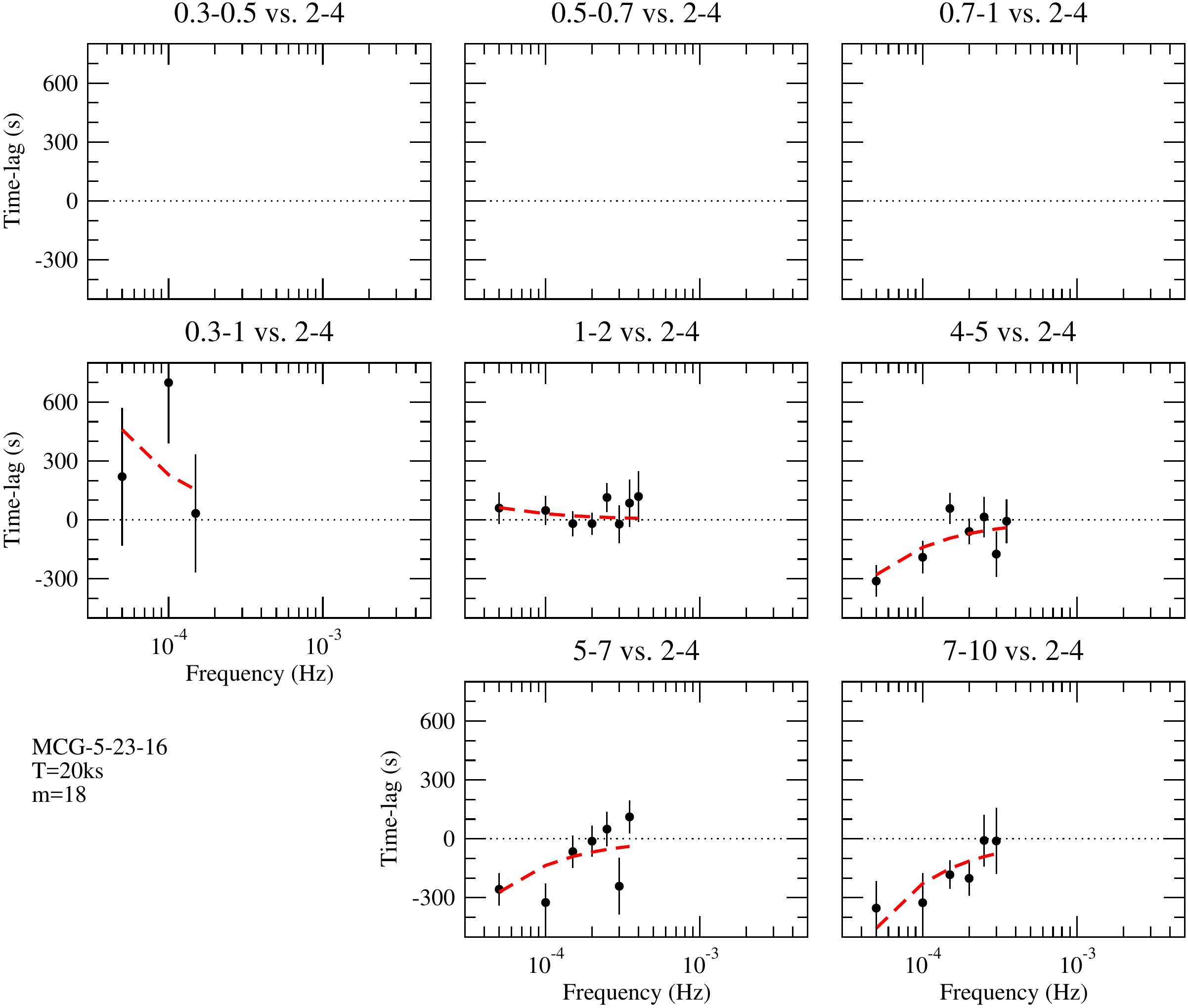}
 \caption{As in Fig. \ref{figb1}, for MCG--5-23-16.}
\label{figb15}
\end{figure*}

\begin{figure*}
 \includegraphics[width=350pt]{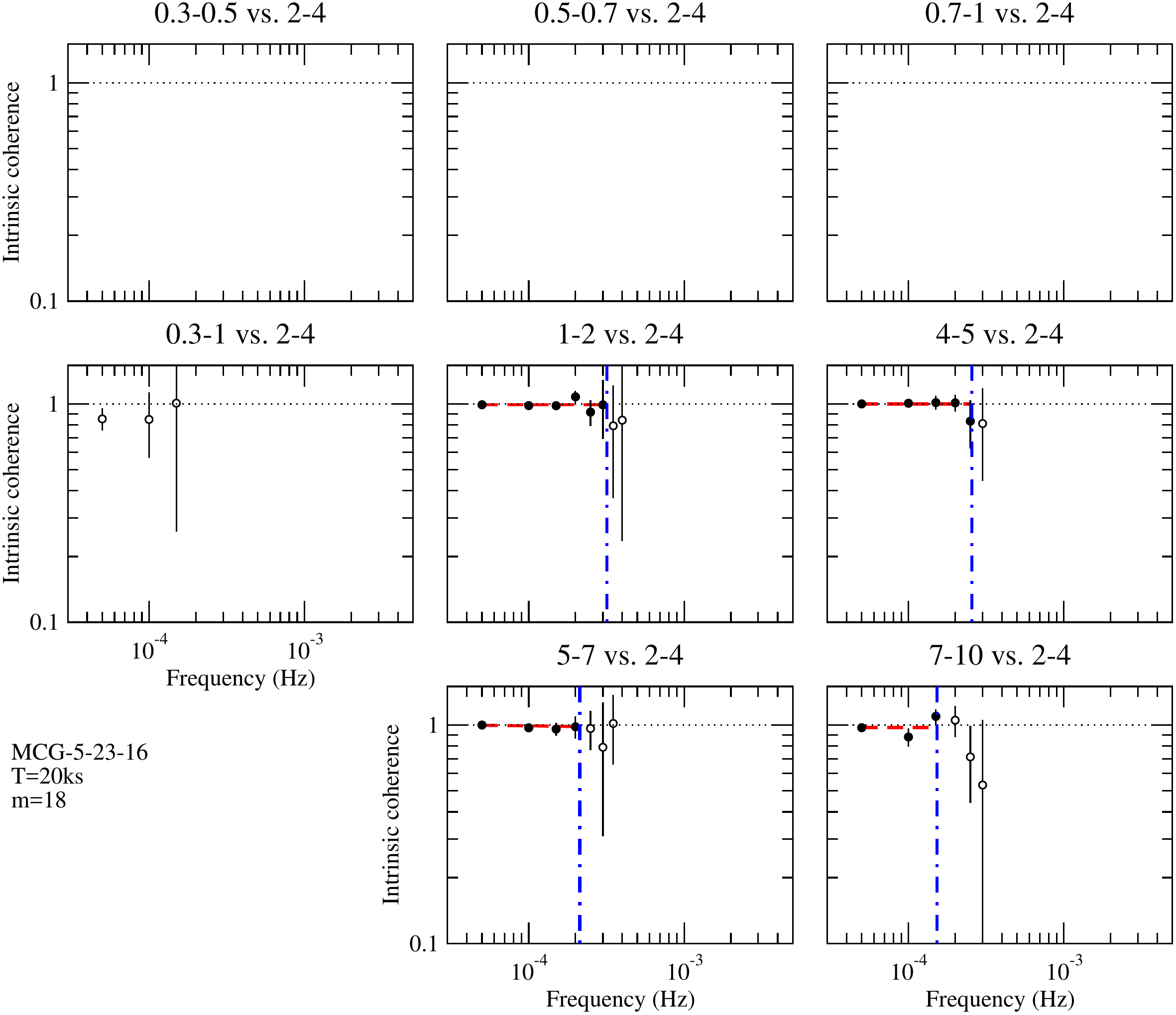}
 \caption{As in Fig. \ref{figb2}, for MCG--5-23-16.}
\label{figb16}
\end{figure*}

\begin{figure*}
 \includegraphics[width=350pt]{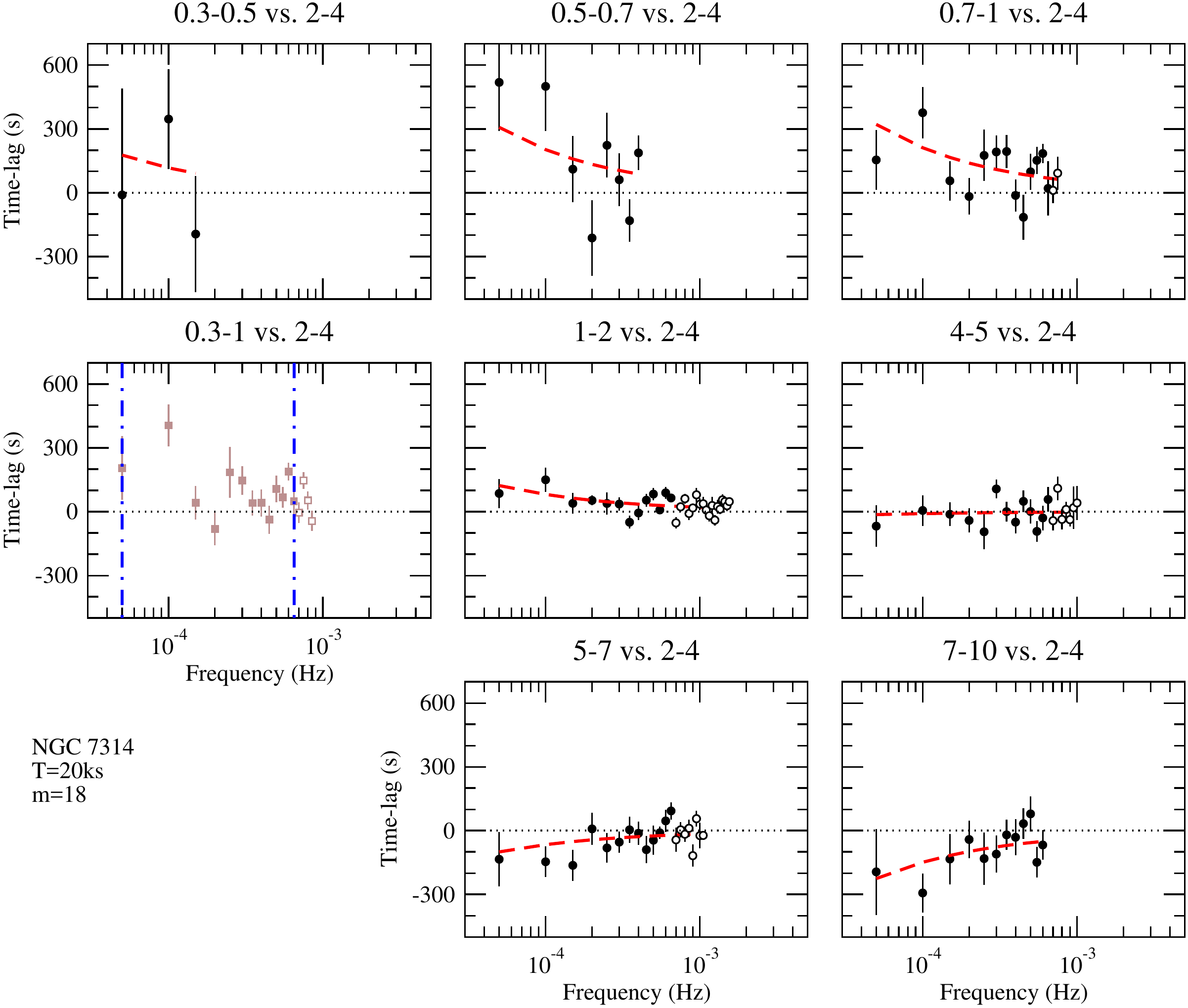}
 \caption{As in Fig. \ref{figb1}, for NGC 7314.}
\label{figb17}
\end{figure*}

\begin{figure*}
 \includegraphics[width=350pt]{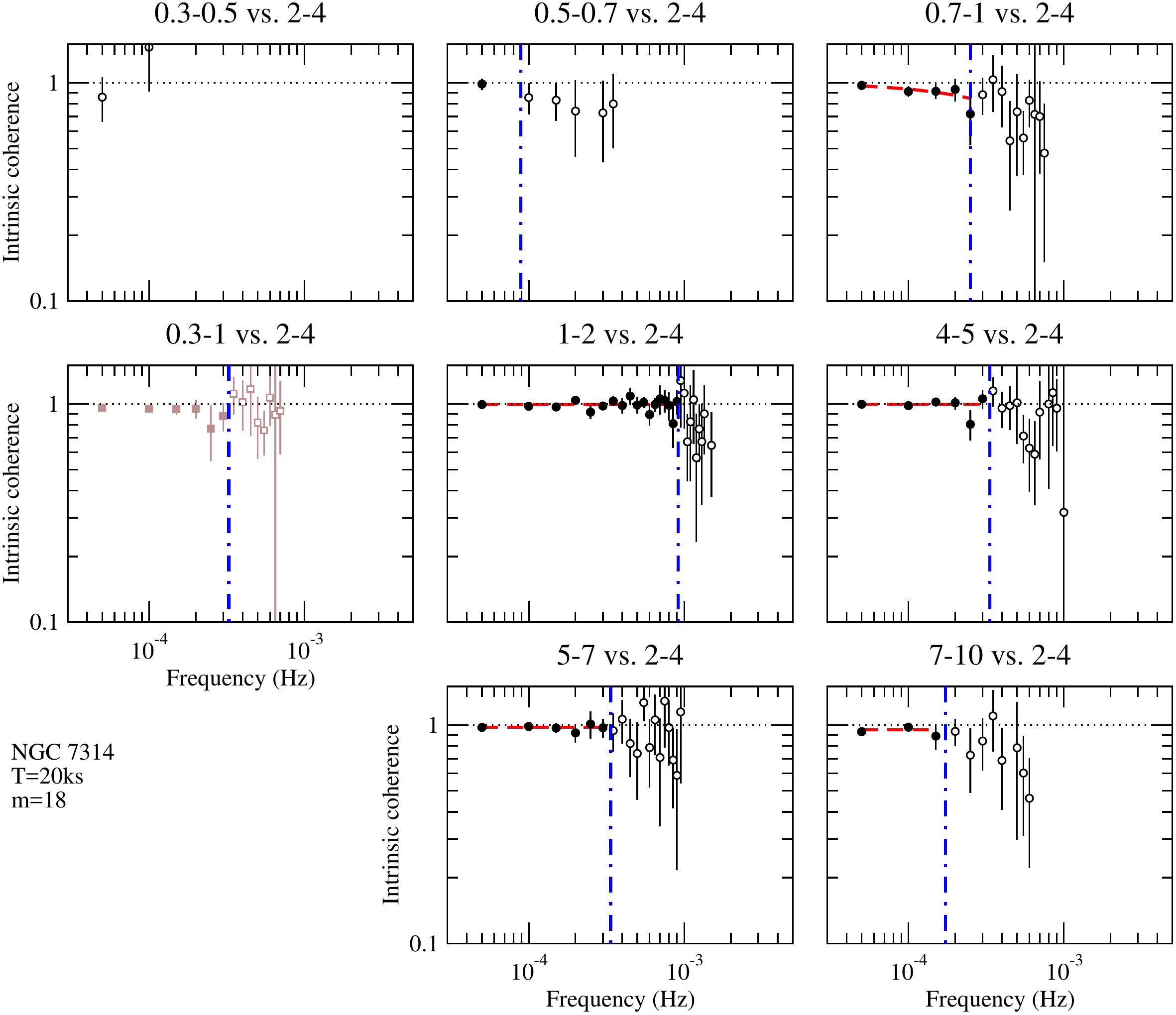}
 \caption{As in Fig. \ref{figb2}, for NGC 7314.}
\label{figb18}
\end{figure*}

\begin{figure*}
 \includegraphics[width=350pt]{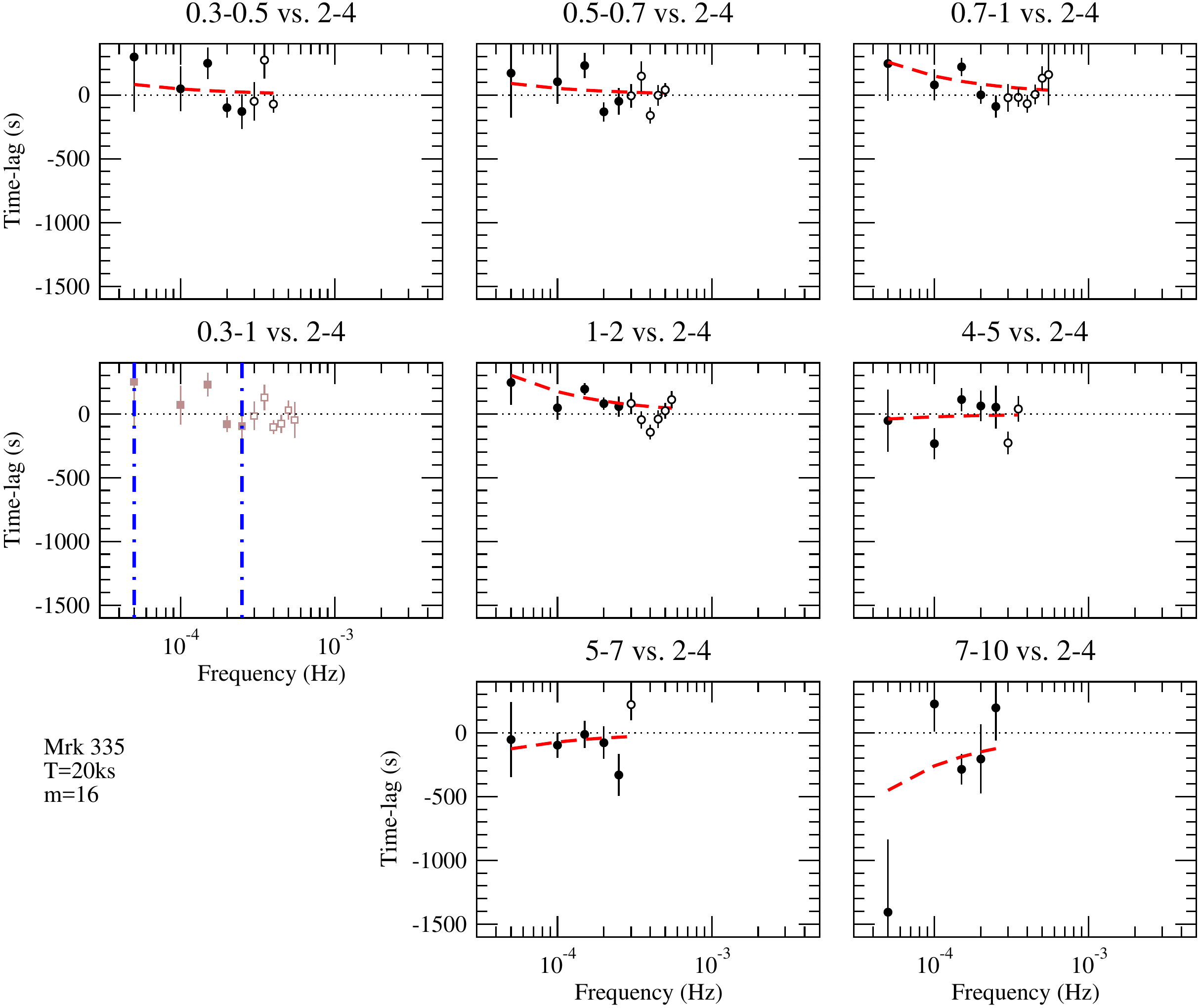}
 \caption{As in Fig. \ref{figb1}, for Mrk 335.}
\label{figb19}
\end{figure*}

\begin{figure*}
 \includegraphics[width=350pt]{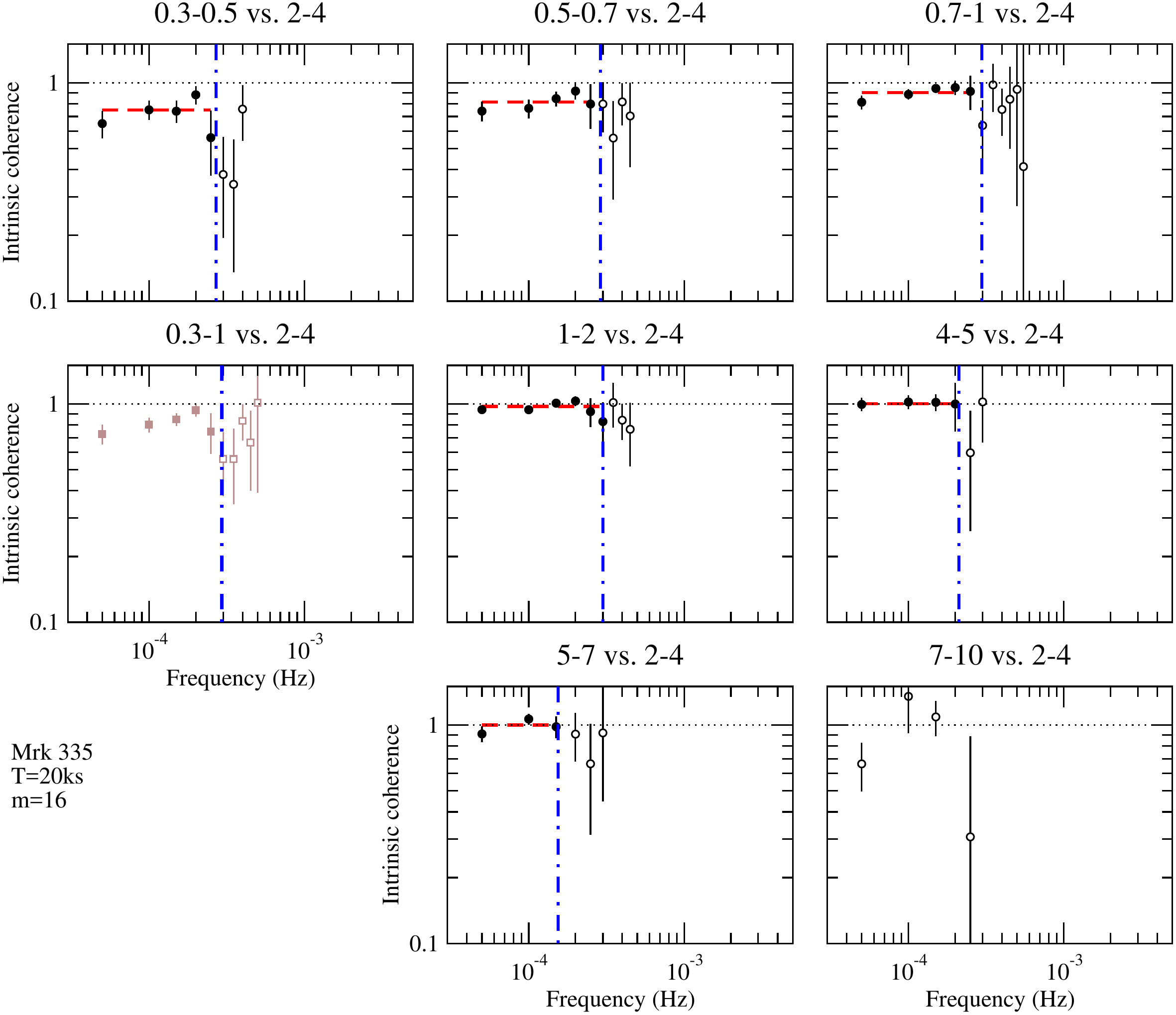}
 \caption{As in Fig. \ref{figb2}, for Mrk 335.}
\label{figb20}
\end{figure*}

%%%%%%%%%%%%%%%%%%%%%%%%%%%%%%%%%%%%%%%%%%%%%%%%%%

% Don't change these lines
\bsp	% typesetting comment
\label{lastpage}
\end{document}